\documentclass[%
manuscript=article,
]{achemso}
\SectionNumbersOn
\usepackage{amsmath,caption,subcaption,siunitx}

\usepackage{xr}
\usepackage{graphicx,float}
\usepackage{dcolumn}
\usepackage{bm,braket}

\usepackage{siunitx,booktabs}
\usepackage{soul,setspace}
\usepackage[dvipsnames]{xcolor}
\definecolor{calmblue}{RGB}{149, 186, 241}
\definecolor{uclagold}{rgb}{1.0, 0.7, 0.0}

\usepackage[colorlinks=true, citecolor=BrickRed, linkcolor=MidnightBlue, urlcolor=MidnightBlue]{hyperref}

\title{Correlated reaction coordinate motion produces non-additive rate enhancement for electron and energy transfer in multi-acceptor structures}

\author{Hanggai Nuomin}%
\affiliation{%
Department of Chemistry, Duke University,
	Durham, North Carolina 27708, United States%
}%
\author{Feng-Feng Song}
\affiliation{%
    The Institute for Solid State Physics, The University of Tokyo, Kashiwa, Chiba 277-8581, Japan%
}
\author{Peng Zhang}%
\affiliation{%
	Department of Chemistry, Duke University,
	Durham, North Carolina 27708, United States%
}%
\author{David N. Beratan}%
\email{david.beratan@duke.edu.}
\affiliation{%
	Department of Chemistry, Duke University,
	Durham, North Carolina 27708, United States%
}
\alsoaffiliation{%
	Department of Physics, Duke University,
	Durham, North Carolina 27708, United States%
}
\alsoaffiliation{%
	Department of
	Biochemistry, Duke University,
	Durham, North Carolina 27710, United States%
}

\begin{document}
\begin{abstract}
Molecular structures with multiple donor, bridge, or acceptor units can display quantum interference effects that influence electron and energy transfer (ET and EnT) rates. Recent experiments found a 4–5-fold increase in ET rates for donor-acceptor structures with two acceptors compared to one.  This result is surprising: simple classical or quantum analysis suggests a factor of two rate enhancement.
We analyze the coupling interactions in multiple acceptor systems and find that rate enhancements beyond additive effects arise from acceptor-acceptor interactions that:  1) shift the reaction free energy, 2) change the donor-acceptor couplings, and 3) alter the reaction-coordinate motion. Consideration of these effects explains the observed rates in multi-acceptor systems and suggests strategies to tailor energy and electron transfer kinetics.

\end{abstract}

\maketitle
\section{Introduction}
Electron transfer (ET) and energy transfer (EnT)  play a central  role in the molecular  sciences~\cite{strumpfer2012quantum,jakowetz2016controls,bakulin2012role}. 
Most models for ET/EnT involve one donor, one acceptor, and one mobile electron or exciton. 
Quantum coherence and many-body particle correlations can play a central role in the mechanisms of ET and EnT~\cite{lin2015two,phelan2019quantum,terai2023correlated,feskov2016efficiency,yuly2020universal,cina2023dynamics,fulton1961vibronic,sahu2025isolating}.
Recent experiments have found  unusual changes in ET rates  as multiple acceptors were introduced to the  structures~\cite{hartnett2017influence,phelan2019quantum,phelan2019quantum1,loong2022photoinduced,bancroft2021charge}.
For example, ET from a photoexcited anthracene donor to two chemically equivalent 1,4-benzoquinone acceptors was 4-5 times faster than transfer to a single acceptor at low temperatures~\cite{phelan2019quantum,phelan2019quantum1}. 
A 3-fold  ET rate enhancement was reported at ambient temperatures for a zinc porphyrin donor linked to two naphthalene-1,8:4,5-bis(dicarboximide) (NDI) acceptors compared to analogs with  one acceptor~\cite{bancroft2021charge}. 
Perylene-(NDI)\textsubscript{4} charge-transfer compounds were reported to have ET rates four times faster than the corresponding single-NDI species~\cite{fisher2023fundamental,fisher2024long}.
Interestingly, these observed rates are not just proportional to the number of acceptors, as would be expected from classical rate theory or from a theory for coherently mixed multi-acceptor states.  In the classical case, the multiple quenching channels would produce donor states with effective transfer rates linear in the acceptor concentration.  In the case of coherently mixed acceptor states, the coupling would scale as $N/\sqrt{N}$, and the rate would scale with $N$, as in the classical case.
The observed rate dependencies on acceptor number are likely caused by  interactions among the acceptors, which break the simplifying assumptions used in the simple analysis just mentioned.~\cite{phelan2019quantum,phelan2019quantum1,loong2022photoinduced,bancroft2021charge,fisher2023fundamental}  Understanding the detailed origins of  multi-acceptor effects on the transfer kinetics is the aim of this study.

Theoretical analysis of ET and EnT in structures with multiple donor (D), bridge, and acceptor (A) units~\cite{jang2004multichromophoric,jang2007multichromophoric,jang2011theory,scholes2001adapting,scholes2003long,scholes2000mechanism,taylor2018generalised}, and on  multi-particle transport have been described previously.~\cite{zarea2013decoherence,yuly2020universal,yuly2021efficient,powell2017redfield,lin2015two}
Yet, the mechanism(s) that produce the novel rate dependencies on acceptor number are poorly understood.~\cite{phelan2019quantum,phelan2019quantum1,hartnett2017influence,bancroft2021charge,lin2023ultrafast} Here, we explore two factors derived from the AA interactions that influence the  non-adiabatic transport.  These include AA interactions that: (1) change the reaction free energy and  may produce new coupling pathways, and (2) may lead to correlated  nuclear motion that can alter the nature of the activated-state crossing. 
When the final state is delocalized across many acceptors, a description of the DA coupling in terms of one lowest-order coupling pathway is not adequate, since many pathway amplitudes of similar magnitude must be summed. We describe a strategy to capture the key higher-order pathway interactions, as well as reaction coordinate effects that are characteristic of multi-acceptor structures.



The analysis described here uses a partial pathway summation approach to capture quantum interference in the effective coupling between initial and final states~\cite{mattuck1992guide}. This strategy allows dissection of  electronic, vibronic, and driving force effects on the transfer rates. 
Using this approach, we provide a unified framework to describe ET/EnT rates  in  multi-acceptor systems~\cite{phelan2019quantum,phelan2019quantum1,fisher2023fundamental,fisher2024long}.

Our simulations find that multi-acceptor systems are expected to show significantly accelerated ET/EnT rates compared to those of single-acceptor systems when: 
(1) strong electronic coupling exists between acceptors that changes the driving force (compared to that in single acceptor structures) and when these interactions establish many coupling pathways of similar magnitude, 
(2) low temperatures that preserve the vibrationally assisted electronic tunneling pathways between acceptors
and 
(3) the reorganization energy associated from AA structural changes on oxidation/reduction (reorganization energy effects) are large.
The enhanced AA reorganization energy in multi-A systems arises from, e.g., large changes in the dipole moment as a result of electron or energy transfer between acceptors. 
The enhanced AA reorganization energy can therefore produce non-additive transfer rates in systems with multiple acceptors.

We begin with an overview of non-adiabatic ET/EnT in multi-A structures. 
We describe coupling in multi-A structures using a vibronically coupled electron tunneling model~\cite{onuchic1986some,onuchic1991electron,coalson1994nonequilibrium,evans1998photoinduced}.
The model captures both the electronic interactions between donor and acceptor as well as  the correlated nuclear motion between the two donor-acceptor reaction coordinates.
We then explore how multiple acceptors and the
 interactions between them influence ET/EnT dynamics in the presence of system-bath interactions.
Finally, we develop guidelines for the design of  multi-acceptor ET/EnT systems based on the consideration of the key physical effects that arise in multi-A structures.

\section{Methods}
\label{sec:methods}
\begin{figure}
    \centering
    \includegraphics[width=\linewidth]{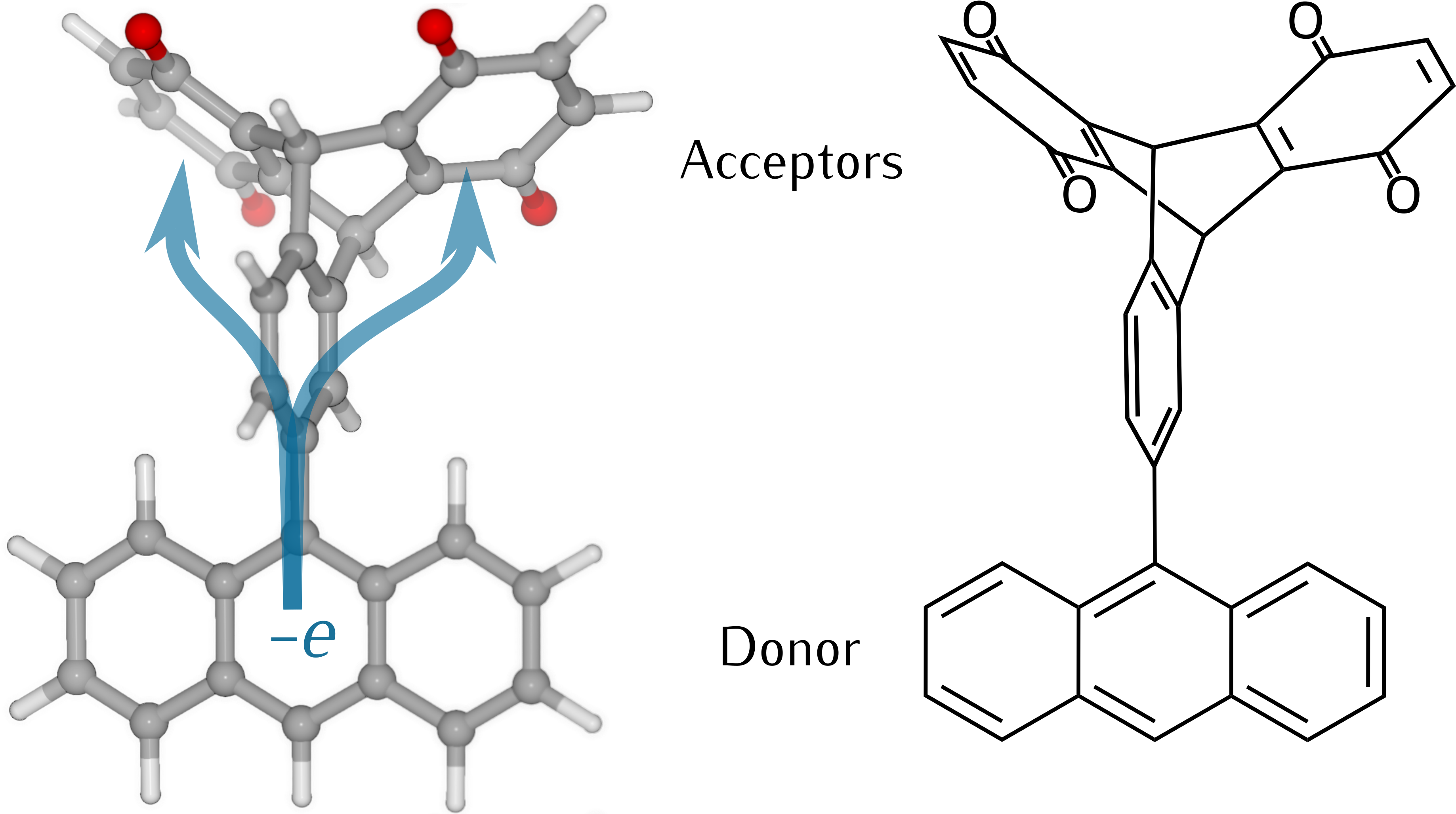}
    \caption{The donor-two-acceptor molecule studied in Ref.~\citenum{phelan2019quantum}. The donor is anthracene, the two acceptors are 1,4-benzoquinone. The donor and acceptors are connected by a 1,2-diisopropylbenzene linker.}
    \label{fig:molecule}
\end{figure}

Our model Hamiltonian is motivated by multi-A charge transfer structures found in the literature~\cite{phelan2019quantum,phelan2019quantum1}. A representative multi-A structure (Fig.~\ref{fig:molecule}) consists of one anthracene donor and two 1,4-benzoquinone acceptors. We focus on a model with three electronic states, corresponding to one donor-localized state and two acceptor-localized states: ($\ket{\mathrm{D}}$, $\ket{\mathrm{A}^1}$, and $\ket{\mathrm{A}^2}$).  Each of the three states is coupled to two sets of vibrational modes. One set of vibrational states is associated mainly with the donor and the first acceptor, while the other set is associated primarily  with vibrations of the donor and the second acceptor. 
Importantly, the first set of vibrational modes can enable electron/energy transfer to the second acceptor.  That is, the \emph{reaction coordinate} for ET/EnT to the second acceptor includes vibrational contributions from the first acceptor, and vice versa.
The reaction coordinates for ET/EnT include contributions from both sets of modes because the two acceptor species have strong electronic interactions.
The correlation between the charge transfer reaction coordinate motion  produces significant non-additive rate changes as the number of acceptors varies,
by amplifying contributions of higher order coupling pathways ({\it vide infra}).

\subsection{A Model for DAA Structures}
\label{sec:model-electronic}
We assume that the electronic interaction ($\hat{V}$) between the donor and acceptor diabatic electronic states are independent of nuclear degrees of freedom (i.e., we make a Condon approximation). The model Hamiltonian for a one-donor/two-acceptor ($\mathrm{DA^1A^2}$) system is:
\begin{align}
    \hat{H} = & \hat{H}_0 + \hat{V} \label{eq:hamiltonian}\\
    \hat{H}_0 = & \ket{\mathrm{D}}\bra{\mathrm{D}}\otimes \hat{H}_{\mathrm{D}}^{(n)} + \ket{\mathrm{A^1}}\bra{\mathrm{A^1}}\otimes H_{\mathrm{A^1}}^{(n)} \\
              & + \ket{\mathrm{A^2}}\bra{\mathrm{A^2}}\otimes H_{\mathrm{A^2}}^{(n)} \notag \\
    \hat{V} = & V_{\mathrm{DA^1}}\ket{\mathrm{D}}\bra{\mathrm{A^1}} + V_{\mathrm{DA^2}}\ket{\mathrm{D}}\bra{\mathrm{A^2}} \\
              & + V_{\mathrm{A^1A^2}}\ket{\mathrm{A^1}}\bra{\mathrm{A^2}} + h.c. \notag
\end{align}
The nuclear Hamiltonians, $\hat{H}_D^{(n)}$, $\hat{H}_{\mathrm{A^1}}^{(n)}$, and $\hat{H}_{\mathrm{A^2}}^{(n)}$, describe nuclear motion  when the electron is  in the $\ket{\mathrm{D}}$ (donor), $\ket{\mathrm{A^1}}$ (acceptor 1), and $\ket{\mathrm{A^2}}$ (acceptor 2) electronic states, respectively. The electronic couplings between the diabatic states are $V_{\mathrm{DA^1}}$, $V_{\mathrm{DA^2}}$, and $V_{\mathrm{A^1A^2}}$.  
The acceptor-acceptor coupling ($V_{\mathrm{A^1A^2}}$) influences the driving force for ET/EnT, enables non-additive ET/EnT coupling pathways (see the caption of Fig.~\ref{fig:paths} for a definition of pathways), and mixes the two sets of vibrational states.
The reaction free energy for ET/EnT from a single donor to a single acceptor (i.e.,  D to A\textsuperscript{1} or,  equivalently, D to A\textsuperscript{2}) is defined as $E_{\mathrm{D}}-E_{\mathrm{A}^1}$.
Here, A\textsuperscript{1} and A\textsuperscript{2} are assumed to be degenerate throughout: $E_{\mathrm{A}^1} = E_{\mathrm{A}^2}$.

The coupling between donor and acceptor diabatic electronic states is assumed to be sufficiently weak to produce non-adiabatic ET/EnT~\cite{evans1998photoinduced,newton1991quantum}.
The assumption of weak coupling is supported by quantum chemical analysis in the anthracene-1,4-benzoquinone donor-acceptor structure in Fig.~\ref{fig:molecule}, with coupling values of $\sim$0.01\,eV (see Sec.~\ref{sec:estimation-of-electronic-parameters}).
In the weak coupling regime, a perturbative approach can be used to estimate the effective DA coupling. A bridge Green's function strategy is particularly effective~\cite{skourtis2004inelastic}. This strategy can account for bridge-mediated superexchange, even when explicit bridge states are not enumerated~\cite{skourtis1999theories}.
The partial summation theory of \emph{vibronic} pathways used in our study offers a different perspective. Here, we define vibronic pathways as coupling pathways that include the nuclear Franck-Condon overlaps 
between the vibrational states coupled to the transfer reaction~\cite{skourtis1999theories}. These pathways account for how the vibrational energy levels of the donor and acceptor influence the efficiency of electron or energy transfer.

The dephasing of vibronic electron transfer pathways can enable electron transfer to occur in otherwise symmetry-forbidden structures~\cite{skourtis2004inelastic}.
However, the influence of multiple donors, bridges, and acceptors on the transfer rate is poorly understood when ET/EnT is symmetry allowed.
The analysis described here uses a partial summation of the donor survival amplitude after propagating along ET/EnT pathways with multiple acceptors. All vibronic pathways are summed coherently (i.e., no vibronic pathway dephasing is included) to compute the ET/EnT rate. The pathway expansion allows us to identify the origins of non-additive rate effects derived from multiple acceptors, especially those caused by AA reorganization energy effects that accelerate ET/EnT. 
The AA reorganization energy produces pathway interference effects that lead to non-additive transfer rates as the number of acceptors increases.

\paragraph{Electronic Interactions.} We set the  effective DA electronic coupling interactions in the DAA structure to be equal:  $V_{\mathrm{DA^1}}=V_{\mathrm{DA^2}}=V_{\mathrm{DA}}$, (i.e., the two sets of pathways are equivalent), motivated by the two-acceptor experiments of Phelan et al.~\cite{phelan2019quantum}, 

\paragraph{Vibrational Hamiltonian.}
We describe the nuclear motion coupled to ET/EnT using the shifted harmonic oscillator model of  vibronic coupling.
For the two-acceptor system introduced in Sec.~\ref{sec:methods}, it is useful to separate the vibrations into two sets, $\sum_{n}\omega_n a_n^\dagger a_n$ and $\sum_{n}\omega_n b_n^\dagger b_n$. 
These sets of vibrational states correspond to the donor-acceptor-1 unit and the donor-acceptor-2 unit, respectively.
The correlation between the motion of these two reaction coordinates influences the rate, as described below. 
The correlation between the reaction coordinates describes the extent to which vibrational modes are shared between them, and whether those shared modes are in or out of phase. Positive correlation means that the motions of  modes coupled to both A\textsuperscript{1} and A\textsuperscript{2} are in phase, while negative correlation means that the modes coupled to A\textsuperscript{1} and A\textsuperscript{2} are out of phase.

The nuclear Hamiltonians are given by
\begin{equation}
    \begin{aligned}
        \hat{H}^{(n)}_\mathrm{D} = & E_\mathrm{D} + \sum_n \omega_n \hat{a}_n^\dagger \hat{a}_n + \sum_n \omega_n \hat{b}_n^\dagger \hat{b}_n 
        \\
        \hat{H}_{\mathrm{A}^1}^{(n)} = & E_\mathrm{A^1} + \lambda_\mathrm{DA^1} + \sum_{n} \omega_n a^\dagger_{n} a_{n} + \sum_{n} \omega_n b^\dagger_{n}b_{n} \\
            &+ \alpha\sum_n s_n (a^\dagger_n + a_n)  + \beta \sum_n s_n (b^\dagger_{n} + b_{n}) \\
        H_{\mathrm{A}^2}^{(n)} = & E_\mathrm{A^2} + \lambda_{\mathrm{DA}^2} + \sum_{n} \omega_n a^\dagger_{n} a_{n} + \sum_{n} \omega_n b^\dagger_{n}b_{n} \\
            & + \beta \sum_n s_n (a^\dagger_n + a_n)
                   + \alpha \sum_n s_n (b^\dagger_{n} + b_{n}). 
    \end{aligned}
    \label{eq:hamiltonian-nuclei-3}
\end{equation}
Here, $E_\mathrm{D}$, $E_\mathrm{A^1}$, and $E_{\mathrm{A^2}}$ are the electronic energies of the $\mathrm{D}$, $\mathrm{A^1}$, and $\mathrm{A}^2$ states, respectively; $\omega_n$ and $s_n$ are the frequencies and the vibronic couplings of the $n^\mathrm{th}$  oscillator. The vibronic couplings are all placed on the $\mathrm{A^1}$ and $\mathrm{A^2}$ states.  Thus, the oscillators are centered at their equilibrium positions for the D\textsuperscript{*}
state (the photo-excited donor) at $t=0$.  The two independent sets of oscillators are defined to have identical frequencies  
$\{\omega_n\}$ and vibronic couplings $\{s_n\}$, to reflect the symmetry of the two  acceptors.
The oscillators in group I ($\sum_n\hat{a}_n^\dagger\hat{a}_n$) are coupled mainly to acceptor $\mathrm{A^1}$ (largely the reaction coordinate for the $\mathrm{D} {\to} \mathrm{A^1}$ ET/EnT), and group II ($\sum_n\hat{b}_n^\dagger\hat{b}_n$) is coupled mainly to the acceptor $\mathrm{A^2}$. 
The parameters $\alpha$ and $\beta$
describe how strongly the two reaction coordinates are shared, as well as  the correlation between the motion of the two reaction coordinates.
For example, $\alpha=1,\beta=0$ corresponds to two independent reaction coordinates,
and $\alpha=\beta=\sqrt{2}/2$ represents two fully positively correlated reaction coordinates (i.e., $\mathrm{D}{\to}{\mathrm{A}^1}$ and $\mathrm{D}{\to}{\mathrm{A}^2}$ reactions share a single reaction coordinate).
The values $\alpha$ and $\beta$ can vary from $-1$ to $1$ and are constrained by $\alpha^2 + \beta^2=1$; this condition guarantees that the reorganization energies associated with ET/EnT  from $\mathrm{D}$ to $\mathrm{A}^1$ and from $\mathrm{D}$ to $\mathrm{A}^2$ are identical.
The reaction coordinates can be represented as two unit vectors in the $\alpha$-$\beta$ plane, as shown in Fig.~\ref{fig:illustration}. The angle between these vectors defines the correlation between the two reaction coordinates.
A metric for the correlation between the two reaction coordinates for  $\mathrm{D \to A^1}$ and $\mathrm{D \to A^2}$ ET/EnT is~\cite{tiwari2017electronic,najbar1996solvent}
$\sum_{i,j}\braket{\hat{x}_{i,\mathrm{A^1}}(t)\hat{x}_{j,\mathrm{A^2}}(t)}\big|_{t=0}$. This metric is the value of a two-time correlation function for nuclear positions of acceptors 1 and 2 at time zero, where $x_{i,\mathrm{A^1}}$ is the $i$-th vibrational mode component of the reaction coordinate that modulates ET/EnT from the donor to acceptor 1, and $x_{i,\mathrm{A^2}}$ is defined similarly for acceptor 2.

The vibronic couplings $\{s_n\}$ and vibrational frequencies $\{\omega_n\}$ are approximated with a Lorentzian spectral density
$
        J(\omega)=\pi\sum_n s_n^2\delta(\omega-\omega_n)=2\lambda \frac{\eta \omega \Omega^2}{(\Omega^2 -\omega^2)^2 + \eta^2 \omega^2}
$
where $\lambda$ is the reorganization energy, $\Omega$ is the mean
frequency of the spectral density distribution, and $\eta$ is a friction coefficient~\cite{xie2013calculation,onuchic1986some}.
With the spectral density (hence the vibronic couplings $\{s_n\}$ and vibrational frequencies $\{\omega_n\}$) specified, the correlation between  vibrational Hamiltonians $H_\mathrm{A^1}^{n}$ and $H_\mathrm{A^2}^{n}$ is determined by the scaling factors $\alpha$ and $\beta$. Examples of correlation between the two reaction coordinates are presented below, and these examples are used in the rate  simulations of  Sec.~\ref{sec:results}.

There are three diabatic electronic states in our model: $\ket{\mathrm{D}}$, $\ket{\mathrm{A}^1}$, and $\ket{\mathrm{A}^2}$. 
We define the acceptor-acceptor reorganization energy as the energy cost to move an electron from A\textsuperscript{1} to A\textsuperscript{2} with all nuclear coordinates fixed. With the current Hamiltonians [see Eq.~\eqref{eq:hamiltonian}-~\eqref{eq:hamiltonian-nuclei-3}], this reorganization energy is 
\begin{equation}
    \lambda^\mathrm{reorg}_{\mathrm{A^1},\mathrm{A^2}} = \sum_n \Big[\frac{(\alpha s_n-\beta s_n)^2}{\omega_n} + \frac{(\beta s_n-\alpha s_n)^2}{\omega_n}\Big].
    \label{eq:reorganization-energy-definition}
\end{equation}
The acceptor-acceptor reorganization energy depends on the values of $\alpha$ and $\beta$; this reorganization energy reflects the correlation between the two reaction coordinates.  As such, the acceptor-acceptor reorganization energy plays a large role in determining the rate difference as the number of acceptors changes. We find ({\it vide infra}) that larger acceptor-acceptor reorganization energies 
enhance non-additive rate effects.
In the next section, we describe four representative cases of reaction coordinate correlation.

\subsection{Reaction coordinate correlation}
\label{sec:model-vibration}
\begin{figure}
    \centering
    \includegraphics[width=.8\linewidth]{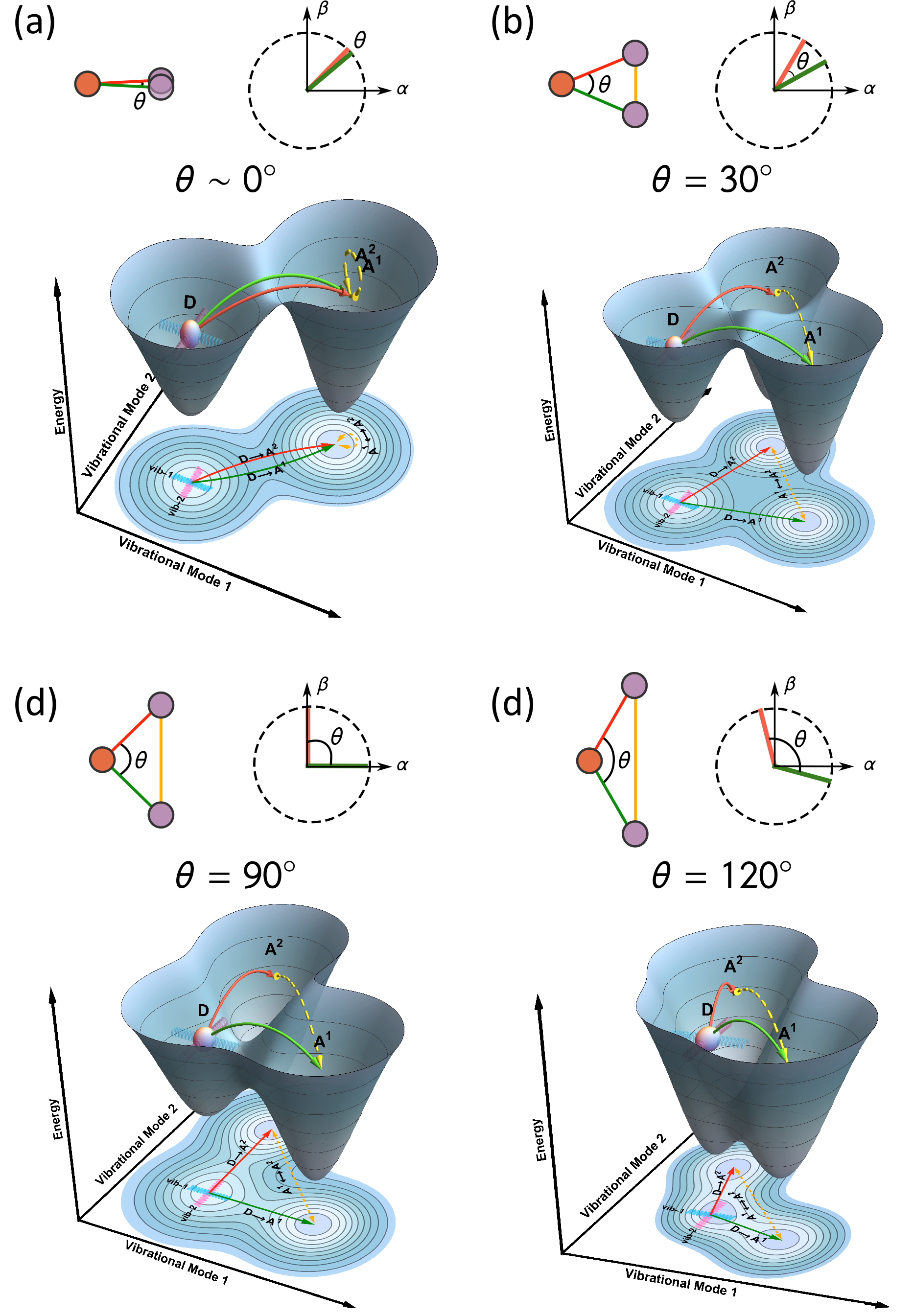}
    \caption{Four cases of donor-acceptor vibrations. Each has a different angle $\theta$ between the  reaction coordinates ($\mathrm{D}{\to}\mathrm{A}^1$ in green and $\mathrm{D}{\to}\mathrm{A}^2$ in red). The corresponding $\alpha$ and $\beta$ values for the two reaction coordinates are represented on the unit circle in the $\alpha$-$\beta$ plane.
    (a) $\theta=0\si{\degree}$. Fully reaction coordinates (i.e., two reaction coordinates move in phase upon ET/EnT to $\mathrm{A}^1$ and $\mathrm{A}^2$ simultaneously.)
    (b) $\theta=30\si{\degree}$. Partially correlated.
    (c) $\theta=90\si{\degree}$. Uncorrelated reaction coordinates for the two ET/EnT reactions, i.e., the two reaction coordinates are orthogonal to each other.
    (d) $\theta=120\si{\degree}$. Negatively correlated reaction coordinates for the two ET/EnT reactions.%
    }
    \label{fig:illustration}
\end{figure}

Using the model Hamiltonian in Eqs.~\eqref{eq:hamiltonian}-\eqref{eq:hamiltonian-nuclei-3}, we study the influence of correlated reaction coordinate motions on ET/EnT reactions with two acceptors. We explore four kinds of  reaction coordinate correlation (Fig.~\ref{fig:illustration}).

\paragraph{Fully, positively correlated reaction coordinates for the two acceptors.}
When $\alpha=\beta$, the vibrations of $\mathrm{A^1}$ and $\mathrm{A^2}$ are fully correlated.
That is, the two ET/EnT pathways ($\mathrm{D} \to \mathrm{A}^1$ and $\mathrm{D} \to \mathrm{A}^2$) share a single reaction coordinate (see  Fig.~\ref{fig:illustration}(a)). In this case, the vibrational Hamiltonians for the two acceptors, $\hat{H}{\mathrm{A}^1}^{(n)}$ and $\hat{H}{\mathrm{A}^2}^{(n)}$, are identical and there are no shifts in the equilibrium positions of the oscillators when a vertical transition occurs between the potential energy surfaces of the $\mathrm{A}^1$ and $\mathrm{A}^2$ states.
The reorganization energy for transfer between the two acceptors is $\lambda^\text{reorg}_{\mathrm{A^1},\mathrm{A^2}}=0$.

For identical donor-acceptor couplings ($V_{D\mathrm{A}^1}=V_{\mathrm{DA^2}}$, see Sec.\ref{sec:model-electronic}) and identical acceptor vibrational Hamiltonians ($H_{\mathrm{A}^1}^{(n)}=H_{\mathrm{A^2}}^{(n)}$, in the same Hilbert space), the two acceptors share the same vibrational spectrum and equilibrium geometries.
Eq.~\eqref{eq:hamiltonian} for $\mathrm{DA^1A^2}$ can be transformed into an effective two-state Hamiltonian, using symmetric and anti-symmetric combinations of the acceptor electronic states: $\ket{+}=\frac{1}{\sqrt{2}}(\ket{\mathrm{A^1}}+\ket{\mathrm{A^2}})$ and $\ket{-}=\frac{1}{\sqrt{2}}(\ket{\mathrm{A^1}}-\ket{\mathrm{A^2}})$. In this special regime, a two-acceptor ET/EnT model can be constructed with the new basis and solved analytically, where the Hamiltonian $\hat{H}_0+\hat{V}$ is
\begin{subequations}
    \label{eq:two-state}
    \begin{align}
    H_0 = & \ket{\mathrm{D}}\bra{\mathrm{D}}\otimes \hat{H}_{\mathrm{D}}^{(n)} + \ket{+}\bra{+}\otimes \hat{H}_{\mathrm{A^1}}^{(n)} \\
        & + \ket{-}\bra{-}\otimes \hat{H}_{\mathrm{A^2}}^{(n)} \notag \\
    V = & \sqrt{2}V_{\mathrm{DA}}\big(\ket{\mathrm{D}}\bra{+}+\ket{+}\bra{\mathrm{D}}\big) + h.c. \\
        & + V_{\mathrm{DA}} \big(\ket{+}\bra{+} - \ket{-}\bra{-}\big). \notag 
\end{align}
\end{subequations}
The electronic state $\ket{\mathrm{D}}$ does not interact with the anti-symmetric combination of the acceptor electronic states $\ket{-}$, and there is no interaction between the two effective acceptor electronic states, $\ket{+}$ and $\ket{-}$. The electronic coupling between $\ket{\mathrm{D}}$ and $\ket{+}$ is $\sqrt{2}V_{\mathrm{DA}}$. The electronic energy of $\ket{+}$ is larger than that of $\ket{\mathrm{A^1}}$ or $\ket{\mathrm{A^2}}$ by  $V_{\mathrm{DA^1}}$ (if $V_{\mathrm{DA}^1} >0$). In the effective two-state (D and $\ket{+}$) system, the driving force for the $\mathrm{D} {\to} \ket{+}$ transfer is less than for ET/EnT from D to A\textsuperscript{1} (or A\textsuperscript{2}). The vibrational Hamiltonians ($\hat{H}_{\mathrm{D}}^{(n)}$, $\hat{H}_{\mathrm{A^1}}^{(n)}$, and $\hat{H}_{\mathrm{A^2}}^{(n)}$) describe vibrational degrees of freedom associated with both the donor-acceptor molecule and the solvent. The nuclear shifts associated with the donor electronic excited state (D\textsuperscript{*}) are taken to be zero, and the initial vibrational states are assumed to be in a thermal distribution on the D\textsuperscript{*} potential energy surface.



\paragraph{Uncorrelated reaction coordinates.}
The motions of the two sets of nuclear modes (i.e., the two donor-to-acceptor reaction coordinates) are uncorrelated when $\alpha=1$ and $\beta=0$ (or vice versa). This means the two reaction coordinates have no overlap and are orthogonal to each other, as depicted in Fig.~\ref{fig:illustration}(c). 
In this regime, the two sets of vibrations, $\{\hat{a}^\dagger_n \hat{a}_n\}$ and $\{\hat{b}^\dagger_n \hat{b}_n\}$, are independent of each other. The $\mathrm{D} {\to} \mathrm{A^1}$  reaction coordinate is orthogonal to the $\mathrm{D} {\to} \mathrm{A^2}$ reaction coordinate. The $\mathrm{A^1} {\to} \mathrm{A^2}$ reaction coordinate is defined as the difference between the $\mathrm{D} {\to} \mathrm{A^1}$ and $\mathrm{D} {\to} \mathrm{A^2}$ reaction coordinates. The reorganization energy for the transfer from $\mathrm{A^1}$ to $\mathrm{A^2}$ (or vice versa) is $2\lambda$. In this case of uncorrelated reaction coordinates,
we expect the rate to depend strongly on the acceptor-acceptor vibrational reorganization. The acceptor-acceptor coupling $V_\mathrm{A^1A^2}$ will also  change the reaction free energy compared to that for a single acceptor.

The uncorrelated case described above divides the parameter regime of reaction-coordinate correlation and acceptor-acceptor reorganization energy into two categories: (1) a partially correlated regime, where the correlation between the two reaction coordinates for $\mathrm{D}{\to}\mathrm{A^1}$ and $\mathrm{D}{\to}\mathrm{A^2}$ reactions is weaker than in the fully correlated case, but stronger than in the uncorrelated case (the acceptor-acceptor reorganization energy lies between that of the fully correlated and uncorrelated cases); and (2) the anti-correlated case, where the two reaction coordinates have opposite phases during inter-acceptor ET, producing a reorganization energy greater than that of the uncorrelated case. These two categories are described below.

\paragraph{Partially correlated reaction coordinates.}
The partially correlated regime lies between the fully positively correlated and the uncorrelated cases. In this regime, the two reaction coordinates share vibrational modes, and the angle between the two reaction coordinates in the $\alpha$-$\beta$ plane falls between 0\textdegree and  90\textdegree, as indicated in Fig.~\ref{fig:illustration}(b).
Here, $\alpha \neq \beta$, with both $\alpha$ and $\beta$ greater than zero.
The acceptor-acceptor reorganization energy can have values in the range between $0\lambda$ and $2\lambda$, where the two extreme values correspond to  positively fully correlated and to uncorrelated cases, respectively.
This intermediate coupling scenario is referred to as the partially correlated case, as the two donor-to-acceptor ET/EnT reaction coordinates share vibrational modes.

In our simulations, we use the parameters $\alpha = \sqrt{3/4}$ and $\beta = \sqrt{1/4}$. The acceptor-acceptor reorganization energy for $\mathrm{A}^1$ to $\mathrm{A^2}$ ET/EnT is $(2 - \sqrt{3})\lambda \approx 0.27 \lambda$, as derived from Eq.\eqref{eq:reorganization-energy-definition}. This reorganization energy is much smaller than the donor-acceptor reorganization energy, suggesting a small influence of the acceptor-acceptor reorganization energy on the rate. However, as we approach the uncorrelated limit, the influence of acceptor-acceptor reorganization energy becomes larger. In the current case of partially correlated reaction coordinates., as shown in the numerical simulations below, the only appreciable difference in ET/EnT rates between one and two acceptor systems arises from the acceptor-acceptor electronic coupling, $V_\mathrm{A^1A^2}$, which shifts the reaction free energy ($\Delta G$) by $|V_\mathrm{A^1A^2}|$. just as in the exactly solvable case (i.e., the fully correlated regime, above).

\paragraph{Anti-correlated reaction coordinates.}
When  $\alpha>0 $ and $\beta<0$, the motion of the reaction coordinates is anti-correlated, i.e., the shared vibrational modes
move out of phase with each other.
The correlation function $\sum_{ij}\braket{\hat{x}_i^{(\mathrm{A}^1)} \hat{x}_i^{(\mathrm{A}^2)}}_\beta < 0$ and the two reaction coordinates move out of phase. 
That means that the equilibrium positions of the $\mathrm{A}^1$ and $\mathrm{A}^2$ potential energy wells are on the opposite side of the donor position (see Fig.~\ref{fig:illustration}(d)).
This scenario is illustrated in Fig.~\ref{fig:illustration}(d), where the two unit vectors represent  reaction coordinates with an angle greater than 90\textdegree.
The acceptor-acceptor reorganization energy is substantial since the equilibrium positions of the A\textsuperscript{1} and A\textsuperscript{2} potential wells are separated by a large distance.
In our simulations, we used 
$\alpha=\frac{\sqrt{2+\sqrt{3}}}{2} (\approx 0.97)$ and $\beta=-\frac{\sqrt{2-\sqrt{3}}}{2} (\approx -0.26)$, such that the reorganization energy for the $\mathrm{A}^1\leftrightarrow\mathrm{A}^2$ transfer is $3\lambda$, which is a representative case in the negatively correlated regime.
This regime yields the largest acceptor-acceptor reorganization energy. The numerical simulations presented below indicate that large acceptor-acceptor reorganization energy plays a crucial role in determining the non-additive rate enhancement. As a result, the anti-correlated case exhibits the most significant rate enhancement.

In summary,  the vibrations of the two acceptors in the fully correlated case are in phase, sharing the same reaction coordinate.  This situation eliminates any acceptor-acceptor vibrational shifts as a result of A\textsuperscript{1}-A\textsuperscript{2} interaction. As such, the fully correlated case can be solved analytically and it does not display an dependence of the rate on the number of acceptors, apart from effects of shifting the reaction free energy as a consequence of acceptor-acceptor interactions. In contrast to the fully correlated regime, the  cases of uncorrelated, partially correlated, and anti-correlated motion produce nonzero acceptor-acceptor reorganization energies,  leading to rate behaviors that can deviate from Marcus theory. The anti-correlated motion case produces the largest acceptor-acceptor reorganization energy and the numerical simulations find  the most significant non-additive rate enhancement for the transfer rates.

\subsection{Rate theory}
We now  calculate the transfer rates 
for the four regimes of  reaction coordinate correlation.
The density matrix for a system that is prepared with an electron on the donor is $\hat{\rho}_0=\ket{\mathrm{D}}\bra{\mathrm{D}}\otimes \frac{e^{-\beta \hat{H}_{\mathrm{D}}^{(n)}}}{\mathcal{Z}}$ ($\mathcal{Z}$ is the vibrational partition function), and the time-dependent donor population $\ket{\mathrm{D}}$ is:  
$
    P(t) = \mathrm{Tr}\braket{\mathrm{D}|\hat{U}_I^\dagger(t)\hat{\rho}_0 \hat{U}_I(t)|{\mathrm{D}}}
         = \left\langle \braket{\mathrm{D}|\hat{U}_I^\dagger(t)|\mathrm{D}}\braket{\mathrm{D}|\hat{U}_I(t)|\mathrm{D}} \right\rangle_\beta$, where
$\braket{\cdots}_\beta$ denotes a  thermal average (trace) over the vibrational states $\frac{e^{-\beta \hat{H}_{\mathrm{D}}}}{\mathcal{Z}}$. $\hat{U}_I(t)$ is the time-evolution operator in the interaction picture: $\hat{U}_I(t) =  \mathcal{T}e^{-i\int_0^t d\tau\,V_I(\tau)}$
and 
$\hat{V}_I(t)=e^{i\hat{H}_0t}\hat{V}e^{-i\hat{H}_0t}$ ($\mathcal{T}$ is the time-ordering operator). The vibronic partition function for the initial state is associated with vibrations that are in equilibrium for the initial donor localized electronic state $\ket{\mathrm{D}}$. This description of the  transfer process assumes a vibrational distribution that is initially in thermal equilibrium, as in the Marcus formulation~\cite{izmaylov2011nonequilibrium,sun2016nonequilibrium,coalson1994nonequilibrium}. By performing a perturbative expansion of the propagator $\hat{U}_I(t)$, the transition amplitude $\braket{\mathrm{D}|\hat{U}_I(t)|\mathrm{D}}$ is calculated from a sum of vibronic propagation pathways~\cite{skourtis1999theories,skourtis2004inelastic,xiao2009turning}.

\begin{figure}
    \centering
    \includegraphics[width=\linewidth]{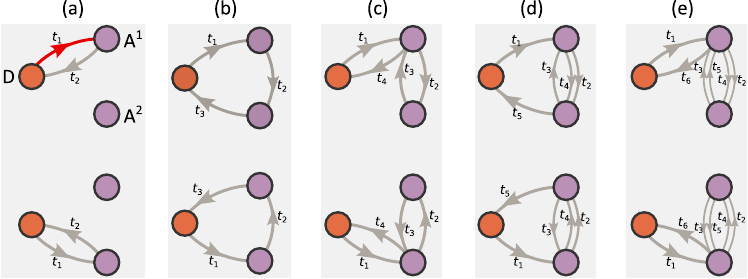}
    \caption{Leading-order pathways for an electron to remain on $\mathrm{D}$ in a $\mathrm{DA^1A^2}$ structure. Each pathway contributes to the electronic propagation. The electron starts on $\ket{\mathrm{D}}$ (orange) at time zero, goes through an acceptor site, and returns to $\ket{D}$ at time $t$. The total propagator is obtained by summing the perturbative expansion, including both electronic and vibronic states (the vibrational degrees of freedom are traced out afterwards). The numeric labels adjacent to the arrows indicate the time at which the system experiences a vibronic interaction, which prompts the electron to transfer from one electronic state to another. For example, the red pathway (upper left) for the electron to jump from $\mathrm{D}$ to $ \mathrm{A^1}$ refers to the interaction $e^{i\hat{H}_{\mathrm{D}}t_1}\ket{\mathrm{D}}\bra{\mathrm{A}^1}e^{-i\hat{H}_{\mathrm{A}^1}t_1}$ (see Sec.~\ref{sec:SI-correction-terms} in SI for additional details). In other words, these pathways account for vibrations, and are therefore vibronic.%
    }
    \label{fig:paths}
\end{figure}

Using the perturbative expansion of the propagator in the donor population expression
, we find that a limited number of coupling pathways contribute to the (time-dependent) population of the donor electronic state $\ket{\mathrm{D}}$ (see the SI for details), which ultimately determines the ET/EnT rate.
Fig.~\ref{fig:paths} shows the electronic amplitudes on $\ket{\mathrm{D}}$ after the system undergoes multiple vibronic transitions between the $\mathrm{D}$, $\mathrm{A}^1$, and $\mathrm{A}^2$ states.
These effective electronic amplitudes incorporate contributions from the vibrational degrees of freedom via Franck-Condon overlap terms (see also Fig.~\ref{fig:SI-path} in SI).

In the \emph{non-adiabatic} ET/EnT described here, the lowest order description of the donor-acceptor coupling would include two pathways  $\mathrm{D} {\to} \mathrm{A}^1 {\to} \mathrm{D}$ and $\mathrm{D} {\to} \mathrm{A^2} {\to} \mathrm{D}$ (Fig.~\ref{fig:paths}a). When just these two paths are included, the population dynamics and the transfer rate is written as the sum of two separate processes: the transfer from $\mathrm{D}$ to $\mathrm{A}^1$ and the transfer from $\mathrm{D}$ to $\mathrm{A}^2$. This approximation is called the ``statistical sum'' or  incoherent sum. There are higher-order pathways than the two ``classical'' pathways, such as those that contain more than two arrows in Fig.~\ref{fig:paths}.  The non-additive effect on the rates arise from acceptor-acceptor interactions, which are third- and higher-order effective electronic coupling interactions. 
The non-additive rate enhancements observed experimentally in two-acceptor systems arise from these higher-order coupling pathways.
The time-dependent donor amplitude can be computed by considering all possible coupling pathways~\cite{mattuck1992guide,xiao2009turning}. However, many of these paths have little influence on the transfer, since their values are very small. By computing the relative strengths of the coupling paths, we can assess their contributions to the transfer rate.

Donor-to-acceptor coupling pathways and their interferences are the next focus of our analysis.  We are most interested in contributions to the donor-acceptor coupling that involves amplitude propagation through just one of the two acceptors.
Pathways that are second order in a donor-acceptor (DA) coupling provide the lowest order contribution to the amplitude for an electron to remain on the donor. 
Examples of pathway couplings of different orders in the DA and AA couplings are indicated below:
\begin{equation*}
\resizebox{\linewidth}{!}{%
$
\begin{aligned}
\text{2\textsuperscript{nd} order overall ($n_\text{DA}{+}n_\text{AA}{=}2{+}0$, Fig.~\ref{fig:paths}a):} \quad
& \mathrm{D} {\to} \mathrm{A}^1 {\to} \mathrm{D} 
& \quad 
& \mathrm{D} {\to} \mathrm{A}^2 {\to} \mathrm{D} \\
\text{3\textsuperscript{rd} order overall ($n_\text{DA}{+}n_\text{AA}{=}2{+}1$, Fig.~\ref{fig:paths}b):} \quad
& \mathrm{D} {\to} \mathrm{A}^1 {\to} \mathrm{A}^2 {\to} \mathrm{D} 
& \quad 
& \mathrm{D} {\to} \mathrm{A}^2 {\to} \mathrm{A}^1 {\to} \mathrm{D} \\
\text{4\textsuperscript{th} order overall ($n_\text{DA}{+}n_\text{AA}{=}2{+}2$, Fig.~\ref{fig:paths}c):} \quad
& \mathrm{D} {\to} \mathrm{A}^1 {\to} \mathrm{A}^2 {\to} \mathrm{A}^1 {\to} \mathrm{D} 
& \quad 
& \mathrm{D} {\to} \mathrm{A}^2 {\to} \mathrm{A}^1 {\to} \mathrm{A}^2 {\to} \mathrm{D} \\
\text{5\textsuperscript{th} order overall ($n_\text{DA}{+}n_\text{AA}{=}2{+}3$, Fig.~\ref{fig:paths}d):} \quad
& \mathrm{D} {\to} \mathrm{A}^1 {\to} \mathrm{A}^2 {\to} \mathrm{A}^1 {\to} \mathrm{A}^2 {\to} \mathrm{D} 
& \quad 
& \mathrm{D} {\to} \mathrm{A}^2 {\to} \mathrm{A}^1 {\to} \mathrm{A}^2 {\to} \mathrm{A}^1 {\to} \mathrm{D} \\
\vdots & 
\end{aligned}
$
}
\end{equation*}
The first two paths ($\mathrm{D} {\to} \mathrm{A^1} {\to} \mathrm{D}$ and $\mathrm{D} {\to} \mathrm{A^2} {\to} \mathrm{D}$) include only donor-to-acceptor couplings, without $\mathrm{A^1}$-$\mathrm{A^2}$ interactions. The path $\mathrm{D} {\to} \mathrm{A^1} {\to} \mathrm{A^2} {\to} \mathrm{A^1} {\to} \mathrm{D}$ is second order in the DA coupling and second order in the AA coupling, making it fourth order overall (see the corresponding four arrows in Fig.~\ref{fig:paths}(c)). Additional coupling pathways are shown in Fig.~\ref{fig:paths}. These 
pathways dominate the \emph{effective} DA coupling, since  paths with more than two DA  interactions (i.e., more than two arrows linking D and A) are at least fourth-order overall and are weak (e.g., $\mathrm{D}{\to}\mathrm{A}^1{\to}\mathrm{D}{\to}\mathrm{A}^2{\to}\mathrm{D}$).
These weak high-order  pathways do not contribute significantly to the  DA interaction~\cite{izmaylov2011nonequilibrium,coalson1994nonequilibrium}. Higher-order pathways become  more important when the DA  coupling is larger.

Summing all donor-to-donor pathways that are second order in the DA coupling produces the donor amplitude  
$\braket{\mathrm{D}|U_I(t)|\mathrm{D}}$:
\begin{align}
\label{eq:approx-amplitude}
\resizebox{.9\linewidth}{!}{%
$\begin{aligned}
    & \braket{\mathrm{D}|\hat{U}_I(t)|\mathrm{D}}\\
    \approx \  & 1 + (-i)^2 V_{\mathrm{DA^1}}^2 \int_0^t dt_1\int_0^{t_1} dt_2\,  \hat{\Pi}_{\mathrm{DA^1D}}(t_1,t_2) & \Leftarrow \raisebox{-0.5\totalheight}{\includegraphics[]{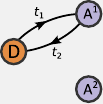}} 
    \\
    & + (-i)^2 V_\mathrm{DA^2}^2 \int_0^t dt_1\int_0^{t_1} dt_2\,  \hat{\Pi}_{\mathrm{DA^2D}}(t_1,t_2)  & \Leftarrow \raisebox{-0.5\totalheight}{\includegraphics[]{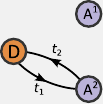}}
    \\
    & + (-i)^3 V_\mathrm{DA^1} V_\mathrm{A^1A^2} V_\mathrm{A^2D} \int_0^t dt_1 \int_0^{t_1} dt_2\, \int_0^{t_2} dt_3\, \hat{\Pi}_{\mathrm{DA^1A^2D}}(t_1,t_2,t_3) & \Leftarrow \raisebox{-0.5\totalheight}{\includegraphics[]{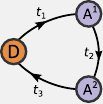}}
    \\
    & + (-i)^3 V_\mathrm{DA^2} V_\mathrm{A^2A^1} V_\mathrm{A^1D} \int_0^t dt_1\int_0^{t_1} dt_2\int_0^{t_2} dt_3\, \hat{\Pi}_\mathrm{DA^2A^1D}(t_1,t_2,t_3)  & \Leftarrow \raisebox{-0.5\totalheight}{\includegraphics[]{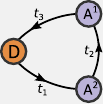}}
    \\
    & + \cdots
\end{aligned}$
}
\end{align}
where the integrands are, for example (see SI for additional details),
\begin{align}
    \hat{\Pi}_\mathrm{DA^1A^2D}(t_1,t_2,t_3) = e^{i\hat{H}_dt_1}e^{-i\hat{H}_\mathrm{A^1}t_1}e^{i\hat{H}_\mathrm{A^1}t_2}e^{-i\hat{H}_\mathrm{A^2}t_2}e^{i\hat{H}_\mathrm{A^2}t_2}e^{-i\hat{H}_dt_2}.
    \label{eq:integrand}
\end{align}
The approximate time-dependent donor population (to second order in the DA coupling) is:
\begin{align}
\label{eq:approx-population}
\resizebox{.9\linewidth}{!}{%
$\begin{aligned}
    P_\mathrm{D}(t) & \approx 1 + 2\mathrm{Re}\Big\{(-i)^2 V_\mathrm{DA^1}^2 \int_0^t dt_1\int_0^{t_1} dt_2\,\Braket{\hat{\Pi}_\mathrm{DA^1D}}_\beta\Big\}  & \Leftarrow \raisebox{-0.5\totalheight}{\includegraphics[]{fig/propagator-1a.pdf}} 
    \\
    & + 2\mathrm{Re}\Big\{(-i)^2 V_\mathrm{DA^2}^2 \int_0^t dt_1\int_0^{t_1} dt_2\, \Braket{\hat{\Pi}_\mathrm{DA^2D}}_\beta\Big\}  
    & \Leftarrow \raisebox{-0.5\totalheight}{\includegraphics[]{fig/propagator-1b.pdf}}
    \\
    & + 2\mathrm{Re}\Big\{(-i)^3 V_\mathrm{DA^1}V_\mathrm{A^1A^2}V_\mathrm{A^2D} \int_0^t dt_1\int_0^{t_1} dt_2 \int_0^{t_2} dt_3\, \Braket{\hat{\Pi}_\mathrm{DA^1A^2D}}_\beta\Big\}  
    & \Leftarrow \raisebox{-0.5\totalheight}{\includegraphics[]{fig/propagator-2a.pdf}}
    \\
    & + 2\mathrm{Re}\Big\{(-i)^3 V_\mathrm{DA^2}V_\mathrm{A^2A^1}V_\mathrm{A^1D} \int_0^t dt_1\int_0^{t_1} dt_2 \int_0^{t_2} dt_3 \, \Braket{\hat{\Pi}_\mathrm{DA^2A^1D}}_\beta\Big\} 
    & \Leftarrow \raisebox{-0.5\totalheight}{\includegraphics[]{fig/propagator-2b.pdf}}
    \\
    & + \cdots
\end{aligned}$
}
\end{align}
where $\mathrm{Re}$ indicates real part. The first two integrals in Eq.~\eqref{eq:approx-population} provide the second-order  contributions to the donor population; these contributions are represented by the diagrams that contain two arrows in Fig.~\eqref{fig:paths} (i.e., the first two diagrams on the left). The third and fourth integrals in Eq.~\eqref{eq:approx-population}  are represented by  diagrams with three arrows in Fig.~\ref{fig:paths}. Taking into account only the first two integrals/diagrams, the transfer rate [$-\frac{\mathrm{d}P_\mathrm{D}(t)}{dt}$] is a simple sum of two independent transfer events (i.e., the transfers from D to A\textsuperscript{1} and D to A\textsuperscript{2}, without considering the propagation between the two acceptors). We consider higher-order terms (e.g., $\mathrm{D}{\to}\mathrm{A}^1{\to}\mathrm{A}^2{\to}\mathrm{A}^1{\to}\mathrm{A}^2{\to}\mathrm{D}$)
that capture  interactions between the two acceptors. There are higher-order pathways that originate from  back-scattering between the acceptor sites and D.  That is,  paths with more than 2 donor state visits contribute to pathway effects (these terms are typically small and are not considered here).
The third and fourth integrals in Eq.~\eqref{eq:approx-population} are first order in the $\mathrm{A}^1\mathrm{A}^2$ coupling.  These integrals  provide first-order corrections (produced by V$_\mathrm{A^1A^2}$ couplings) to the simple sum of $\mathrm{D}$-$\mathrm{A^1}$ and $\mathrm{D}$-$\mathrm{A^2}$ paths.

Assuming exponential decay of the initially prepared reduced donor population with time, the  transfer rate $k$ is computed by differentiating the time-dependent donor population, and evaluating it at  long times~\cite{coalson1994nonequilibrium}, $k\approx-\lim\limits_{t{\to} \infty}\frac{\mathrm{d}P_\mathrm{D}(t)}{dt}$. The  transfer rate is the sum of thermally averaged integrals of operator products, as described by Eq.~\eqref{eq:approx-population}.
The  thermal averages of the operator products appear in the SI. The rate is calculated using these thermal averages with Monte Carlo numerical integration~\cite{lepage2021adaptive}.

\subsection{Estimation of Electronic Parameters for DAA Molecules}
\label{sec:estimation-of-electronic-parameters}
To compare the results of our simulations  with those of  experiments in Ref.~\citenum{phelan2019quantum}, we analyzed the DAA molecule of Ref.~\citenum{phelan2019quantum} (also see Fig.~\ref{fig:molecule}) to estimate the electronic coupling parameters for our dynamics simulations. Density functional theory (DFT) calculations were used to compute the dipole  and  transition dipole moments between the locally excited state and the two charge transfer states($\mathrm{D^+A^{1-}A^{2}}$ and $\mathrm{D^+A^{1}A^{2-}}$). The  DFT calculations were carried out at the cam-b3lyp/def2svp level with Grimme's D3(BJ) dispersion correction. The computed dipole and transition dipole moments allowed us to calculated the diabatic couplings $V_\mathrm{DA^1}=-144.84\,\text{cm}^{-1}$, $V_\mathrm{DA^2}=-67.97\,\text{cm}^{-1}$ and $V_\mathrm{A^1A^2}=575.79\,\text{cm}^{-1}$, using the generalized Mulliken-Hush approach~\cite{subotnik2008constructing}. 
We determined the electronic couplings between donor-acceptor and acceptor-acceptor pairs used in later rate calculations, based on the magnitudes of the diabatic state couplings computed with the above strategy. 

\section{Results and Discussion}
\label{sec:results}
\subsection{ET/EnT to two equivalent acceptors}
\label{sec:et-to-two-equivalent-acceptors}
Using the rate theory described above, we explore structure-function relationships for multi-acceptor  transfer kinetics. The  transfer rates are calculated as a function of reaction free energy at selected temperatures; the effect of correlated reaction-coordinate motion on the ET/EnT rates is also explored.  As discussed in Sec.~\ref{sec:model-vibration}, we consider four cases of  reaction-coordinate motion correlations in the two-acceptor species.  The limits are for reaction coordinates that are: (i) fully and positively correlated, (ii) partially and positively correlated, (iii) uncorrelated, and (iv) negatively correlated.

\begin{figure}
    \centering
    \begin{subfigure}[b]{0.49\textwidth}
     \centering
     \includegraphics[width=\textwidth]{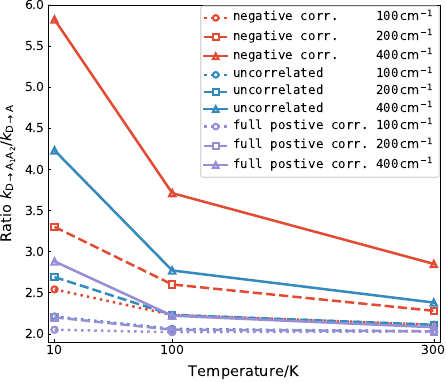}
    \end{subfigure}
    \begin{subfigure}[b]{0.49\textwidth}
     \centering
     \includegraphics[width=\textwidth]{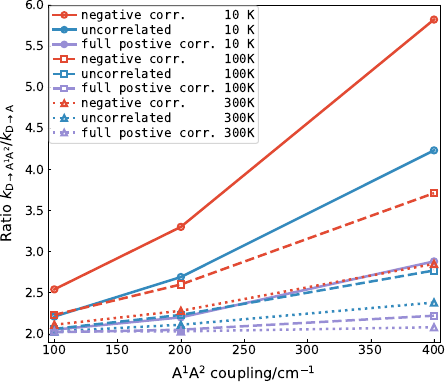}
    \end{subfigure}
    \caption{Transfer rate ratios for two- and one-acceptor  transfer rates. The spectral density and electronic parameters used in the calculations are $\lambda=5245.4\,\mathrm{cm}^{-1}$, $\eta=263.4\,\mathrm{cm}^{-1}$, $\Omega=76.8\,\mathrm{cm}^{-1}$, $V_\mathrm{DA^1}=V_\mathrm{DA^2}=109.7\,\mathrm{cm}^{-1}$. \textbf{First:} Ratios at various $\mathrm{A^1A^2}$ electronic couplings. ${k_{\mathrm{D} {\to}  \mathrm{A^1}\mathrm{A^2}}(\Delta G^{k_\mathrm{max}}_{\mathrm{D} {\to} \mathrm{A^1}\mathrm{A^2}})}/{k_{\mathrm{D} {\to} A}(\Delta G^{k_\mathrm{max}}_{\mathrm{D} {\to} \mathrm{A^1}\mathrm{A^2}})}$. The two-acceptor transfer rates are the maximum values of the function $k_{\mathrm{D} {\to} \mathrm{A^1}\mathrm{A^2}}(\Delta G)$, and the free energy that maximizes $k_{\mathrm{D} {\to} \mathrm{A^1}\mathrm{A^2}}(\Delta G)$ is $\Delta G^{k_\mathrm{max}}_{\mathrm{D} {\to} \mathrm{A^1}\mathrm{A^2}}$. The ratios for the partially positively correlated case are nearly identical to the ratios for the fully positively correlated case, so the data are not shown. \textbf{Second:} ET/EnT rate ratios for one- and two-acceptor rates at various temperatures: ${k_{\mathrm{D} {\to}  \mathrm{A^1}\mathrm{A^2}}(\Delta G^{k_\mathrm{max}}_{\mathrm{D} {\to} \mathrm{A^1}\mathrm{A^2}})}/{k_{\mathrm{D} {\to} A}(\Delta G^{k_\mathrm{max}}_{\mathrm{D} {\to} \mathrm{A^1}\mathrm{A^2}})}$ at 10\,K, 100\,K, and 300\,K.}
    \label{fig:ratio-plot}
    \label{fig:ratio-plot-temp}
\end{figure}

Fig.~\ref{fig:ratio-plot} shows the ratio of  transfer rates for  one-acceptor and two-acceptor systems. The unusually large rate enhancement for the two-acceptor ET/EnT (i.e., beyond a simple factor of 2  enhancement) is reflected in the large transfer rate ratio: $k_{\mathrm{D} {\to} \mathrm{A^1A^2}}/k_{\mathrm{D} {\to} A}=5.82$ when $V_{\mathrm{A^1A^2}}=400\,\mathrm{cm}^{-1}$ and $T=10\,\mathrm{K}$. This ratio is close to the measured ET rate enhancement of 4-fold that is reported in Ref.~\citenum{phelan2019quantum}.
Fig.~\ref{fig:ratio-plot} shows that the rate ratio decreases as the temperature grows, consistent with the experimental findings of Ref.~\citenum{phelan2019quantum}.

For the the one-donor two-acceptor ET molecule studied in Refs.~\citenum{phelan2019quantum1,bancroft2021charge}, the two acceptor units are nearly parallel to each other ($\pi$-stacked), in contrast to the triangular structure of the species reported in Ref.~\citenum{phelan2019quantum}. With nearly parallel charge-transfer dipoles for the molecule in Ref.~\citenum{phelan2019quantum1,bancroft2021charge}, the $\mathrm{A^1}\leftrightarrow\mathrm{A^2}$ ET  reorganization energy is small due to the nearly zero vector difference between the parallel dipoles for the two charge transfer states. As a result, we expect little rate enhancement above the expected factor of two for the two-acceptor system compared to the one-acceptor system,  consistent with the two-fold rate acceleration reported in Refs.~\citenum{phelan2019quantum1,bancroft2021charge}.

We now explore the cases presented in Sec.~\ref{sec:model-vibration}.


\begin{figure}
    \centering
    \includegraphics[width=\linewidth]{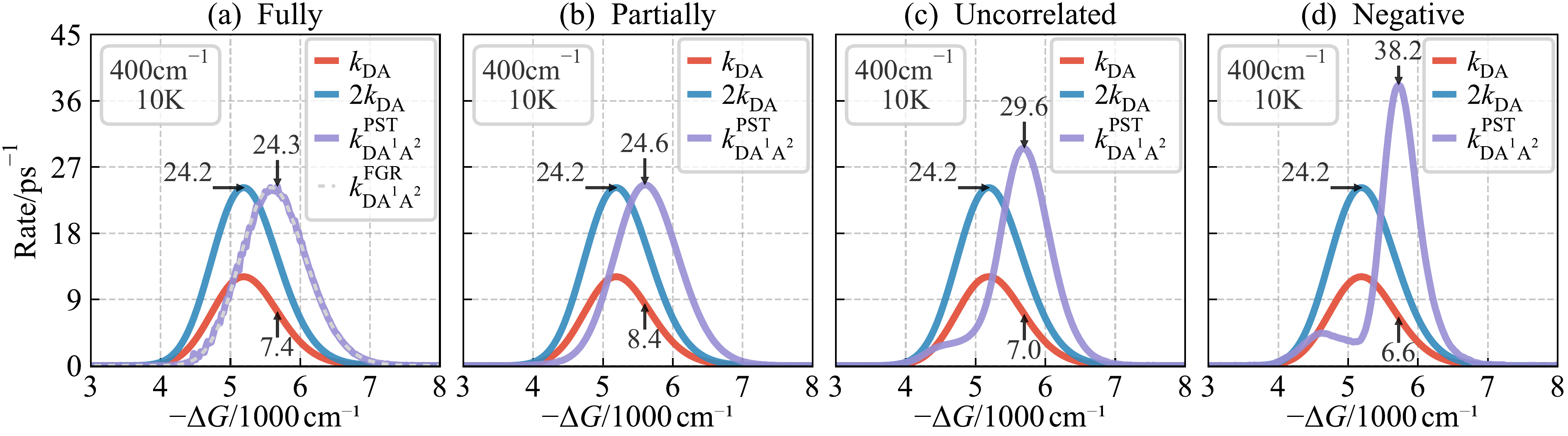}
    \caption{One and two-acceptor transfer rates, $k_{\mathrm{D}\mathrm{A}}$ and $k_{\mathrm{D}\mathrm{A^1A^2}}$, as a function of the free energy for ET/EnT to a single acceptor, $\Delta G=E_{\mathrm{D}} - E_{\mathrm{A}^{1,2}}$. The red line indicates the rate for a one-acceptor system. The light-blue line indicates a statistical sum of two one-acceptor transfers (i.e., twice the red-circles).  The solid purple line ($k_{\mathrm{DA^1A^2}}^\mathrm{PST}$) indicates the rate computed from the perturbative approach (the partial summation approach). The dashed line ($k_{\mathrm{DA^1A^2}}^\mathrm{FGR}$) indicates the golden rule rate based on the effective 2-state system of  Eq.\eqref{eq:two-state}. The electronic couplings are: $V_{\mathrm{DA^1}}=V_{\mathrm{DA^2}}=109.7\,\mathrm{cm}^{-1}$. The temperatures and $V_\mathrm{A^1A^2}$ values are denoted in the figures. The four cases, ``Fully'', ``Partially'', ``Uncorrelated'', and ``Negative'' correspond to the four scenarios of correlation between the two reaction coordinates, as discussed in Sec.~\ref{sec:et-to-two-equivalent-acceptors}.%
    }
    \label{fig:overall-rate-fig}
\end{figure}

\paragraph{Fully correlated reaction coordinates for the two ET/EnT reactions.} 
This case serves as a test of the partial summation rate theory, since the influence of having two acceptors compared to one can be computed exactly. We computed the transfer rates for effective two-state systems (i.e., $\ket{\mathrm{D}}$ and $\ket{+}$) reduced from the DA\textsuperscript{1}A\textsuperscript{2}, provided $\hat{H}_{\mathrm{A^1}}^{(n)}=\hat{H}_{\mathrm{A^2}}^{(n)}$, using the two-state Fermi golden rule framework~\cite{xie2013calculation,coalson1994nonequilibrium}. We compared the Fermi  golden rule results to the results for  DA\textsuperscript{1}A\textsuperscript{2} found using  partial summation approximation described above.
Parameter values used in the analysis appear in the caption of Fig.~\ref{fig:overall-rate-fig}.

The first panel in Fig.~\ref{fig:overall-rate-fig} (titled `Fully')
shows Fermi's golden rule rates derived from the effective two-state system (exploiting molecular symmetry), as well as the rates calculated using  partial summation (see Eq.~\eqref{eq:integrand} and \eqref{eq:approx-population}); the two values are in excellent agreement. 
The rate computed  by partial summation of electronic propagation, in the limit of an infinite-order summation of the $\mathrm{A^1}\leftrightarrow\mathrm{A^2}$ coupling (although the $\mathrm{D}\leftrightarrow\mathrm{A^1}/\mathrm{A^2}$ coupling is always retained to second order), is equivalent to the two-state Fermi's golden rule rate for this  case (i.e., fully correlated acceptor vibrations) (see SI).

The transfer rate $k_{\mathrm{D} {\to} \mathrm{A^1A^2}}$, from one donor ($\mathrm{D}$) to a symmetric combination of two acceptors ($\mathrm{A^1}$ and $\mathrm{A^2}$), only differs from the  sum of rates to each of the acceptors ($2k_{\mathrm{D} {\to} \mathrm{A}}$) because of a free energy offset arising from acceptor-acceptor coupling, $\delta=V_{\mathrm{A^1A^2}}$ (see the first panel of Fig.~\ref{fig:overall-rate-fig}).  This offset produces a shift of the peak value position in the log(rate) vs driving force curve, as expected from Marcus-style rate theories (the maximum rate is found when the driving force  equals the reorganization energy). 
Fig.~\ref{fig:overall-rate-fig}(a) (also Fig.~\ref{fig:equivalent-acceptors} in the SI) shows that the reaction driving force decreases by $\vert V_{\mathrm{A^1}\mathrm{A^2}}\vert$ for the symmetric linear combination of acceptor species, compared to the single acceptor case. The reorganization energy for ET/EnT from $\mathrm{D}$ to the delocalized symmetric $\mathrm{A^1} \mathrm{A^2}$ acceptor state is unchanged from the case of ET/EnT to a single $\mathrm{A^1}$ or $\mathrm{A^2}$ species. The cause is that the electron polarizes the molecule along the same reaction coordinate
for $\mathrm{D{\to} A^1}$ and $\mathrm{D{\to} A^2}$. ET/EnT in this special case of $H_\mathrm{A^1}^{(n)}=H_\mathrm{A^2}^{(n)}$ has a fully positively correlated coupled nuclear modes. Thus, the dependence of the rate on the reaction free energy for the two-acceptor system is twice as large as that for the one-acceptor system: $k_{\mathrm{D} {\to} \mathrm{A^1A^2}}(\Delta G)=2k_{\mathrm{D} {\to} A}(\Delta G -\delta)$.

The partially correlated and uncorrelated cases are discussed in SI.

\paragraph{Transfer rates in $\mathrm{DA^1A^2}$ structures with negatively correlated acceptor vibrations.}
The most interesting case is when 
the two reaction coordinates are oriented in opposite directions, forming an angle $\theta > \pi$, while the acceptor-acceptor reorganization energy exceeds $2\lambda$. For the purposes of numerical simulations, we use
$\alpha=\frac{\sqrt{2+\sqrt{3}}}{2} (\approx 0.97)$ and $\beta=-\frac{\sqrt{2-\sqrt{3}}}{2} (\approx -0.26)$; the reorganization energy for the $\mathrm{A^1}\mathrm{A^2}$ ET/EnT is $3\lambda$. The computed rates as a function of driving force are shown in the fourth panel (labeled 'Negative') of Fig.~\ref{fig:overall-rate-fig} (see also Fig.~\ref{fig:anti-correlated-acceptors}), where we observe the most distinct rate profile as a function of the driving force, along with the most pronounced rate enhancement compared to single-donor ET/EnT.

ET/EnT rate enhancements are always found to grow with $V_{\mathrm{A^1A^2}}$, and also to grow as the temperature decreases (see SI). A larger acceptor-acceptor coupling ($V_{\mathrm{A^1A^2}}$) further splits the energies  of the symmetric and anti-symmetric acceptor superposition states $\ket{+}=(\ket{\mathrm{A^1}}+\ket{\mathrm{A^2}})/\sqrt{2}$ and $\ket{+}=(\ket{\mathrm{A^1}}-\ket{\mathrm{A^2}})/\sqrt{2}$. Strong acceptor-acceptor electronic coupling, combined with a large acceptor-acceptor reorganization energy, also enhances the contribution of non-additive coupling pathway effects.
Low temperatures or, conversely, the presence of high-frequency vibrations, help preserve the contributions from  higher order vibrationally averaged electron coupling pathways
that give rise to non-additive rate behavior. 
This may be understood with the following logic. In the high temperate regime, the vibrational modes are effectively classical, and we argue in the SI that, for two-acceptor ET/EnT systems, classical vibrations cannot produce rate enhancements greater than a factor of 2 times the rate for single acceptors (i.e., the incoherent sum obtained from  the two lowest-order couplings in Fig.~\ref{fig:paths} dominate). One can use a generalized Marcus theory~\cite{taylor2018generalised} to prove more rigorously that the upper limit for the two-acceptor system is 2 time the rate for the single acceptor system in the  high-temperature limit.
Mathematically, the expression for the non-additive higher-order pathways includes a factor, $\coth(\omega/k_\text{B}T)$, which suppresses their contributions at high temperatures.

Tables~\ref{tab:ratio100cm}, ~\ref{tab:ratio200cm}, and \ref{tab:ratio400cm} (see SI) indicate that the peaks of the two-acceptor transfer rates (as a function of the driving force, $-\Delta G$) are larger than the sum of the two independent $\mathrm{D} {\to} \mathrm{A}$ rates (at a driving force that maximizes the two-acceptor rates). The simulations find that the maximum rate enhancement for the two-acceptor  system vs the  sum of $\mathrm{D} {\to} \mathrm{A^1}$ and $\mathrm{D} {\to} \mathrm{A^2}$ rates  arises under the conditions: (i) a large $\mathrm{A^1} {\to} \mathrm{A^2}$ reorganization energy (relative to the donor-acceptor electronic coupling) allows significant contributions of non-additive pathways 
that distinguish between the symmetric $\ket{+}$ and anti-symmetric $\ket{-}$ states. 
This distinction between the symmetric and anti-symmetric combinations of acceptor states is evident in the two peaks of the rate versus free energy plot in Fig.~\ref{fig:overall-rate-fig}, (ii) a large electronic coupling occurs between $\mathrm{A^1}$ and $\mathrm{A^2}$ compared to the donor-acceptor electronic couplings, (iii) $k_\mathrm{B}T$ is small compared to the acceptor-acceptor coupling (see Fig.~\ref{fig:ratio-plot}). 

An earlier theory of  electron transfer rate acceleration at low temperature~\cite{phelan2019quantum} focused on the influence of enhanced donor-acceptor couplings ($V_\mathrm{DA}$) and electron-vibration couplings ($V_\mathrm{SB}$). In the two-acceptor structure, it was suggested that the symmetric superposition of $\ket{\mathrm{A^1}}$ and $\ket{\mathrm{A^2}}$ to form $\ket{+}$ would cause the electronic coupling of the DA structure to increases by a factor of $\sqrt{2}$.  This result is consistent with our analysis for the fully correlated case. Furthermore, they argued that, at low temperatures, the electronic-vibrational coupling also increases by a factor of $\sqrt{2}$ due to the superposition.
These two effects
enhance the ET/EnT rate by four-fold~\cite{phelan2019quantum} in the two acceptor system, compared to the one acceptor system, based on the rate formula $k\propto V_\mathrm{DA}^2 V_\mathrm{SB}^2 /(\Delta G)^2$  that is valid in the weak system-bath coupling regime at low temperatures~\citenum{phelan2019quantum}. The theoretical prediction of Phelan et al. shows that the upper limit of the ET rate enhancement is four-fold, while the theory that we developed shows that the rate enhancement can be more than 4-fold (e.g., in the negatively correlated case of Fig.~\ref{fig:overall-rate-fig}).



The two-acceptor system described in Ref.~\citenum{phelan2019quantum} is a small rigid molecule with an near equilateral triangular  arrangement of $\mathrm{D}$, $\mathrm{A^1}$ and $\mathrm{A^2}$ (see Fig.1(E) of Ref.~\citenum{phelan2019quantum}). The dipole moment vector changes upon formation of the charge-separated states ${\mathrm{D} {\to} \mathrm{A^1}}$ or ${\mathrm{D} {\to} \mathrm{A^2}}$, so the change in the dipole moment for $\mathrm{A^1} {\to} \mathrm{A^2}$ ET/EnT can be substantial. This leads to a significant reorganization energy for $\mathrm{A^1} {\to} \mathrm{A^2}$ ET/EnT, compared to the DA coupling $V_{DA}$. This significant $\mathrm{A^1}$-$\mathrm{A^2}$ reorganization energy at low temperatures is the likely source of the 4- to 5-fold enhancement of the two-acceptor transfer rate reported in Ref.~\citenum{phelan2019quantum}, compared to the one-acceptor transfer rate.

As described in Sec.~\ref{sec:model-vibration}, the correlation between the vibrations associated with the two electronic states ($\ket{\mathrm{A^1}}$ and $\ket{\mathrm{A^2}}$) determines the acceptor-acceptor reorganization energy and, equivalently, the acceptor-acceptor reorganization energy indicates the correlation (positive or negative) between the vibrations of the two acceptors.

\subsection{ET/EnT to three equivalent acceptors}
The conditions for ET/EnT rate enhancement can be extended to structures with more than two acceptors. In such multi-acceptor constructs, we anticipate that the rate enhancement could be further amplified because of the increased number of pathways and the potential for more complex correlations among the reaction coordinates.
To show the effectiveness of the rate enhancement conditions discovered above, we studied a one-donor three-acceptor system (see SI for details). Following the same approach as for the 2-acceptor system, we study ET/EnT in two representative scenarios, where the vibrations of the three acceptors are (1) fully correlated and (2) anti-correlated.

\paragraph{Fully positively correlated reaction coordinates for ET/EnT to three acceptors.}
Full correlation among the  vibrational modes of the three acceptors implies that $H^{(n)}_{\mathrm{A^1}}=H^{(n)}_{\mathrm{A^2}}=H^{(n)}_\mathrm{A_3}$. In this case, by rotating the electronic basis, the one-donor three-acceptor system can be transformed into an effective one-donor one-acceptor system where the donor is coupled to the highest-energy eigenstate of the acceptor electronic Hamiltonian because of the orbital parity effect (see SI).
The reaction free energy, $\Delta G = E_\mathrm{D} - E_\mathrm{A}$, of the effective two-state system is reduced by $2V_\mathrm{AA}$. As such, the rate vs. driving force relationship will be shifted by $2V_\mathrm{AA}$ compared to the curve for three noninteracting acceptors (i.e., when there is no electronic couplings among the three acceptors).
In this fully correlated case, there is no rate enhancement beyond factor of three (apart from a free energy shift), as the system behaves as a two-state system with one acceptor and a modified reaction free energy.

\paragraph{Negatively correlated reaction coordinates for ET/EnT to a three-unit acceptor.}
The maximum reaction rates (vs. driving force) occur when the vibrations of the three acceptors are negatively correlated ($\lambda_\mathrm{A^1A^2}^\text{reorg} = \lambda_\mathrm{A^1A^3}^\text{reorg} = \lambda_\mathrm{A^2A^3}^\text{reorg} = 3\lambda_\mathrm{DA^1}^\text{reorg}$). In this case, the reaction coordinates that couple the donor and the three acceptors move  
out of phase and
define a large AA reorganization energy. The large AA reorganization energy can be realized, for example, when the dipole moments of $\mathrm{D^+A^{1,-}A^{2}A^{3}}$,  $\mathrm{D^+A^{1}A^{2,-}}$, and $\mathrm{D^+A^{1}A^{2}A^{3,-}}$ have a large difference,
since the motion of the acceptor vibrations reflecting changes in the electronic charge distribution that result from transfer.
This scenario is relevant to the ET experiments~\cite{phelan2019quantum,phelan2019quantum1,fisher2023fundamental,fisher2024long,bancroft2021charge,lin2023ultrafast} 
performed on the 3-acceptor analog of the 2- and 4-acceptor slip-stacked structures. 
In these structures, the angle between the dipole moments of the  charge transfer states ($\mathrm{D^+A^{1,-}A^{2}A^{3}}$,  $\mathrm{D^+A^{1}A^{2,-}}A^{3}$, and $\mathrm{D^+A^{1}A^{2}}A^{3,-}$) is 
approximately $120^\circ$ (see Fig.~\ref{fig:illustration}), defining a large AA reorganization energy.
In this case, the rate enhancement $\frac{k_{\mathrm{D{\to} A^1A^2A^3}}}{3k_\mathrm{D{\to} A}}$ for the driving force that maximizes $k_{\mathrm{D{\to} A^1A^2A^3}}$, is expected to be larger than that for the two-acceptors system, since the vibronic pathway interference effect is more pronounced when more acceptors are added to the structure.

\begin{figure}
    \centering
    \includegraphics[width=\linewidth]{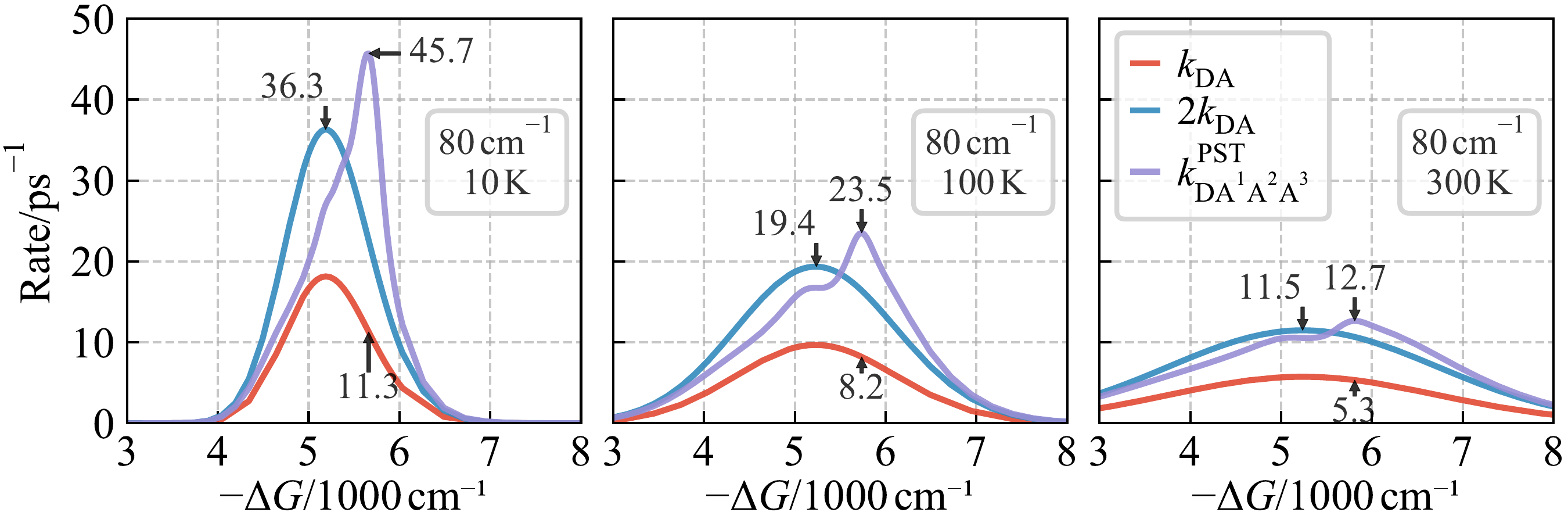}
    \caption{ET/EnT rates for a donor coupled to three acceptors. 
    $k_{\mathrm{D} {\to} \mathrm{A}}$ and $k_{\mathrm{D} {\to} \mathrm{A^1A^2A^3}}$, as a function of the free energy for ET/EnT to a single acceptor, $\Delta G=E_{\mathrm{D}} - E_{\mathrm{A}^{1/2/3}}$. The red line indicates the rate for a one-acceptor system. The blue line indicates a statistical sum of two one-acceptor transfers (i.e., twice the red line).  The solid purple line ($k_{\mathrm{DA^1A^2}}^\mathrm{PST}$) indicates the rate obtained from the partial summation approach.
    The electronic couplings are: $V_{\mathrm{DA^{1/2/3}}}=109.7\,\mathrm{cm}^{-1}$. The temperatures are are $10\, \mathrm{K}$, $100\, \mathrm{K}$, and $300\, \mathrm{K}$. The value of $V_\mathrm{A^1A^2}=V_\mathrm{A^1A^3}=V_\mathrm{A^2A^3}$ is $80\,\mathrm{cm}^{-1}$, as indicated in the figure. The reorganization energy for ET/EnT from donor to an acceptor is the same as the 2-acceptor case: $\lambda^\mathrm{reorg}_{\mathrm{D},\mathrm{A^{1/2/3}}}=5245.4\,\mathrm{cm}^{-1}$.}
    \label{fig:three-acceptors-full-correlation}
\end{figure}

For a relatively small electronic coupling among the three acceptors, $V_\mathrm{A^1A^2}=V_\mathrm{A^1A^3}=V_\mathrm{A^2A^3}=80\,\mathrm{cm}^{-1}$, the transfer rate from the donor to three acceptors with identical energies is shown in Fig.~\ref{fig:three-acceptors-full-correlation}. 
At 10\,K, the rate is 4-fold larger than the single-acceptor case. For comparison, in the two-acceptor structure, a similar 4-fold rate enhancement is achieved  with a much larger acceptor-acceptor coupling of $250\,\mathrm{cm}^{-1}$ (Fig.~\ref{fig:ratio-plot}). This indicates that the three-acceptor constructs have greater likelihood for rate enhancements, even with smaller electronic couplings between the acceptors.

These findings show that extending ET/EnT systems to include more than two acceptors can significantly enhance transfer rates. In particular, the three-acceptor system exhibits a greater potential for rate enhancement compared to the two-acceptor system. This is attributed to the interplay between reaction-coordinate correlations and the acceptor-acceptor electronic couplings. In the three-acceptor system, the additional acceptor expands the number of constructively interfering ET/EnT pathways, which can amplify the effects of reaction-coordinate correlations. Even with smaller acceptor-acceptor electronic couplings, the three anti-correlated reaction coordinates of the three donor-acceptor pairs can amplify the contributions of non-additive pathway statistics
and enhance transfer rates at some driving forces.
This highlights the promise of multi-acceptor architectures for boosting charge or energy transfer efficiency by leveraging increased vibrational and electronic interactions, as well as exploiting non-additive pathway effects.

\section{Conclusions}

We have examined ET/EnT in multi-acceptor systems. Our findings show how the combined  effects of electronic delocalization and vibronic coupling can produce enhanced transfer rates, leading to strategies to manipulate transport rates in multi-acceptor assemblies.

We identified the  conditions under which significant rate enhancements occur in these systems. Specifically, sufficiently strong acceptor-acceptor electronic coupling is needed to create delocalized acceptor states, which can interact more effectively with the donor. Maintaining low temperatures is crucial for preserving the quantum nature of the vibrational modes and preserving
delocalization among the multiple acceptors,
such as symmetric and anti-symmetric combinations, which allow electron delocalization to influence the effective driving force. Additionally, low temperatures sustain the contributions of non-additive vibronic pathway effects.
Moreover, significant reorganization energy associated with the acceptors amplifies the effect of electronic delocalization by increasing the contributions of non-additive vibronic pathways (see SI). In this regime, our calculations predict 5-6-fold rate enhancements for two-acceptor constructs, relative to single-acceptor systems.

The simulation results for the DAA system were compared to experimental observations, and good agreement was found. This comparison supports our  approach and highlights its ability to capture the dependence of rate enhancements on acceptor-acceptor coupling, acceptor-acceptor reorganization energy, and temperature. 

For transfer constructs with more than two acceptors, we found that the potential for rate enhancement is even greater than in the two-acceptor case. The additional acceptor(s) increase  electronic delocalization and amplify the effects  of reaction-coordinate correlation on the Franck-Condon overlap, producing enhanced transfer rates even when acceptor-acceptor electronic couplings are relatively small. These findings demonstrate the unique advantages of multi-acceptor architectures in boosting ET/EnT efficiency and provide a strategy to  design systems with even greater rate enhancements.

To support these findings, we developed a general rate theory capable of describing multi-donor and multi-acceptor constructs. The theory is based on a partial summation of vibronic propagation pathways and operates in the golden rule regime, accounting for vibrational dynamics without explicitly including environmental decoherence. The computational implementation of this theory is  accessible~\cite{github}, providing a  tool for further studies of  ET/EnT assemblies and an  intuitive framework for designing and interpreting multi-acceptor architectures ~\cite{phelan2019quantum,phelan2019quantum1,fisher2023fundamental,fisher2024long,williams2024charge}.

While our analysis has focused on  non-adiabatic transfer, it will also be useful to explore regimes with stronger donor–acceptor and acceptor-acceptor couplings, where the non-adiabatic approximation no longer holds. Increasing the number of donors and acceptors expands the number of  vibronic pathways~\cite{brouwer1993two,yonemoto1994photoinduced,willemse1995reversible,polander2011benzothiadiazole,kwon2012unique,lin2019bending,zhao2020recent,wang2023charge}. As the number of pathways grows, we expect additional richness in the rate behavior and enhanced contributions from higher-order vibronic processes. Our theoretical  and computational framework~\cite{github} can readily be used to explore  adiabatic or intermediate dynamical regimes, providing a foundation to investigate new physical phenomena and design principles in strongly coupled multi-donor/multi-acceptor structures.



\begin{acknowledgement}
    This material is based on work supported by the Department of Energy under Grant No. DE-SC0019400.
\end{acknowledgement}

\bibliography{abbreviated} 

\end{document}


\title{Supporting Information \\ Condensed phase dynamics  of multi-acceptor electron transfer}

\maketitle
\tableofcontents
\newpage


\section{Additional Figures and Tables}
\subsection{Vibronic Pathways}
\begin{figure}[h]
    \centering
    \includegraphics[width=0.5\linewidth]{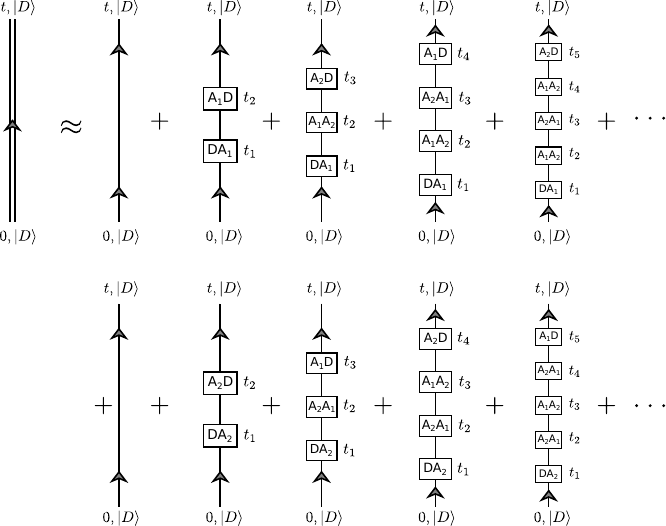}
    \caption{The overall propagator describing the survival probability amplitude of the donor state is expanded to a series of perturbative propagators by the order of electronic couplings.}
    \label{fig:SI-path}
\end{figure}

\subsection{Rate profiles}
\begin{figure}
    \centering
    \includegraphics[width=.5\linewidth]{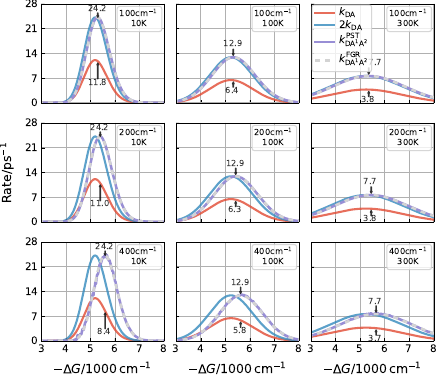}
    \caption{One and two-acceptor electron transfer rates, $k_{\mathrm{D}\mathrm{A}}$ and $k_{\mathrm{D}\mathrm{A^1A^2}}$, as a function of the free energy for electron transfer to a single acceptor, $\Delta G=E_{\mathrm{D}} - E_{\mathrm{A}^{1,2}}$. The red-circles indicate the rate for a one-acceptor system. The light-blue diamonds indicate a statistical sum of two one-acceptor transfers (i.e., twice the red-circles).  The solid purple line ($k_{\mathrm{DA^1A^2}}^\mathrm{PST}$) indicates the rate obtained from the perturbative approach (partial summation theory). The dashed line ($k_{\mathrm{DA^1A^2}}^\mathrm{FGR}$) indicates the golden rule rate based on the effective 2-state system of  Eq.\eqref{eq:two-state}. The electronic couplings are: $V_{\mathrm{DA^1}}=V_{\mathrm{DA^2}}=109.7\,\mathrm{cm}^{-1}$. The temperatures and $V_\mathrm{A^1A^2}$ values are denoted in the figures. }
    \label{fig:equivalent-acceptors}
\end{figure}

\begin{figure}
    \centering
    \includegraphics[width=.5\linewidth]{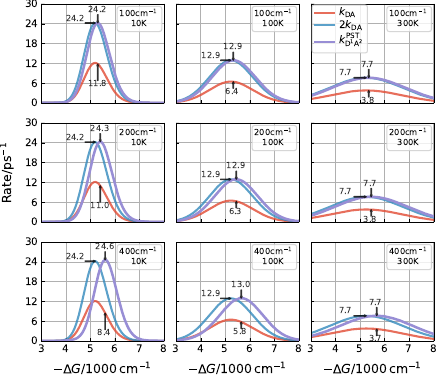}
    \caption{Electron-transfer rate as a function of free energy when the oscillators coupled to the two acceptors are partially positively correlated. The bath parameters and the definitions indicated in the legends are the same as those in the caption of Fig.~\ref{fig:equivalent-acceptors}. Compared to the fully correlated  (i.e., $H_{\mathrm{A^1}}^{(n)}=H_{\mathrm{A^2}}^{(n)}$) , the rate maximum of the partially correlated system is computed to be larger than the simple sum of rates for two non-interacting acceptors, the blue-diamonds, $2k_{\mathrm{D{\to} A}}$). Similar shifts of the rate maxima in  Fig.~\ref{fig:equivalent-acceptors} also appear in the correlated model.
    }
    \label{fig:partially-correlated-acceptors}
\end{figure}

\begin{figure}
    \centering
    \includegraphics[width=.5\linewidth]{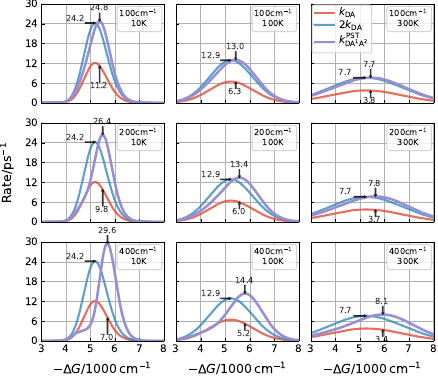}
    \caption{Electron transfer rates as a function of free energy when the vibrations on the two acceptors are uncorrelated. The bath parameters and  legend definitions are the same as in the caption of Fig.~\ref{fig:equivalent-acceptors}.
    }
    \label{fig:independent-acceptors}
\end{figure}

\begin{figure}
    \centering
    \includegraphics[width=.5\linewidth]{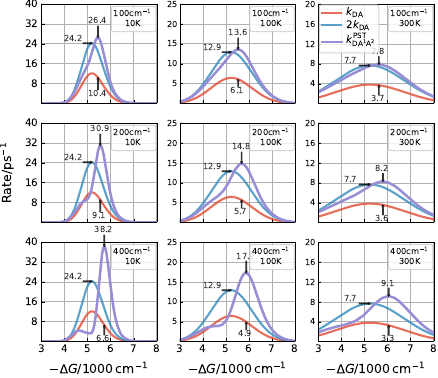}
    \caption{Computed electron transfer rates as a function of free energy for inversely correlated vibrations on the two acceptors. The bath parameters and definitions used in the legends are the same as in the caption of Fig.~\ref{fig:equivalent-acceptors}.
    }
    \label{fig:anti-correlated-acceptors}
\end{figure}

\FloatBarrier

\begin{table}
    \centering
    \begin{tabular}{cccccc}
    \hline
        Case/Rate & $\Delta G$ ($\mathrm{cm}^{-1}$) & $\lambda^\mathrm{reorg}_{\mathrm{A^1,A^2}}$ ($\mathrm{cm}^{-1}$) & $k_{\mathrm{D} {\to} \mathrm{A^1A^2}}$($\mathrm{ps}^{-1}$) & $k_{\mathrm{D} {\to} A}$ ($\mathrm{ps}^{-1}$) & $\frac{k_{\mathrm{D} {\to} \mathrm{A^1A^2}}}{k_{\mathrm{D} {\to} A}}$ \\
        \hline $T=10\,\mathrm{K}$\\  \hline 
        Full Corr.   & 5299  & $0$       & 24.2   & 11.8  &  2.05 \\
        Partial Corr.& 5299  & $1405$    & 24.2   & 11.8  &  2.05 \\
        Uncorr.      & 5374  & $10490$   & 24.8   & 11.2  &  2.21 \\
        Inv. Corr.   & 5449  & $15736$   & 26.4   & 10.4  &  2.54\\ \hline
        $T=100\,\mathrm{K}$ \\ \hline 
        Full Corr.   & 5324  & $0$       & 12.9   & 6.4  &  2.02 \\
        Partial Corr.& 5324  & $1405$    & 12.9   & 6.4  &  2.02 \\
        Uncorr.      & 5424  & $10490$   & 13.0   & 6.3  &  2.06 \\
        Inv. Corr.   & 5524  & $15736$   & 13.6   & 6.1  &  2.23 \\ 
        \hline $T=300\,\mathrm{K}$ \\ \hline 
        Full Corr.   & 5349  & $0$       & 7.7   & 3.8  &  2.03 \\
        Partial Corr.& 5349  & $1405$    & 7.7   & 3.8  &  2.03 \\
        Uncorr.      & 5424  & $10490$   & 7.7   & 3.8  &  2.03 \\
        Inv. Corr.   & 5599  & $15736$   & 7.8   & 3.7  &  2.11\\ \hline
    \end{tabular}
    \caption{The rates of one-acceptor and two-acceptor electron transfer for $V_{\mathrm{A^1A^2}}=100\,\mathrm{cm}^{-1}$ and $V_\mathrm{DA^1}=V_\mathrm{DA^2}=109.7\,\mathrm{cm}^{-1}$. The bath parameters appear in the caption of Fig.~\ref{fig:equivalent-acceptors}. The value of $k_{\mathrm{D} {\to} \mathrm{A^1A^2}}$ is the maximum two-acceptor transfer rates for various $\Delta G$ values. $k_{\mathrm{D} {\to} A}$ is the one-acceptor transfer rate where $\Delta G$ maximizes $k_{\mathrm{D} {\to} \mathrm{A^1A^2}}$.}
    \label{tab:ratio100cm}
\end{table}

\begin{table}
    \centering
    \begin{tabular}{cccccc}
    \hline
        Case/Rate & $\Delta G$ ($\mathrm{cm}^{-1}$) & $\lambda^\mathrm{reorg}_{\mathrm{A^1,A^2}}$ ($\mathrm{cm}^{-1}$) & $k_{\mathrm{D} {\to} \mathrm{A^1A^2}}$($\mathrm{ps}^{-1}$) & $k_{\mathrm{D} {\to} A}$ ($\mathrm{ps}^{-1}$) & $\frac{k_{\mathrm{D} {\to} \mathrm{A^1A^2}}}{k_{\mathrm{D} {\to} A}}$ \\
        \hline $T=10\,\mathrm{K}$\\  \hline 
        Full Corr.   & 5399  & $0$       & 24.2   & 11.0 &  2.20 \\
        Partial Corr.& 5399  & $1405$    & 24.3   & 11.0 &  2.21 \\
        Uncorr.      & 5499  & $10490$   & 26.4   & 9.8  &  2.69 \\
        Inv. Corr.   & 5549  & $15736$   & 30.0   & 9.1  &  3.30 \\ 
        \hline $T=100\,\mathrm{K}$ \\ \hline 
        Full Corr.   & 5424  & $0$       & 12.9   & 6.3  &  2.05 \\
        Partial Corr.& 5449  & $1405$    & 12.9   & 6.3  &  2.05 \\
        Uncorr.      & 5574  & $10490$   & 13.4   & 6.0  &  2.23 \\
        Inv. Corr.   & 5674  & $15736$   & 14.8   & 5.7  &  2.60 \\ 
        \hline $T=300\,\mathrm{K}$ \\ \hline 
        Full Corr.   & 5449  & $0$       & 7.7   & 3.8  &  2.03 \\
        Partial Corr.& 5449  & $1405$    & 7.7   & 3.8  &  2.03 \\
        Uncorr.      & 5624  & $10490$   & 7.8   & 3.7  &  2.11 \\
        Inv. Corr.   & 5798  & $15736$   & 8.2   & 3.6  &  2.28 \\ \hline
    \end{tabular}
    \caption{The rates for one- and two-acceptor electron transfer when $V_{\mathrm{A^1A^2}}=200\,\mathrm{cm}^{-1}$ and $V_{\mathrm{DA^1}}=V_{\mathrm{DA^2}}=109.7\,\mathrm{cm}^{-1}$. The bath parameters are indicated in the caption of Fig.~\ref{fig:equivalent-acceptors}. The values of the rates are also indicated in Fig.~\ref{fig:equivalent-acceptors}, \ref{fig:partially-correlated-acceptors} \ref{fig:independent-acceptors}, and \ref{fig:anti-correlated-acceptors}. The value of $k_{\mathrm{D} {\to} \mathrm{A^1A^2}}$ is the maximum of the two-acceptor transfer rate for the different $\Delta G$ values. $k_{\mathrm{D} {\to} A}$ is the one-acceptor transfer rate when $\Delta G$ maximizes $k_{\mathrm{D} {\to} \mathrm{A^1A^2}}$.}
    \label{tab:ratio200cm}
\end{table}

\begin{table}
    \centering
    \begin{tabular}{cccccc}
    \hline
        Case/Rate & $\Delta G$ ($\mathrm{cm}^{-1}$) & $\lambda^\mathrm{reorg}_{\mathrm{A^1,A^2}}$ ($\mathrm{cm}^{-1}$) & $k_{\mathrm{D} {\to} \mathrm{A^1A^2}}$($\mathrm{ps}^{-1}$) & $k_{\mathrm{D} {\to} A}$ ($\mathrm{ps}^{-1}$) & $\frac{k_{\mathrm{D} {\to} \mathrm{A^1A^2}}}{k_{\mathrm{D} {\to} A}}$ \\
        \hline $T=10\,\mathrm{K}$\\  \hline 
        Full Corr.   & 5599  & $0$       & 24.2   & 8.4  &  2.88 \\
        Partial Corr.& 5599  & $1405$    & 24.6   & 8.4  &  2.93 \\
        Uncorr.      & 5699  & $10490$   & 29.6   & 7.0  &  4.23 \\
        Inv. Corr.   & 5724  & $15736$   & 38.2   & 6.6  &  5.79 \\ 
        \hline $T=100\,\mathrm{K}$ \\ \hline 
        Full Corr.   & 5623  & $0$       & 12.9   & 5.8  &  2.22 \\
        Partial Corr.& 5649  & $1405$    & 13.0   & 5.8  &  2.24 \\
        Uncorr.      & 5798  & $10490$   & 14.4   & 5.2  &  2.77 \\
        Inv. Corr.   & 5898  & $15736$   & 17.8   & 4.8  &  3.71 \\ 
        \hline $T=300\,\mathrm{K}$ \\ \hline 
        Full Corr.   & 5649  & $0$       & 7.7   & 3.7  &  2.08 \\
        Partial Corr.& 5674  & $1405$    & 7.7   & 3.7  &  2.08 \\
        Uncorr.      & 5923  & $10490$   & 8.1   & 3.4  &  2.38 \\
        Inv. Corr.   & 6023  & $15736$   & 9.4   & 3.3  &  2.85 \\ \hline
    \end{tabular}
    \caption{The rates of one-acceptor and two-acceptor electron transfer when $V_{\mathrm{A^1A^2}}=400\,\mathrm{cm}^{-1}$ and $V_{\mathrm{DA^1}}=V_{\mathrm{DA^2}}=109.7\,\mathrm{cm}^{-1}$. The bath parameters appear in the caption of Fig.~\ref{fig:equivalent-acceptors}.
    The value of $k_{\mathrm{D} {\to} \mathrm{A^1A^2}}$ is the maximum two-acceptor transfer rate for different $\Delta G$ values. $k_{\mathrm{D} {\to} A}$ is the transfer rate for the one acceptor case, when $\Delta G$ maximizes $k_{\mathrm{D} {\to} \mathrm{A^1A^2}}$.}
    \label{tab:ratio400cm}
\end{table}

\subsection{Electron transfer rates in $\mathrm{DA^1A^2}$ structures with partially correlated reaction coordinates.}
Here, we explore electron transfer with partially correlated vibrations on  $\ket{\mathrm{A^1}}$ and $\ket{\mathrm{A^2}}$. Taking $\alpha=\sqrt{3/4}$ and $\beta=\sqrt{1/4}$, the reorganization energy for $\mathrm{A^1}$ to $\mathrm{A^2}$ electron transfer is $(2 - \sqrt{3})\lambda \approx 0.27 \lambda$, from Eq.\eqref{eq:reorganization-energy-definition}.

The computed rates of electron transfer from D to A, as a function of reaction free energy ($\Delta G=E_\mathrm{D}-E_{\mathrm{A^1}}=E_\mathrm{D}-E_\mathrm{A^2}$), are shown in the second panel of Fig.~\ref{fig:overall-rate-fig}. The relationship between the rate and driving force in the partially correlated case is similar to that found in the fully correlated case shown in the first panel of Fig.~\ref{fig:overall-rate-fig}.  Partial correlation, however, produces a slightly faster rate for the two-acceptor system when $V_{\mathrm{A^1A^2}}=400\,\mathrm{cm}^{-1}$ and $T=10\,\mathrm{K}$ ($24.6\,\mathrm{ps}^{-1}$ vs. $24.2\,\mathrm{ps}^{-1}$), as shown in the first panel (titled `Fully') and the second panel (titled `Partially') in Fig.~\ref{fig:overall-rate-fig}. 
The increased rate arises from enhanced Franck-Condon overlap between the donor and acceptor, and also from the free energy shift that arises from the increased driving force that arises from the electronic coupling between the two acceptors.

\subsection{Electron transfer rates in $\mathrm{DA^1A^2}$ structures with uncorrelated acceptor vibrations.} 
The acceptor nuclear motion is uncorrelated when $\alpha=1$ and $\beta=0$. The reorganization energy for acceptor to acceptor electron transfer is $2\lambda$, and the ET/EnT rates are shown in the third panel (titled `Uncorrelated') of Fig.~\ref{fig:overall-rate-fig} vs. driving force. A more pronounced rate enhancement is observed compared to the partially correlated case, driven by a larger inter-acceptor reorganization energy that 
amplifies the contributions of non-classical/non-statistical vibronic pathways, modifying the rate profile. This can be clearly seen in the expressions of these pathways (see SI).

\section{A Classical model for two-acceptor electron transfer}
\label{sec:SI-classical-model}

A model of two-acceptor electron transfer (ET) mediated by a classical vibrational
environment can be described by the following Hamiltonian that
involves random vibrational coordinates $\mathbf{x}$. The Hamiltonian is similar to Hopfield's
treatment of one-donor one-acceptor electron transfer. The donor-acceptor
coupling $V$ is assumed to be constant. The energies of the donor
and the two acceptors are functions of random vibrational coordinates
$\mathbf{x}$.

\[
\hat{H}=\begin{pmatrix}E_{D}(\mathbf{x}) & V & V\\
V & E_{A_{1}}(\mathbf{x}) & V_{12}\\
V & V_{12} & E_{A_{2}}(\mathbf{x})
\end{pmatrix}
\]

We can rewrite the Hamiltonian using the superposition acceptor states
($\ket{+}$ and $\ket{-}$) by partially diagonalizing the Hamiltonian in the subspace of the
two acceptors. The result is

\[
\hat{H}'=\begin{pmatrix}E_{D} & V_{+}(\mathbf{x}) & V_{-}(\mathbf{x})\\
V_{+}(\mathbf{x}) & E_{+}(\mathbf{x}) & 0\\
V_{-}(\mathbf{x}) & 0 & E_{-}(\mathbf{x})
\end{pmatrix}
\]

In this representation, the coupling $V_{12}$ between the two acceptors $A_{1}$ and
$A_{2}$ is removed. Thus, the total electron transfer rate is a sum
of the transfer rates from the donor to the first acceptor and second
acceptor. However, the price to pay is that the donor-acceptor (donor
to the superposition acceptor states $|+\rangle$ and $|-\rangle$)
couplings $V_{+}(\mathbf{x})$ and $V_{-}(\text{x})$ are no longer
constant. By studying the dependence of the couplings on the vibrational
coordinates, we can intuitively understand the reason for ET rate enhancement.
Now, we give the expression of $V_{+}(\mathbf{x})$ and $V_{-}(\text{x})$.
They are functions of $V$, $E_{A_{1}}(\mathbf{x})$ and $E_{A_{2}}(\mathbf{x})$:
\begin{align*}
V_{+}(\mathbf{x})= & \frac{V\left(E_{A_{1}}-E_{A_{2}}+\sqrt{\left(E_{A_{1}}-E_{A_{2}}\right){}^{2}+4V_{12}^{2}}+2V_{12}\right)}{V_{12}\sqrt{8-\frac{8\left(E_{A_{1}}-E_{A_{2}}\right)}{E_{A_{1}}-E_{A_{2}}-\sqrt{\left(E_{A_{1}}-E_{A_{2}}\right){}^{2}+4V_{12}^{2}}}}}=\frac{V\left(\Delta(\mathbf{x})+\sqrt{\Delta^{2}(\mathbf{x})+4V_{12}^{2}}+2V_{12}\right)}{2\sqrt{2}V_{12}\sqrt{1-\frac{1}{1-\sqrt{1+4V_{12}^{2}/\Delta^{2}(\mathbf{x})}}}}\\
V_{-}(\mathbf{x})= & \frac{V\left(E_{A_{1}}-E_{A_{2}}-\sqrt{\left(E_{A_{1}}-E_{A_{2}}\right){}^{2}+4V_{12}^{2}}-2V_{12}\right)}{V_{12}\sqrt{8-\frac{8\left(E_{A_{1}}-E_{A_{2}}\right)}{E_{A_{1}}-E_{A_{2}}+\sqrt{\left(E_{A_{1}}-E_{A_{2}}\right){}^{2}+4V_{12}^{2}}}}}=\frac{V\left(\Delta(\mathbf{x})-\sqrt{\Delta^{2}(\mathbf{x})+4V_{12}^{2}}-2V_{12}\right)}{2\sqrt{2}V_{12}\sqrt{1-\frac{1}{1+\sqrt{1+4V_{12}^{2}/\Delta^{2}(\mathbf{x})}}}}
\end{align*}
where we can define $\Delta(\mathbf{x})=E_{A_{1}}(\mathbf{x})-E_{A_{2}}(\mathbf{x})$.
For clarity, we do not explicitly write out the dependence of $E_{A_{1}}$
and $E_{A_{2}}$ on the stochastic coordinates $\mathbf{x}$ and one should bear
in mind that $E_{A_{1}}$ and $E_{A_{2}}$ are stochastic variables since they are
functions of stochastic coordinates $\mathbf{x}$. 

The electron transfer now has two channels: one from $\ket{D}$ to $\ket{\uparrow}$, the other from $\ket{D}$ to $\ket{\downarrow}$. The overall rate is the sum of the rates of the two channels. The rates depend on the couplings $V_+$ and $V_-$ quadratically, so it is helpful to calculate $V_+^2$ and $V_-^2$:
\begin{align}
    V_+^2(\mathbf{x}) & = V^2 \left(1+\frac{1}{\sqrt{\left[\frac{\Delta(\mathbf{x})}{2V_{12}}\right]^2 + 1}}\right) \\
    V_-^2(\mathbf{x}) & = V^2 \left(1-\frac{1}{\sqrt{\left[\frac{\Delta(\mathbf{x})}{2V_{12}}\right]^2 + 1}}\right)
\end{align}


\section{A kinetic model for two-acceptor electron transfer}
Give a population vector $\vec{P}(t)=(P_D, P_{A_1}, P_{A_2})^T$, the time evolution of $\vec{P}(t)$ is governed by
\begin{align}
    \frac{d}{dt}\vec{P}(t) = 
    \begin{pmatrix}
        -k_{DA_1}-k_{DA_2} & k_{A_1D} & k_{A_2D} \\
        k_{DA_1} & -k_{A_1D}-k_{A_1A_2} & k_{A_2A_1} \\
        k_{DA_2} & k_{A_1A_2} & -k_{A_2D}-k_{A_2A_1}
    \end{pmatrix}
    \vec{P}(t) =: \mathbf{M}\vec{P}(t).
\end{align}
Assuming $k_{DA_1}=k_{DA_2}=k_1$, $k_{A_1D}=k_{A_2D} = k_2$, and $k_{A_1A_2}=k_{A_2A_1}=k_3$, this model can be reduced to a two state problem by diagonalizing the right-bottom $2\times2$ block of the $\mathbf{M}$:
\begin{align}
    \mathbf{M} = 
    \begin{pmatrix}
        -2k_1 & k_2 & k_2 \\
        k_1 & -k_2-k_3 & k_3 \\
        k_1 & k_3 & -k_2-k_3
    \end{pmatrix} \to 
    \begin{pmatrix}
        -2k_1 & \sqrt{2}k_2 & 0 \\
        \sqrt{2}k_1 & -k_2 & 0 \\
        0 & 0 & -k_2-k_3
    \end{pmatrix}
\end{align}

\section{Three-acceptor Hamiltonian}
The Hamiltonian is 
\begin{align}
    \hat{H} &= \hat{H}_0 + \hat{V} \label{eq:hamiltonian-3-acceptors}\\
    \hat{H}_0 &= \ket{D}\bra{D}\otimes \hat{H}^{(n)}_D + \ket{\mathrm{A^1}}\bra{\mathrm{A^1}}\otimes H^{(n)}_{\mathrm{A^1}} + \ket{\mathrm{A^2}}\bra{\mathrm{A^2}}\otimes H^{(n)}_{\mathrm{A^2}} + \ket{A_3}\bra{A_3}\otimes H^{(n)}_{A_3}\notag \\
    \hat{V} &= V_{DA}\sum_{i=1}^{3}(\ket{D}\bra{A_i} + \ket{A_i}\bra{D}) + V_{AA}\sum_{i,j=1}^3 \ket{A_i}\bra{A_j}. \notag
\end{align}
The nuclear Hamiltonians are
\begin{align*}
    \hat{H}^{(n)}_\mathrm{D} = & E_\mathrm{D} + \sum_n \omega_n \hat{a}_n^\dagger \hat{a}_n + \sum_n \omega_n \hat{b}_n^\dagger \hat{b}_n\\
    \hat{H}_{\mathrm{A}^1}^{(n)} = & E_\mathrm{A^1} + \lambda_\mathrm{DA^1} + \sum_{n} \omega_n a^\dagger_{n} a_{n} + \sum_{n} \omega_n b^\dagger_{n}b_{n}  + \sum_{n} \omega_n c^\dagger_{n}c_{n} + 
        \alpha\sum_n s_n (a^\dagger_n + a_n) + 
        \beta \sum_n s_n (b^\dagger_{n} + b_{n}) +
        \gamma \sum_n s_n (c^\dagger_{n} + c_{n}) 
        \\
    H_{\mathrm{A}^2}^{(n)} = & E_\mathrm{A^2} + \lambda_{\mathrm{DA}^2} + \sum_{n} \omega_n a^\dagger_{n} a_{n} + \sum_{n} \omega_n b^\dagger_{n}b_{n}  + \sum_{n} \omega_n c^\dagger_{n}c_{n} +
        \beta\sum_n s_n (a^\dagger_n + a_n) + 
        \alpha \sum_n s_n (b^\dagger_{n} + b_{n}) +
        \gamma \sum_n s_n (c^\dagger_{n} + c_{n}) \\ 
    H_{\mathrm{A}^3}^{(n)} = & E_\mathrm{A^3} + \lambda_{\mathrm{DA}^2} + \sum_{n} \omega_n a^\dagger_{n} a_{n} + \sum_{n} \omega_n b^\dagger_{n}b_{n}  + \sum_{n} \omega_n c^\dagger_{n}c_{n} +
        \gamma\sum_n s_n (a^\dagger_n + a_n) + 
        \beta \sum_n s_n (b^\dagger_{n} + b_{n}) +
        \alpha \sum_n s_n (c^\dagger_{n} + c_{n}).
\end{align*}
where $\alpha=-\sqrt{\frac{2}{3}}$, $\beta=\frac{1}{\sqrt{6}}$, and $\gamma=\frac{1}{\sqrt{6}}$.

\section{Second-order perturbation theory for a two-acceptor electron transfer
model}

In this section, we derive the rate expression using the second-order
perturbation theory. This scenario corresponds to the ``leading-order''
approximation in the paper. 

The Hamiltonian of a one-donor two-acceptor electron transfer system
is:

\[
\hat{H}_{0}=\hat{h}_{a}\left|a\right\rangle \left\langle a\right|+\hat{h}_{b}\left|b\right\rangle \left\langle b\right|+\hat{h}_{c}\left|c\right\rangle \left\langle c\right|
\]
\[
V=g_{ab}\left|a\right\rangle \left\langle b\right|+g_{ac}\left|a\right\rangle \left\langle c\right|+g_{bc}\left|b\right\rangle \left\langle c\right|+h.c.
\]

\begin{align*}
V_{I}(t)= & g_{ab}e^{i\hat{h}_{a}t}e^{-i\hat{h}_{b}t}\left|a\right\rangle \left\langle b\right|+g_{ab}^{*}e^{i\hat{h}_{b}t}e^{-i\hat{h}_{a}t}\left|b\right\rangle \left\langle a\right|\\
+ & g_{ac}e^{i\hat{h}_{a}t}e^{-i\hat{h}_{c}t}\left|a\right\rangle \left\langle c\right|+g_{ac}^{*}e^{i\hat{h}_{c}t}e^{-i\hat{h}_{a}t}\left|c\right\rangle \left\langle a\right|\\
+ & g_{bc}e^{i\hat{h}_{b}t}e^{-i\hat{h}_{c}t}\left|b\right\rangle \left\langle c\right|+g_{bc}^{*}e^{i\hat{h}_{c}t}e^{-i\hat{h}_{b}t}\left|c\right\rangle \left\langle b\right|
\end{align*}

\[
\hat{U}_{I}(t)=1-i\int_{0}^{t}\hat{V}_{I}(t)dt_{1}-\int_{0}^{t}dt_{1}\int_{0}^{t_{1}}dt_{2}\hat{V}_{I}(t_{1})\hat{V}_{I}(t_{2})+\cdots
\]

The population of the state $\left|a\right\rangle $ is (the Schrodinger-picture
propagator: $e^{-iH_{0}t}\hat{U}_{I}(t)$)
\begin{align*}
P_{a}(t) & =\mathrm{Tr}\left[\rho_{n}\left|a\right\rangle \left\langle a\right|\hat{U}_{I}^{\dagger}(t)e^{iH_{0}t}\left|a\right\rangle \left\langle a\right|e^{-iH_{0}t}\hat{U}_{I}(t)\right]\\
 & =\mathrm{Tr}\left[\rho_{n}\left|a\right\rangle \left\langle a\right|\hat{U}_{I}^{\dagger}(t)\left|a\right\rangle \left\langle a\right|\hat{U}_{I}(t)\right]\\
 & =\mathrm{Tr}_{n}\left[\rho_{n}\left\langle a\right|\hat{U}_{I}^{\dagger}(t)\left|a\right\rangle \left\langle a\right|\hat{U}_{I}(t)\left|a\right\rangle \right]
\end{align*}
The quantity in the middle is
\begin{align*}
\left\langle a\right|\hat{U}_{I}(t)\left|a\right\rangle  & =1-i\int_{0}^{t}\left\langle a\right|\hat{V}_{I}(t)\left|a\right\rangle dt'\\
 & -\int_{0}^{t}dt_{1}\int_{0}^{t_{1}}dt_{2}\left\langle a\right|\hat{V}_{I}(t_{1})\hat{V}_{I}(t_{2})\left|a\right\rangle \\
 & +\cdots
\end{align*}
The expectation value in the first integral is
\[
\left\langle a\right|\hat{V}_{I}(t)\left|a\right\rangle =0.
\]
To calculate the expectation in the second integral, the operator
product $\hat{V}_{I}(t')\hat{V}_{I}(t'')$ needs to be calculated
\begin{align*}
\hat{V}_{I}(t_{1})\hat{V}_{I}(t_{2}) & =g_{ab}^{2}e^{i\hat{h}_{a}t_{1}}e^{-i\hat{h}_{b}(t_{1}-t_{2})}e^{-i\hat{h}_{a}t_{2}}\left|a\right\rangle \left\langle a\right|\\
 & +g_{ac}^{2}e^{i\hat{h}_{a}t_{1}}e^{-i\hat{h}_{c}(t_{1}-t_{2})}e^{-i\hat{h}_{a}t_{2}}\left|a\right\rangle \left\langle a\right|\\
 & +\cdots\left|b\right\rangle \left\langle b\right|+\cdots\left|c\right\rangle \left\langle c\right|.
\end{align*}
Thus, the integral is
\begin{equation}
\int_{0}^{t}dt_{1}\int_{0}^{t_{1}}dt_{2}\left(g_{ab}^{2}e^{i\hat{h}_{a}t_{1}}e^{-i\hat{h}_{b}(t_{1}-t_{2})}e^{-i\hat{h}_{a}t_{2}}+g_{ac}^{2}e^{i\hat{h}_{a}t_{1}}e^{-i\hat{h}_{c}(t_{1}-t_{2})}e^{-i\hat{h}_{a}t_{2}}\right)\label{eq:1}
\end{equation}
each term of which is identical to a 2-state expression (i.e. when
the third state doesn't exist). This is the rate constant of the electron
transfer reaction. 

\section{Higher-order correction in terms of $g_{bc}$}
\label{sec:SI-correction-terms}
\subsection{Infinite-order correction when $\hat{h}_{b}=\hat{h}_{c}$}

We can further assume $\hat{h}_{b}=\hat{h}_{c}$ and calculate higher-order
corrections in terms of $g_{bc}$ to the second-order rate above.
The first correction is order-3 for $\hat{V}_{I}(t)$ and we only
pick the terms that are order-2 for $g_{ab}$ and order-1 for $g_{ac}$:

\begin{align*}
 & (-i)^{3}\int_{0}^{t}dt_{1}\int_{0}^{t_{1}}dt_{2}\int_{0}^{t_{2}}dt_{3}\left\langle a\right|V_{I}(t_{1})V_{I}(t_{2})V_{I}(t_{3})\left|a\right\rangle \\
\sim & (-i)^{3}\int_{0}^{t}dt_{1}\int_{0}^{t_{1}}dt_{2}\int_{0}^{t_{2}}dt_{3}\underbrace{g_{ab}e^{i\hat{h}_{a}t_{1}}e^{-i\hat{h}_{b}t_{1}}g_{bc}e^{i\hat{h}_{b}t_{2}}e^{-i\hat{h}_{c}t_{2}}g_{ac}^{*}e^{i\hat{h}_{c}t_{3}}e^{-i\hat{h}_{a}t_{3}}}_{ab|bc|c^{*}a^{*}}\\
 & +(-i)^{3}\int_{0}^{t}dt_{1}\int_{0}^{t_{1}}dt_{2}\int_{0}^{t_{2}}dt_{3}\underbrace{g_{ac}e^{i\hat{h}_{a}t_{1}}e^{-i\hat{h}_{c}t_{1}}g_{bc}^{*}e^{i\hat{h}_{c}t_{2}}e^{-i\hat{h}_{b}t_{2}}g_{ab}^{*}e^{i\hat{h}_{b}t_{3}}e^{-i\hat{h}_{a}t_{3}}}_{ac|c^{*}b^{*}|b^{*}a^{*}}
\end{align*}
The first integral is (after using $\hat{h}_{b}=\hat{h}_{c}$)
\begin{align*}
 & (-i)^{3}\int_{0}^{t}dt_{1}\int_{0}^{t_{1}}dt_{2}\int_{0}^{t_{2}}dt_{3}g_{ab}g_{bc}g_{ac}^{*}e^{i\hat{h}_{a}t_{1}}e^{-i\hat{h}_{b}t_{1}}e^{i\hat{h}_{b}t_{3}}e^{-i\hat{h}_{a}t_{3}}\\
= & (-i)^{3}g_{ab}g_{bc}g_{ac}^{*}\int_{0}^{t}dt_{1}e^{i\hat{h}_{a}t_{1}}e^{-i\hat{h}_{b}t_{1}}{\color{yellow}{\color{red}\int_{0}^{t_{1}}dt_{2}\int_{0}^{t_{2}}dt_{3}}}e^{i\hat{h}_{b}t_{3}}e^{-i\hat{h}_{a}t_{3}}\\
= & (-i)^{3}g_{ab}g_{bc}g_{ac}^{*}\int_{0}^{t}dt_{1}e^{i\hat{h}_{a}t_{1}}e^{-i\hat{h}_{b}t_{1}}{\color{red}\int_{0}^{t_{1}}dt_{3}\int_{t_{3}}^{t_{1}}dt_{2}}e^{i\hat{h}_{b}t_{3}}e^{-i\hat{h}_{a}t_{3}}\\
= & (-i)^{3}g_{ab}g_{bc}g_{ac}^{*}\int_{0}^{t}dt_{1}e^{i\hat{h}_{a}t_{1}}e^{-i\hat{h}_{b}t_{1}}\int_{0}^{t_{1}}dt_{3}(t_{1}-t_{3})e^{i\hat{h}_{b}t_{3}}e^{-i\hat{h}_{a}t_{3}}\\
= & (-i)^{3}g_{ab}g_{bc}g_{ac}^{*}\int_{0}^{t}dt_{1}\int_{0}^{t_{1}}dt_{3}(t_{1}-t_{3})e^{i\hat{h}_{a}t_{1}}e^{-i\hat{h}_{b}(t_{1}-t_{3})}e^{-i\hat{h}_{a}t_{3}}\\
= & (-i)^{3}g_{ab}g_{bc}g_{ac}^{*}\int_{0}^{t}dt_{1}\int_{0}^{t_{1}}dt_{2}(t_{1}-t_{2})e^{i\hat{h}_{a}t_{1}}e^{-i\hat{h}_{b}(t_{1}-t_{2})}e^{-i\hat{h}_{a}t_{2}}
\end{align*}
which is almost the same as the integral in \ref{eq:1} except an
additional time difference term $(t_{1}-t_{2})$. From the second
line to the third line the integration order is changed. In the last
line $t_{3}$ is replaced with $t_{2}$.

Continuing this process to calculate higher-order corrections in terms
of $g_{bc}$ up until the infinite order will take full effect of
the coupling between $\left|b\right\rangle $ and $\left|c\right\rangle $
into account. The expression of the correction is
\begin{align*}
\mathrm{correction}= & g_{ab}g_{bc}g_{ac}^{*}\int_{0}^{t}dt_{1}\int_{0}^{t_{1}}dt_{2}e^{-ig_{bc}(t_{1}-t_{2})}e^{i\hat{h}_{a}t_{1}}e^{-i\hat{h}_{b}(t_{1}-t_{2})}e^{-i\hat{h}_{a}t_{2}}\\
+ & g_{ac}g_{bc}^{*}g_{ac}^{*}\int_{0}^{t}dt_{1}\int_{0}^{t_{1}}dt_{2}e^{-ig_{bc}(t_{1}-t_{2})}e^{i\hat{h}_{a}t_{1}}e^{-i\hat{h}_{b}(t_{1}-t_{2})}e^{-i\hat{h}_{a}t_{2}}.
\end{align*}

Adding this correction to the rate equation in \ref{eq:1}, we have
the final expression of the Fermi's golden rule electron transfer
rate constant for a two-acceptor system
\begin{align*}
 & \int_{0}^{t}dt_{1}\int_{0}^{t_{1}}dt_{2}\left(g_{ab}^{2}e^{i\hat{h}_{a}t_{1}}e^{-i\hat{h}_{b}(t_{1}-t_{2})}e^{-i\hat{h}_{a}t_{2}}+g_{ac}^{2}e^{i\hat{h}_{a}t_{1}}e^{-i\hat{h}_{b}(t_{1}-t_{2})}e^{-i\hat{h}_{a}t_{2}}\right)\\
+ & \int_{0}^{t}dt_{1}\int_{0}^{t_{1}}dt_{2}g_{ab}^{*}g_{ac}e^{ig_{bc}(t_{1}-t_{2})}e^{i\hat{h}_{a}t_{1}}e^{-i\hat{h}_{b}(t_{1}-t_{2})}e^{-i\hat{h}_{a}t_{2}}\\
+ & \int_{0}^{t}dt_{1}\int_{0}^{t_{1}}dt_{2}g_{ab}^{*}g_{ac}e^{ig_{bc}(t_{1}-t_{2})}e^{i\hat{h}_{a}t_{1}}e^{-i\hat{h}_{b}(t_{1}-t_{2})}e^{-i\hat{h}_{a}t_{2}}.
\end{align*}
Further assuming $g_{ab}=g_{ac}$, the rate expression above is reduced
to

\[
2\int_{0}^{t}dt_{1}\int_{0}^{t_{1}}dt_{2}\left(g_{ab}^{2}e^{i\hat{h}_{a}t_{1}}e^{-i(\hat{h}_{b}+g_{ab})(t_{1}-t_{2})}e^{-i\hat{h}_{a}t_{2}}\right).
\]

\subsection{General cases when $\hat{h}_{a}$, $\hat{h}_{b}$ and $\hat{h}_{c}$
are harmonic}

Using the polaron transformation, we remove the diagonal linear vibronic
couplings (LVC). The 3-state LVC Hamiltonian is
\begin{align*}
H_{0} & =\left(E_{a}+\omega a^{\dagger}a+s_{a}(a^{\dagger}+a)\right)\left|a\right\rangle \left\langle a\right|\\
 & +\left(E_{b}+\omega a^{\dagger}a+s_{b}(a^{\dagger}+a)\right)\left|b\right\rangle \left\langle b\right|\\
 & +\left(E_{c}+\omega a^{\dagger}a+s_{c}(a^{\dagger}+a)\right)\left|c\right\rangle \left\langle c\right|\\
V & =g_{12}\left|a\right\rangle \left\langle b\right|+g_{13}\left|a\right\rangle \left\langle c\right|+g_{23}\left|b\right\rangle \left\langle c\right|+h.c.
\end{align*}
The polaron transformation uses the unitary operator 
\[
U_{p}=e^{\sum_{i}(a^{\dagger}-a)\frac{s_{i}}{\omega}\left|i\right\rangle \left\langle i\right|}.
\]
Noticing $e^{\lambda(a^{\dagger}-a)}ae^{-\lambda(a^{\dagger}-a)}=a-\lambda$
and applying $U_{p}$ to $H_{0}$:
\begin{align*}
\tilde{H}_{0}= & U_{p}H_{0}U_{p}^{\dagger}\\
= & \left(E_{a}+\omega(a^{\dagger}-\frac{s_{a}}{\omega})(a-\frac{s_{a}}{\omega})+s_{a}(a^{\dagger}+a-2\frac{s_{a}}{\omega})\right)\left|a\right\rangle \left\langle a\right|\\
+ & \left(E_{b}+\omega(a^{\dagger}a-\frac{s_{b}}{\omega}(a^{\dagger}+a)+\frac{s_{b}^{2}}{\omega^{2}})+s_{b}(a^{\dagger}+a-2\frac{s_{b}}{\omega})\right)\left|b\right\rangle \left\langle b\right|\\
+ & \left(E_{c}+\omega a^{\dagger}a-\frac{s_{c}^{2}}{\omega}\right)\left|c\right\rangle \left\langle c\right|\\
= & \sum_{i}\left(E_{i}+\omega a^{\dagger}a-s_{i}^{2}/\omega\right)\left|i\right\rangle \left\langle i\right|.
\end{align*}
Applying $U_{p}$ to $V$:
\begin{align*}
\tilde{V}= & U_{p}VU_{p}^{\dagger}\\
= & g_{ab}\left|a\right\rangle \left\langle b\right|e^{(a^{\dagger}-a)(s_{a}-s_{b})/\omega}\\
+ & g_{ac}\left|a\right\rangle \left\langle c\right|e^{(a^{\dagger}-a)(s_{a}-s_{c})/\omega}\\
+ & g_{bc}\left|b\right\rangle \left\langle c\right|e^{(a^{\dagger}-a)(s_{b}-s_{c})/\omega}+h.c.
\end{align*}
To use the perturbation theory, first note
\begin{align*}
\tilde{V}_{I}(t) & =e^{i\tilde{H}_{0}t}\tilde{V}e^{-i\tilde{H}_{0}t}\\
 & =g_{ab}\left|a\right\rangle \left\langle b\right|e^{i\left((E_{a}-\frac{s_{a}^{2}}{\omega})-(E_{b}-\frac{s_{b}^{2}}{\omega})\right)t}e^{(a^{\dagger}e^{i\omega t}-ae^{-i\omega t})(s_{a}-s_{b})/\omega}\\
 & +g_{ba}\left|b\right\rangle \left\langle a\right|e^{-i\left((E_{a}-\frac{s_{a}^{2}}{\omega})-(E_{b}-\frac{s_{b}^{2}}{\omega})\right)t}e^{-(a^{\dagger}e^{i\omega t}-ae^{-i\omega t})(s_{a}-s_{b})/\omega}\\
 & +g_{ac}\left|a\right\rangle \left\langle c\right|e^{i\left((E_{a}-\frac{s_{a}^{2}}{\omega})-(E_{c}-\frac{s_{c}^{2}}{\omega})\right)t}e^{(a^{\dagger}e^{i\omega t}-ae^{-i\omega t})(s_{a}-s_{c})/\omega}\\
 & +g_{ca}\left|c\right\rangle \left\langle a\right|e^{-i\left((E_{a}-\frac{s_{a}^{2}}{\omega})-(E_{c}-\frac{s_{c}^{2}}{\omega})\right)t}e^{-(a^{\dagger}e^{i\omega t}-ae^{-i\omega t})(s_{a}-s_{c})/\omega}\\
 & +g_{bc}\left|b\right\rangle \left\langle c\right|e^{i\left((E_{b}-\frac{s_{b}^{2}}{\omega})-(E_{c}-\frac{s_{c}^{2}}{\omega})\right)t}e^{(a^{\dagger}e^{i\omega t}-ae^{-i\omega t})(s_{b}-s_{c})/\omega}\\
 & +g_{cb}\left|c\right\rangle \left\langle b\right|e^{-i\left((E_{b}-\frac{s_{b}^{2}}{\omega})-(E_{c}-\frac{s_{c}^{2}}{\omega})\right)t}e^{-(a^{\dagger}e^{i\omega t}-ae^{-i\omega t})(s_{b}-s_{c})/\omega}.
\end{align*}
Then, we consider, as we did in Section 1, the population on $\left|1\right\rangle $
and let the initial nuclear density matrix be $\rho_{n}=e^{\beta a^{\dagger}a}/\mathcal{Z}$.
\begin{align*}
P_{a}(t) & =\mathrm{Tr}\left[\rho_{n}\left|a\right\rangle \left\langle a\right|\underbrace{U_{p}^{\dagger}\hat{U}_{I}^{\dagger}(t)e^{iH_{0}t}U_{p}}_{U_{S}^{\dagger}(t)}\left|a\right\rangle \left\langle a\right|\underbrace{U_{p}^{\dagger}e^{-iH_{0}t}\hat{U}_{I}(t)U_{p}}_{U_{S}(t)}\right]\\
 & =\mathrm{Tr}_{n}\left[\rho_{n}\left(e^{-(a^{\dagger}-a)\frac{s_{a}}{\omega}}\left\langle a\right|\hat{U}_{I}^{\dagger}(t)\left|a\right\rangle e^{(a^{\dagger}-a)\frac{s_{a}}{\omega}}\right)\left(e^{-(a^{\dagger}-a)\frac{s_{1}}{\omega}}\left\langle a\right|\hat{U}_{I}^{\dagger}(t)\left|a\right\rangle e^{(a^{\dagger}-a)\frac{s_{a}}{\omega}}\right)\right]\\
 & \approx1-\int_{0}^{t}dt_{1}\int_{0}^{t_{1}}dt_{2}\mathrm{Tr}_{n}\left[\rho_{n}e^{-(a^{\dagger}-a)\frac{s_{a}}{\omega}}\left\langle a\right|\tilde{V}_{I}(t_{1})\tilde{V}_{I}(t_{2})\left|a\right\rangle e^{(a^{\dagger}-a)\frac{s_{a}}{\omega}}\right]\\
 & \;\;\;\;\;\,\,\,-\int_{0}^{t}dt_{1}\int_{0}^{t_{1}}dt_{2}\mathrm{Tr}_{n}\left[\rho_{n}e^{-(a^{\dagger}-a)\frac{s_{a}}{\omega}}\left\langle a\right|\tilde{V}_{I}^{\dagger}(t_{2})\tilde{V}_{I}^{\dagger}(t_{1})\left|a\right\rangle e^{(a^{\dagger}-a)\frac{s_{a}}{\omega}}\right]\\
 & =1-2\mathrm{Re}\left\{ \int_{0}^{t}dt_{1}\int_{0}^{t_{1}}dt_{2}\mathrm{Tr}_{n}\left[\rho_{n}e^{-(a^{\dagger}-a)\frac{s_{a}}{\omega}}\left\langle a\right|\tilde{V}_{I}(t_{1})\tilde{V}_{I}(t_{2})\left|a\right\rangle e^{(a^{\dagger}-a)\frac{s_{a}}{\omega}}\right]\right\} 
\end{align*}
Keep the terms involving $g_{ac}$ and $g_{ac}$ up to the second-order,
higer-order corrections in terms of $g_{bc}$ can be made to the integral
inside the Re function. They are listed in \ref{tab:High-order-corrections-1}.
\begin{table}[h]
\caption{High-order corrections in terms of $g_{bc}$ to the integral inside
$\mathrm{Re}(\cdot)$. We use a convention: $f_{mn}(t)=e^{i\left((E_{m}-\frac{s_{m}^{2}}{\omega})-(E_{n}-\frac{s_{n}^{2}}{\omega})\right)t}e^{(a^{\dagger}e^{i\omega t}-ae^{-i\omega t})(s_{m}-s_{n})/\omega}$
($m,n=a,b,c$). \label{tab:High-order-corrections-1}}

\centering{}%
\begin{tabular}{cc|c}
\hline 
Order & Pathway & $\mathrm{Tr}_{n}\left[\rho_{n}e^{-(a^{\dagger}-a)\frac{s_{1}}{\omega}}\left(\mathrm{Expression}\right)e^{(a^{\dagger}-a)\frac{s_{1}}{\omega}}\right]$\tabularnewline
\hline 
1 & ab-bc-ca & $(\textrm{-}i)^{3}\int_{0}^{t}dt_{1}\int_{0}^{t_{1}}dt_{2}\int_{0}^{t_{2}}dt_{3}f_{ab}(t_{1})f_{bc}(t_{2})f_{ca}(t_{3})$\tabularnewline
2 & ab-bc-cb-ba & $(\textrm{-}i)^{4}\int_{0}^{t}dt_{1}\int_{0}^{t_{1}}dt_{2}\int_{0}^{t_{2}}dt_{3}\int_{0}^{t_{3}}dt_{4}f_{ab}(t_{1})f_{bc}(t_{2})f_{cb}(t_{3})f_{ba}(t_{4})$\tabularnewline
3 & ab-bc-cb-bc-ca & $(\textrm{-}i)^{5}\int_{0}^{t}dt_{1}\int_{0}^{t_{1}}dt_{2}\int_{0}^{t_{2}}dt_{3}\int_{0}^{t_{3}}dt_{4}\int_{0}^{t_{4}}dt_{5}f_{ab}(t_{1})f_{bc}(t_{2})f_{cb}(t_{3})f_{bc}(t_{4})f_{ca}(t_{5})$\tabularnewline
4 & ab-bc-cb-bc-cb-ba & $(\textrm{-}i)^{6}\int_{0}^{t}dt_{1}\int_{0}^{t_{1}}dt_{2}\int_{0}^{t_{2}}dt_{3}\int_{0}^{t_{3}}dt_{4}\int_{0}^{t_{4}}dt_{5}\int_{0}^{t_{5}}dt_{6}f_{ab}(t_{1})f_{bc}(t_{2})f_{cb}(t_{3})f_{bc}(t_{4})f_{cb}(t_{4})f_{ba}(t_{6})$\tabularnewline
\hline 
\end{tabular}
\end{table}
By changing the integration order, we have a new table, \ref{tab:The-integrals}.
The integration order is changed such that the inetgrals most inside
are done first, so that the entire integrals can be calcualted from
inside to outside, one by one. This way, the correction terms can
be regarded as a correction to $h_{b}$ and $h_{c}$ in \ref{eq:1}.
This is ensured by the first integral (over the variable $t_{1}$)
and the last integral (ovet the integration variable $t_{n\textrm{-}max}$)
sandwiching the integrals inside. 
\begin{table}[h]
\caption{The integrals in Table I after changing integration order. $\hat{f}_{mn}(t)=e^{i\left((E_{m}-\frac{s_{m}^{2}}{\omega})-(E_{n}-\frac{s_{n}^{2}}{\omega})\right)t}e^{(a^{\dagger}e^{i\omega t}-ae^{-i\omega t})(s_{m}-s_{n})/\omega}$
($m,n=a,b,c$). \label{tab:The-integrals}}

\centering{}%
\begin{tabular}{cc|c}
\hline 
Order & Pathway & $\mathrm{Tr}_{n}\left[\rho_{n}e^{-(a^{\dagger}-a)\frac{s_{1}}{\omega}}\left(\mathrm{Expression}\right)e^{(a^{\dagger}-a)\frac{s_{1}}{\omega}}\right]$\tabularnewline
\hline 
1 & ab-bc-ca & $(\textrm{-}i)^{3}\int_{0}^{t}dt_{1}\int_{0}^{t_{1}}dt_{3}f_{ab}(t_{1})\left(\int_{t_{3}}^{t_{1}}dt_{2}f_{bc}(t_{2})\right)f_{ca}(t_{3})$\tabularnewline
2 & ab-bc-cb-ba & $(\textrm{-}i)^{4}\int_{0}^{t}dt_{1}\int_{0}^{t_{1}}dt_{4}f_{ab}(t_{1})\left[\int_{t_{4}}^{t_{1}}dt_{3}\left(\int_{t_{3}}^{t_{1}}dt_{2}f_{bc}(t_{2})\right)f_{cb}(t_{3})\right]f_{ba}(t_{4})$\tabularnewline
3 & ab-bc-cb-bc-ca & $(\textrm{-}i)^{5}\int_{0}^{t}dt_{1}\int_{0}^{t_{1}}dt_{5}f_{ab}(t_{1})\left\{ \int_{t_{5}}^{t_{1}}dt_{4}\left[\int_{t_{4}}^{t_{1}}dt_{3}\left(\int_{t_{3}}^{t_{1}}dt_{2}f_{bc}(t_{2})\right)f_{cb}(t_{3})\right]f_{bc}(t_{4})\right\} f_{ca}(t_{5})$\tabularnewline
4 & ab-bc-cb-bc-cb-ba & $(\textrm{-}i)^{6}\int_{0}^{t}dt_{1}\int_{0}^{t_{1}}dt_{6}f_{ab}(t_{1})\left(\int_{t_{6}}^{t_{1}}dt_{5}\left\{ \int_{t_{5}}^{t_{1}}dt_{4}\left[\int_{t_{4}}^{t_{1}}dt_{3}\left(\int_{t_{3}}^{t_{1}}dt_{2}f_{bc}(t_{2})\right)f_{cb}(t_{3})\right]f_{bc}(t_{4})\right\} f_{cb}(t_{5})\right)f_{ba}(t_{6})$\tabularnewline
\hline 
\end{tabular}
\end{table}
 As a verification of the formula given in \ref{tab:The-integrals},
we can set $f_{bc}=1$ which implies $h_{b}=h_{c}$. The correction
in order 2, for example, gives $\frac{(t_{1}-t_{4})^{2}}{2!}$; the
correction in order 3 gives $\frac{(t_{1}-t_{5})^{3}}{6}$. These
results confirms the correctness of the correction terms in \ref{tab:The-integrals}.

\subsubsection{Calculation of the trace}

One way to proceed is to calculate the trace of the operators inside
the integral first. 

For the first-order correction, the trace of the operators is
\[
\mathrm{Tr}\left[\rho_{n}e^{-(a^{\dagger}-a)\frac{s_{1}}{\omega}}\hat{f}_{ab}(t_{1})\hat{f}_{bc}(t_{2})\hat{f}_{ca}(t_{3})e^{(a^{\dagger}-a)\frac{s_{1}}{\omega}}\right]
\]

The product $f_{ab}(t_{1})f_{bc}(t_{2})f_{ca}(t_{3})$ is
\begin{align*}
 & \exp\left[i\left((E_{a}-\frac{s_{a}^{2}}{\omega})-(E_{b}-\frac{s_{b}^{2}}{\omega})\right)t_{1}+i\left((E_{b}-\frac{s_{b}^{2}}{\omega})-(E_{c}-\frac{s_{c}^{2}}{\omega})\right)t_{2}+i\left((E_{c}-\frac{s_{c}^{2}}{\omega})-(E_{a}-\frac{s_{a}^{2}}{\omega})\right)t_{3}\right]\times\\
 & e^{-(a^{\dagger}-a)\frac{s_{1}}{\omega}}\times\\
 & {\color{teal}e^{(a^{\dagger}e^{i\omega t_{1}}-ae^{-i\omega t_{1}})(s_{a}-s_{b})/\omega}e^{(a^{\dagger}e^{i\omega t_{2}}-ae^{-i\omega t_{2}})(s_{b}-s_{c})/\omega}e^{(a^{\dagger}e^{i\omega t_{3}}-ae^{-i\omega t_{3}})(s_{c}-s_{a})/\omega}}\\
 & e^{(a^{\dagger}-a)\frac{s_{1}}{\omega}}
\end{align*}
The part in green is, since $e^{A}e^{B}e^{C}=e^{A+B}e^{C}e^{[A,B]/2}=e^{A+B+C}e^{[A,B]/2}e^{[A,C]/2+[B,C]/2}$,
\begin{align*}
 & \exp\Big\{(a^{\dagger}e^{i\omega t_{1}}-ae^{-i\omega t_{1}})(s_{a}-s_{b})/\omega+\\
 & (a^{\dagger}e^{i\omega t_{2}}-ae^{-i\omega t_{2}})(s_{b}-s_{c})/\omega+\\
 & (a^{\dagger}e^{i\omega t_{3}}-ae^{-i\omega t_{3}})(s_{c}-s_{a})/\omega\Big\}\\
\times & \exp\left\{ \frac{(s_{a}-s_{b})(s_{b}-s_{c})}{\omega^{2}}\left[i\sin\omega(t_{1}-t_{2})\right]\right\} \\
\times & \exp\left\{ \frac{(s_{a}-s_{b})(s_{c}-s_{a})}{\omega^{2}}\left[i\sin\omega(t_{1}-t_{3})\right]\right\} \\
\times & \exp\left\{ \frac{(s_{b}-s_{c})(s_{c}-s_{a})}{\omega^{2}}\left[i\sin\omega(t_{2}-t_{3})\right]\right\} 
\end{align*}
where we use $[a^{\dagger}e^{i\omega t_{1}}-ae^{-i\omega t_{1}},a^{\dagger}e^{i\omega t_{2}}-ae^{-i\omega t_{2}}]=[a^{\dagger},-a]e^{i\omega(t_{1}-t_{2})}+[-a,a^{\dagger}]e^{-i\omega(t_{1}-t_{2})}=e^{i\omega(t_{1}-t_{2})}-e^{-i\omega(t_{1}-t_{2})}=2i\sin\omega(t_{1}-t_{2})$.

After applying the shift operator $e^{-(a^{\dagger}-a)\frac{s_{a}}{\omega}}(\cdot)e^{(a^{\dagger}-a)\frac{s_{a}}{\omega}}$,
the product $f_{ab}(t_{1})f_{bc}(t_{2})f_{ca}(t_{3})$ is (w/out all
the time factors)
\begin{align*}
 & \exp\Big\{\left[(a^{\dagger}+\frac{s_{a}}{\omega})e^{i\omega t_{1}}-(a+\frac{s_{a}}{\omega})e^{-i\omega t_{1}}\right](s_{a}-s_{b})/\omega+\\
 & \left[(a^{\dagger}+\frac{s_{a}}{\omega})e^{i\omega t_{2}}-(a+\frac{s_{a}}{\omega})e^{-i\omega t_{2}}\right](s_{b}-s_{c})/\omega+\\
 & \left[(a^{\dagger}+\frac{s_{a}}{\omega})e^{i\omega t_{3}}-(a+\frac{s_{a}}{\omega})e^{-i\omega t_{3}}\right](s_{c}-s_{a})/\omega\Big\}\\
= & {\color{red}\exp\Bigg\{ a^{\dagger}\left(\frac{s_{a}-s_{b}}{\omega}e^{i\omega t_{1}}+\frac{s_{b}-s_{c}}{\omega}e^{i\omega t_{2}}+\frac{s_{c}-s_{a}}{\omega}e^{i\omega t_{3}}\right)}\\
 & {\color{red}-a\left(\frac{s_{a}-s_{b}}{\omega}e^{-i\omega t_{1}}+\frac{s_{b}-s_{c}}{\omega}e^{-i\omega t_{2}}+\frac{s_{c}-s_{a}}{\omega}e^{-i\omega t_{3}}\right)\Bigg\}}\\
\times & \exp2i\left\{ \frac{s_{a}(s_{a}-s_{b})}{\omega^{2}}\sin\omega t_{1}+\frac{s_{a}(s_{b}-s_{c})}{\omega^{2}}\sin\omega t_{2}+\frac{s_{a}(s_{c}-s_{a})}{\omega^{2}}\sin\omega t_{3}\right\} \\
\equiv & \exp\left\{ {\color{red}\tilde{f}(t_{1},t_{2},t_{3})}\right\} \exp\left\{ \cdots\right\} 
\end{align*}
The trace of the operator part over the initial nuclear density matrix
$\rho_{n}$ is
\begin{align*}
 & \mathrm{Tr}\left[\rho_{n}e^{\tilde{f}(t_{1},t_{2},t_{3})}\right]\\
= & e^{\frac{1}{2}\left\langle \tilde{f}(t_{1},t_{2},t_{3})^{2}\right\rangle }\\
= & \exp\left\{ \frac{-1}{2}\left\langle a^{\dagger}a+aa^{\dagger}\right\rangle \left|\frac{s_{a}-s_{b}}{\omega}e^{i\omega t_{1}}+\frac{s_{b}-s_{c}}{\omega}e^{i\omega t_{2}}+\frac{s_{c}-s_{a}}{\omega}e^{i\omega t_{3}}\right|^{2}\right\} \\
= & \exp\left\{ \frac{-1}{2}\coth\left(\frac{\beta\omega}{2}\right)\left|\frac{s_{a}-s_{b}}{\omega}e^{i\omega t_{1}}+\frac{s_{b}-s_{c}}{\omega}e^{i\omega t_{2}}+\frac{s_{c}-s_{a}}{\omega}e^{i\omega t_{3}}\right|^{2}\right\} .
\end{align*}
Summerizing the results we have:

\paragraph{The first-order correction}

(using $\Delta_{ij}\equiv s_{i}-s_{i}$ and $\tilde{E}_{i}\equiv E_{i}-s_{i}^{2}/\omega$)
\begin{align*}
(\textrm{-}i)^{3}\int_{0}^{t}dt_{1}\int_{0}^{t_{1}}dt_{2}\int_{0}^{t_{2}}dt_{3} & e^{i\left(\tilde{E}_{a}-\tilde{E}_{b}\right)t_{1}+i\left(\tilde{E}_{b}-\tilde{E}_{c}\right)t_{2}+i\left(\tilde{E}_{c}-\tilde{E}_{a}\right)t_{3}}\\
\times & \exp\left\{ \frac{\Delta_{ab}\Delta_{bc}}{\omega^{2}}\left[i\sin\omega(t_{1}-t_{2})\right]\right\} \\
\times & \exp\left\{ \frac{\Delta_{ab}\Delta_{ca}}{\omega^{2}}\left[i\sin\omega(t_{1}-t_{3})\right]\right\} \\
\times & \exp\left\{ \frac{\Delta_{bc}\Delta_{ca}}{\omega^{2}}\left[i\sin\omega(t_{2}-t_{3})\right]\right\} \\
\times & \exp2i\left\{ \frac{s_{a}\Delta_{ab}}{\omega^{2}}\sin\omega t_{1}+\frac{s_{a}\Delta_{bc}}{\omega^{2}}\sin\omega t_{2}+\frac{s_{a}\Delta_{ca}}{\omega^{2}}\sin\omega t_{3}\right\} \\
\times & \exp\left\{ \frac{-1}{2}\coth\left(\frac{\beta\omega}{2}\right)\left|\frac{\Delta_{ab}}{\omega}e^{i\omega t_{1}}+\frac{\Delta_{bc}}{\omega}e^{i\omega t_{2}}+\frac{\Delta_{ca}}{\omega}e^{i\omega t_{3}}\right|^{2}\right\} 
\end{align*}
The last part can be simplified further to be
\begin{align*}
 & \cdots\\
\times & \exp\left\{ -\coth\left(\frac{\beta\omega}{2}\right)\left[\frac{\Delta_{ab}\Delta_{bc}\cos\omega(t_{1}-t_{2})}{\omega^{2}}+\frac{\Delta_{ab}\Delta_{ca}\cos\omega(t_{1}-t_{3})}{\omega^{2}}+\frac{\Delta_{bc}\Delta_{ca}\cos\omega(t_{1}-t_{2})}{\omega^{2}}+\frac{\Delta_{ab}^{2}+\Delta_{bc}^{2}+\Delta_{ca}^{2}}{2\omega^{2}}\right]\right\} 
\end{align*}
The whole expression is
\begin{align*}
(\textrm{-}i)^{3}\int_{0}^{t}dt_{1}\int_{0}^{t_{1}}dt_{2}\int_{0}^{t_{2}}dt_{3} & e^{i\tilde{E}_{a}\left(t_{1}-t_{3}\right)-i\tilde{E}_{b}\left(t_{1}-t_{2}\right)-i\tilde{E}_{c}\left(t_{2}-t_{3}\right)}\\
\times & \exp\left\{ \frac{\Delta_{ab}\Delta_{bc}}{\omega^{2}}\left[-\coth\left(\frac{\beta\omega}{2}\right)\cos\omega(t_{1}-t_{2})+i\sin\omega(t_{1}-t_{2})\right]\right\} \\
\times & \exp\left\{ \frac{\Delta_{ab}\Delta_{ca}}{\omega^{2}}\left[-\coth\left(\frac{\beta\omega}{2}\right)\cos\omega(t_{1}-t_{3})+i\sin\omega(t_{1}-t_{3})\right]\right\} \\
\times & \exp\left\{ \frac{\Delta_{bc}\Delta_{ca}}{\omega^{2}}\left[-\coth\left(\frac{\beta\omega}{2}\right)\cos\omega(t_{2}-t_{3})+i\sin\omega(t_{2}-t_{3})\right]\right\} \\
\times & \exp\left\{ -\coth\left(\frac{\beta\omega}{2}\right)\frac{\Delta_{ab}^{2}+\Delta_{bc}^{2}+\Delta_{ca}^{2}}{2\omega^{2}}\right\} \\
\times & \exp2i\left\{ \frac{s_{a}\Delta_{ab}}{\omega^{2}}\sin\omega t_{1}+\frac{s_{a}\Delta_{bc}}{\omega^{2}}\sin\omega t_{2}+\frac{s_{a}\Delta_{ca}}{\omega^{2}}\sin\omega t_{3}\right\} 
\end{align*}

If $\Delta_{bc}=0$, $\tilde{E_{b}}=\tilde{E}_{c}$, and $s_{a}=0$,
the correction is 
\begin{align*}
(\textrm{-}i)^{3}\int_{0}^{t}dt_{1}\int_{0}^{t_{1}}dt_{2}\int_{0}^{t_{2}}dt_{3} & e^{i(\tilde{E}_{a}-\tilde{E}_{b})\left(t_{1}-t_{3}\right)}\\
\times & \exp\left\{ -\frac{\Delta_{ab}^{2}}{\omega^{2}}\left[-\coth\left(\frac{\beta\omega}{2}\right)\cos\omega(t_{1}-t_{3})+i\sin\omega(t_{1}-t_{3})\right]\right\} \\
\times & \exp\left\{ -\coth\left(\frac{\beta\omega}{2}\right)\frac{\Delta_{ab}^{2}}{\omega^{2}}\right\} \\
= & \exp\left\{ \frac{\Delta_{ab}^{2}}{\omega^{2}}\left[\coth\left(\frac{\beta\omega}{2}\right)\left[1-\cos\omega(t_{1}-t_{3})\right]+i\sin\omega(t_{1}-t_{3})\right]\right\} 
\end{align*}
End

\paragraph{The second-order correction (ab-bc-cb-ba),
\begin{align*}
(\textrm{-}i)^{4}\int_{0}^{t}dt_{1}\int_{0}^{t_{1}}dt_{2}\int_{0}^{t_{2}}dt_{3}\int_{0}^{t_{3}}dt_{4} & e^{i\left(\tilde{E}_{a}-\tilde{E}_{b}\right)t_{1}+i\left(\tilde{E}_{b}-\tilde{E}_{c}\right)t_{2}+i\left(\tilde{E}_{c}-\tilde{E}_{b}\right)t_{3}+i\left(\tilde{E}_{b}-\tilde{E}_{a}\right)t_{4}}\protect\\
\times & \exp\left\{ \frac{\Delta_{ab}\Delta_{bc}}{\omega^{2}}\left[-\coth\left(\frac{\beta\omega}{2}\right)\cos\omega(t_{1}-t_{2})+i\sin\omega(t_{1}-t_{2})\right]\right\} \protect\\
\times & \exp\left\{ \frac{\Delta_{ab}\Delta_{cb}}{\omega^{2}}\left[-\coth\left(\frac{\beta\omega}{2}\right)\cos\omega(t_{1}-t_{3})+i\sin\omega(t_{1}-t_{3})\right]\right\} \protect\\
\times & \exp\left\{ \frac{\Delta_{ab}\Delta_{ba}}{\omega^{2}}\left[-\coth\left(\frac{\beta\omega}{2}\right)\cos\omega(t_{1}-t_{4})+i\sin\omega(t_{1}-t_{4})\right]\right\} \protect\\
\times & \exp\left\{ \frac{\Delta_{bc}\Delta_{cb}}{\omega^{2}}\left[-\coth\left(\frac{\beta\omega}{2}\right)\cos\omega(t_{2}-t_{3})+i\sin\omega(t_{2}-t_{3})\right]\right\} \protect\\
\times & \exp\left\{ \frac{\Delta_{bc}\Delta_{ba}}{\omega^{2}}\left[-\coth\left(\frac{\beta\omega}{2}\right)\cos\omega(t_{2}-t_{4})+i\sin\omega(t_{2}-t_{4})\right]\right\} \protect\\
\times & \exp\left\{ \frac{\Delta_{cb}\Delta_{ba}}{\omega^{2}}\left[-\coth\left(\frac{\beta\omega}{2}\right)\cos\omega(t_{3}-t_{4})+i\sin\omega(t_{3}-t_{4})\right]\right\} \protect\\
\times & \exp\left\{ -\coth\left(\frac{\beta\omega}{2}\right)\frac{\Delta_{ab}^{2}+\Delta_{bc}^{2}+\Delta_{cb}^{2}+\Delta_{ba}^{2}}{2\omega^{2}}\right\} \protect\\
\times & \exp2i\left\{ \frac{s_{a}\Delta_{ab}}{\omega^{2}}\sin\omega t_{1}+\frac{s_{a}\Delta_{bc}}{\omega^{2}}\sin\omega t_{2}+\frac{s_{a}\Delta_{cb}}{\omega^{2}}\sin\omega t_{3}+\frac{s_{a}\Delta_{ba}}{\omega^{2}}\sin\omega t_{3}\right\} 
\end{align*}
}

\paragraph{Change of Variables}

To numerically evaluate the integral $\int_{0}^{\beta}dt_{1}\int_{0}^{t_{1}}dt_{2}\int_{0}^{t_{2}}dt_{3}\int_{0}^{t_{3}}dt_{4}$,
we use the following transformation
\[
\begin{cases}
t_{1}=y_{1} & \int_{0}^{\beta}dy_{1}\int_{0}^{y_{1}}dt_{2}\int_{0}^{t_{2}}dt_{3}\int_{0}^{t_{3}}dt_{4}\\
t_{2}=y_{2}\frac{y_{1}}{\beta} & \int_{0}^{\beta}dy_{1}\int_{0}^{\beta}\frac{y_{1}}{\beta}dy_{2}\int_{0}^{y_{2}\frac{y_{1}}{\beta}}dt_{3}\int_{0}^{t_{3}}dt_{4}\\
t_{3}=y_{3}\frac{y_{2}y_{1}}{\beta^{2}} & \int_{0}^{\beta}dy_{1}\int_{0}^{\beta}\frac{y_{1}}{\beta}dy_{2}\int_{0}^{\beta}\frac{y_{2}y_{1}}{\beta^{2}}dy_{3}\int_{0}^{y_{3}\frac{y_{2}y_{1}}{\beta^{2}}}dt_{4}\\
t_{4}=y_{4}\frac{y_{3}y_{2}y_{1}}{\beta^{3}} & \int_{0}^{\beta}dy_{1}\int_{0}^{\beta}\frac{y_{1}}{\beta}dy_{2}\int_{0}^{\beta}\frac{y_{2}y_{1}}{\beta^{2}}dy_{3}\int_{0}^{\beta}\frac{y_{3}y_{2}y_{1}}{\beta^{3}}dy_{4}
\end{cases}
\]
In summary, the integral $\int_{0}^{\beta}dt_{1}\int_{0}^{t_{1}}dt_{2}\cdots\int_{0}^{t_{n-1}}dt_{n}$
can be transformed to
\[
\int_{0}^{\beta}dy_{1}\int_{0}^{\beta}dy_{2}\cdots\int_{0}^{\beta}dy_{n}\left(\frac{y_{1}}{\beta}\right)^{n-1}\left(\frac{y_{2}}{\beta}\right)^{n-2}\cdots\left(\frac{y_{n-1}}{\beta}\right)^{1}
\]
To transform the integral into a unit volume, we can use
\[
\begin{cases}
t_{1}=\beta x_{1} & \int_{0}^{1}\beta dx_{1}\int_{0}^{\beta x_{1}}dt_{2}\int_{0}^{t_{2}}dt_{3}\int_{0}^{t_{3}}dt_{4}\\
t_{2}=\beta x_{2}x_{1} & \int_{0}^{1}\beta dx_{1}\int_{0}^{1}\beta x_{1}dx_{2}\int_{0}^{\beta x_{2}x_{1}}dt_{3}\int_{0}^{t_{3}}dt_{4}\\
t_{3}=\beta x_{3}x_{2}x_{1} & \int_{0}^{1}\beta dx_{1}\int_{0}^{1}\beta x_{1}dx_{2}\int_{0}^{1}\beta x_{2}x_{1}dx_{3}\int_{0}^{\beta x_{3}x_{2}x_{1}}dt_{4}\\
t_{4}=\beta x_{4}x_{3}x_{2}x_{1} & \int_{0}^{1}\beta dx_{1}\int_{0}^{1}\beta x_{1}dx_{2}\int_{0}^{1}\beta x_{2}x_{1}dx_{3}\int_{0}^{1}\beta x_{3}x_{2}x_{1}dt_{4}
\end{cases}
\]
In summary, the integral $\int_{0}^{\beta}dt_{1}\int_{0}^{t_{1}}dt_{2}\cdots\int_{0}^{t_{n-1}}dt_{n}$
can be transformed to
\[
\int_{0}^{1}dx_{1}\int_{0}^{1}dx_{2}\cdots\int_{0}^{1}dx_{n}\left(\beta x_{1}^{n-1}\right)\left(\beta x_{2}^{n-2}\right)\cdots\left(\beta x_{n-1}^{1}\right)\left(\beta x_{n}^{0}\right)
\]

\section{Evaluation of the term $e^{ih_{1}t_{1}}e^{-ih_{2}(t_{1}-t_{2})}e^{-ih_{1}t_{2}}$ }

Suppose $h_{1}=E_{1}+\sum_{n}\left[\omega_{n}a_{n}^{\dagger}a_{n}+c_{1,n}(a_{n}^{\dagger}+a_{n})+\frac{c_{1,n}^{2}}{\omega_{n}}\right]$
and $h_{2}=E_{2}+\sum_{n}\left[\omega_{n}a_{n}^{\dagger}a_{n}+c_{2,n}(a_{n}^{\dagger}+a_{n})+\frac{c_{2,n}^{2}}{\omega_{n}}\right]$
. Futher we absorb the reorganization energies into the sites energies:
$E_{1}\to E_{1}+\sum_{n}\frac{c_{1,n}^{2}}{\omega_{n}}$ and $E_{2}\to E_{2}+\sum_{n}\frac{c_{2,n}^{2}}{\omega_{n}}$.

Consider the case where only one mode exist. 
\begin{align}
 & e^{ih_{1}t_{1}}e^{-ih_{2}(t_{1}-t_{2})}e^{-ih_{1}t_{2}}\nonumber \\
= & e^{i(E_{1}-E_{2})(t_{1}-t_{2})}\times\nonumber \\
 & e^{i[\omega a^{\dagger}a+c_{1}(a^{\dagger}+a)]t_{1}}e^{-i[\omega a^{\dagger}a+c_{2}(a^{\dagger}+a)]t_{1}}\times\nonumber \\
 & e^{i[\omega a^{\dagger}a+c_{2}(a^{\dagger}+a)]t_{2}}e^{-i[\omega a^{\dagger}a+c_{1}(a^{\dagger}+a)]t_{2}}.\label{eq:a2}
\end{align}
Let the first two exponential factors in the equation above be $u(t_{1})=e^{i[\omega a^{\dagger}a+c_{1}(a^{\dagger}+a)]t_{1}}e^{-i[\omega a^{\dagger}a+c_{2}(a^{\dagger}+a)]t_{1}}$.
Set $b^{\dagger}=a^{\dagger}+\frac{c_{1}}{\omega}$, i.e., $a^{\dagger}=b^{\dagger}-\frac{c_{1}}{\omega}$
\begin{align*}
 & \partial_{t}u(t)\\
= & ie^{i[\omega a^{\dagger}a+c_{1}(a^{\dagger}+a)]t_{1}}[(c_{1}-c_{2})(a^{\dagger}+a)]e^{-i[\omega a^{\dagger}a+c_{1}(a^{\dagger}+a)]t_{1}}u(t_{1})\\
= & ie^{i[\omega a^{\dagger}a+c_{1}(a^{\dagger}+a)+\frac{c_{1}^{2}}{\omega}]t_{1}}[(c_{1}-c_{2})(a^{\dagger}+a)]e^{-i[\omega a^{\dagger}a+c_{1}(a^{\dagger}+a)+\frac{c_{1}^{2}}{\omega}]t_{1}}u(t_{1})\\
= & ie^{i\omega b^{\dagger}bt_{1}}[(c_{1}-c_{2})(b^{\dagger}+b-\frac{2c_{1}}{\omega})]e^{-i\omega b^{\dagger}bt_{1}}u(t_{1})\\
= & i(c_{1}-c_{2})(b^{\dagger}e^{i\omega t_{1}}+be^{-i\omega t_{1}}-\frac{2c_{1}}{\omega})u(t_{1})\\
= & i(c_{1}-c_{2})\left[\left(a^{\dagger}+\frac{c_{1}}{\omega}\right)e^{i\omega t_{1}}+\left(a+\frac{c_{1}}{\omega}\right)e^{-i\omega t_{1}}-\frac{2c_{1}}{\omega}\right]u(t_{1})\\
= & i(c_{1}-c_{2})\left[a^{\dagger}e^{i\omega t_{1}}+ae^{-i\omega t_{1}}+\frac{c_{1}}{\omega}(e^{i\omega t_{1}}+e^{-i\omega t_{1}})-\frac{2c_{1}}{\omega}\right]u(t_{1})\\
= & i(c_{1}-c_{2})\left[a^{\dagger}e^{i\omega t_{1}}+ae^{-i\omega t_{1}}+\frac{2c_{1}}{\omega}\cos(\omega t_{1})-\frac{2c_{1}}{\omega}\right]u(t_{1})
\end{align*}
To obtain a single-exponential expression for $u(t)$, we let $A(t)=i(c_{1}-c_{2})\left[a^{\dagger}e^{i\omega t_{1}}+ae^{-i\omega t_{1}}+\frac{2c_{1}}{\omega}\cos(\omega t_{1})-\frac{2c_{1}}{\omega}\right]$
and write $u(t)$ as a product
\begin{align*}
u(t)= & \lim_{N\to\infty}e^{A(\frac{Nt}{N})\frac{t}{N}}\cdots e^{A(\frac{2t}{N})\frac{t}{N}}e^{A(\frac{t}{N})\frac{t}{N}}\\
= & \lim_{N\to\infty}e^{\sum_{n=1}^{N}A(\frac{nt}{N})\frac{t}{N}}e^{\sum_{k>l}\frac{[A(\frac{kt}{N}),A(\frac{lt}{N})]}{2}\left(\frac{t^{2}}{N^{2}}\right)}\\
= & e^{\int_{0}^{t}d\tau A(\tau)}e^{\int_{0}^{t}d\tau_{1}\int_{0}^{\tau_{1}}d\tau_{2}\frac{[A(\tau_{1}),A(\tau_{2})]}{2}}
\end{align*}
since $e^{\sum_{n}^{N}A_{n}}=e^{A_{N}}\cdots e^{A_{1}}e^{-\sum_{i>j}[A_{i},A_{j}]}$
if $[A_{i},A_{j}]$ is a number. Further, 
\begin{align*}
\frac{[A(\tau_{1}),A(\tau_{2})]}{2} & =\frac{-(c_{1}-c_{2})^{2}}{2}2i\sin[\omega(\tau_{2}-\tau_{1})]
\end{align*}
\[
\int_{0}^{t}d\tau_{1}\int_{0}^{\tau_{1}}d\tau_{2}\frac{[A(\tau_{1}),A(\tau_{2})]}{2}=-i(c_{1}-c_{2})^{2}\frac{\sin\omega t-\omega t}{\omega^{2}}.
\]
Thus, with $\Delta=(c_{1}-c_{2})$,

\begin{align*}
u(t_{1})= & e^{i\int_{0}^{t_{1}}\Delta(a^{\dagger}e^{i\omega\tau}+ae^{-i\omega\tau})d\tau}e^{i\int_{0}^{t_{1}}d\tau\Delta\left(\frac{2c_{1}}{\omega}\cos(\omega t_{1})-\frac{2c_{1}}{\omega}\right)}e^{\int_{0}^{t_{1}}d\tau_{1}\int_{0}^{\tau_{1}}d\tau_{2}\frac{[A(\tau_{1}),A(\tau_{2})]}{2}}\\
= & e^{i\int_{0}^{t_{1}}\Delta(a^{\dagger}e^{i\omega\tau}+ae^{-i\omega\tau})d\tau}e^{i\Delta\left(\frac{2c_{1}\sin(\omega t_{1})-2c_{1}\omega t_{1}}{\omega^{2}}\right)}e^{-i\Delta^{2}\frac{\sin\omega t_{1}-\omega t_{1}}{\omega^{2}}}\\
= & e^{i\int_{0}^{t_{1}}\Delta(a^{\dagger}e^{i\omega\tau}+ae^{-i\omega\tau})d\tau}e^{i\Delta\left(\frac{(2c_{1}-\Delta)\sin(\omega t_{1})-\text{(}2c_{1}-\Delta)\omega t_{1}}{\omega^{2}}\right)}\\
= & e^{i\int_{0}^{t_{1}}\Delta(a^{\dagger}e^{i\omega\tau}+ae^{-i\omega\tau})d\tau}e^{i\Delta\left(\frac{(c_{1}+c_{2})\sin(\omega t_{1})-\text{(}c_{1}+c_{2})\omega t_{1}}{\omega^{2}}\right)}
\end{align*}
and \ref{eq:a2} becomes (without the first trivial factor)
\begin{align*}
 & e^{i\int_{0}^{t_{1}}\Delta(a^{\dagger}e^{i\omega\tau}+ae^{-i\omega\tau})d\tau}e^{-i\int_{0}^{t_{2}}\Delta(a^{\dagger}e^{i\omega\tau}+ae^{-i\omega\tau})d\tau}\\
 & e^{i\Delta\left(\frac{(c_{1}+c_{2})\sin\omega t_{1}-\text{(}c_{1}+c_{2})\omega t_{1}}{\omega^{2}}\right)}e^{-i\Delta\left(\frac{(c_{1}+c_{2})\sin\omega t_{2}-\text{(}c_{1}+c_{2})\omega t_{2}}{\omega^{2}}\right)}\\
= & e^{i\int_{t_{2}}^{t_{1}}\Delta(a^{\dagger}e^{i\omega\tau}+ae^{-i\omega\tau})d\tau}e^{\left[\int_{0}^{t_{1}}\Delta(a^{\dagger}e^{i\omega\tau}+ae^{-i\omega\tau})d\tau,\int_{0}^{t_{2}}\Delta(a^{\dagger}e^{i\omega\tau}+ae^{-i\omega\tau})d\tau\right]/2}\\
 & {\color{red}e^{i\Delta\left(\frac{(c_{1}+c_{2})(\sin\omega t_{1}-\sin\omega t_{2})-\text{(}c_{1}+c_{2})\omega(t_{1}-t_{2})}{\omega^{2}}\right)}}
\end{align*}
because $e^{A}e^{B}=e^{A+B}e^{[A,B]/2}$.

We then calculate the commutator $\left[i\int_{0}^{t_{1}}\Delta(a^{\dagger}e^{i\omega\tau}+ae^{-i\omega\tau})d\tau,-i\int_{0}^{t_{2}}\Delta(a^{\dagger}e^{i\omega\tau}+ae^{-i\omega\tau})d\tau\right]/2$:
\begin{align*}
 & \left[\int_{0}^{t_{1}}\Delta(a^{\dagger}e^{i\omega\tau}+ae^{-i\omega\tau})d\tau,\int_{0}^{t_{2}}\Delta(a^{\dagger}e^{i\omega\tau}+ae^{-i\omega\tau})d\tau\right]/2\\
= & \frac{\Delta^{2}}{2}\int_{0}^{t_{1}}d\tau_{1}\int_{0}^{t_{2}}d\tau_{2}[a^{\dagger}e^{i\omega\tau_{1}}+ae^{-i\omega\tau_{1}},a^{\dagger}e^{i\omega\tau_{2}}+ae^{-i\omega\tau_{2}}]\\
= & \frac{\Delta^{2}}{2}\int_{0}^{t_{1}}d\tau_{1}\int_{0}^{t_{2}}d\tau_{2}([a,a^{\dagger}]e^{-i\omega(\tau_{1}-\tau_{2})}+[a^{\dagger},a]e^{i\omega(\tau_{1}-\tau_{2})})\\
= & \frac{\Delta^{2}}{2}\int_{0}^{t_{1}}d\tau_{1}\int_{0}^{t_{2}}d\tau_{2}(e^{-i\omega(\tau_{1}-\tau_{2})}-e^{i\omega(\tau_{1}-\tau_{2})})\\
= & \frac{\Delta^{2}}{2}\int_{0}^{t_{1}}d\tau_{1}\int_{0}^{t_{2}}d\tau_{2}(2i\sin[\omega(\tau_{2}-\tau_{1})])\\
= & i\Delta^{2}\frac{-\sin\omega(t_{1}-t_{2})+\left(\sin\omega t_{1}-\sin\omega t_{2}\right)}{\omega^{2}}.
\end{align*}
Now \ref{eq:a2} is
\begin{align}
 & e^{i(E_{1}-E_{2})(t_{1}-t_{2})}\nonumber \\
 & e^{i\int_{t_{2}}^{t_{1}}\Delta(a^{\dagger}e^{i\omega\tau}+ae^{-i\omega\tau})d\tau}\nonumber \\
 & e^{i\Delta^{2}\frac{-\sin\left(\omega\left(t_{1}-t_{2}\right)\right)}{\omega^{2}}+i\Delta\frac{\left(\sin\omega t_{1}-\sin\omega t_{2}\right)\left(c_{1}+c_{2}+\Delta\right)}{\omega^{2}}-i\Delta\frac{(c_{1}+c_{2})(t_{1}-t_{2})}{\omega}}\nonumber \\
= & e^{i\left[\left(E_{1}-\frac{c_{1}^{2}}{\omega}\right)-\left(E_{2}-\frac{c_{2}^{2}}{\omega}\right)\right](t_{1}-t_{2})}\nonumber \\
 & e^{i\int_{t_{2}}^{t_{1}}\Delta(a^{\dagger}e^{i\omega\tau}+ae^{-i\omega\tau})d\tau}\nonumber \\
 & {\color{red}e^{i\Delta^{2}\frac{-\sin\omega(t_{1}-t_{2})}{\omega^{2}}+i\Delta\frac{\left(\sin\omega t_{1}-\sin\omega t_{2}\right)\left(2c_{1}\right)}{\omega^{2}}}}\label{eq:a3}
\end{align}
Then we calculate the operator part:
\begin{align*}
 & e^{i\int_{t_{2}}^{t_{1}}\Delta(a^{\dagger}e^{i\omega\tau}+ae^{-i\omega\tau})d\tau}\\
= & e^{\Delta\left[a^{\dagger}\left(e^{i\omega t_{1}}-e^{i\omega t_{2}}\right)-a\left(e^{-i\omega t_{1}}-e^{-i\omega t_{2}}\right)\right]/\omega}.
\end{align*}
The average value $\mathrm{Tr}(\rho e^{-\Delta\left[a^{\dagger}\left(e^{i\omega t_{1}}-e^{i\omega t_{2}}\right)-a\left(e^{-i\omega t_{1}}-e^{-i\omega t_{2}}\right)\right]/\omega})$
is
\begin{align*}
 & \mathrm{Tr}(\rho e^{\Delta\left[a^{\dagger}\left(e^{i\omega t_{1}}-e^{i\omega t_{2}}\right)-a\left(e^{-i\omega t_{1}}-e^{-i\omega t_{2}}\right)\right]/\omega})\\
= & e^{\mathrm{Tr}(\rho\frac{1}{2}\Delta^{2}\left[a^{\dagger}\left(e^{i\omega t_{1}}-e^{i\omega t_{2}}\right)-a\left(e^{-i\omega t_{1}}-e^{-i\omega t_{2}}\right)\right]^{2}/\omega^{2})}\\
= & e^{\frac{-\Delta^{2}}{2\omega^{2}}\left\langle \left(a^{\dagger}a+aa^{\dagger}\right)\left(1-e^{i\omega(t_{1}-t_{2})}-e^{i\omega(t_{2}-t_{1})}+1\right)\right\rangle }\\
= & e^{\frac{-\Delta^{2}}{2\omega^{2}}\left\langle \left(a^{\dagger}a+aa^{\dagger}\right)\left(2-2\cos\omega(t_{1}-t_{2})\right)\right\rangle }\\
= & e^{\frac{-\Delta^{2}}{\omega^{2}}\left[\left(2n(\beta\omega)+1\right)\left(1-\cos\omega(t_{1}-t_{2})\right)\right]}\\
= & e^{\frac{-\Delta^{2}}{\omega^{2}}\left[\left(\coth(\beta\omega/2)\right)\left(1-\cos\omega(t_{1}-t_{2})\right)\right]}
\end{align*}
Summarizing, we have the time-dependent rate $(\tau=t_{1}-t_{2})$:
\begin{align*}
k(t_{1}) & =2\mathrm{Re}\int_{0}^{t_{1}}dt_{2}e^{i\left[\left(E_{1}-\frac{c_{1}^{2}}{\omega}\right)-\left(E_{2}-\frac{c_{2}^{2}}{\omega}\right)\right](t_{1}-t_{2})}e^{-\frac{\Delta^{2}}{\omega^{2}}\left[\left(\coth(\beta\omega/2)\right)\left(1-\cos\omega(t_{1}-t_{2})\right)+i\frac{\sin\left(\omega\left(t_{1}-t_{2}\right)\right)}{\omega^{2}}\right]+i\Delta\frac{\left(2c_{1}\right)\left(\sin\omega t_{1}-\sin\omega t_{2}\right)}{\omega^{2}}}\\
 & =2\mathrm{Re}\int_{0}^{t_{1}}d\tau e^{i\left[\left(E_{1}-\frac{c_{1}^{2}}{\omega}\right)-\left(E_{2}-\frac{c_{2}^{2}}{\omega}\right)\right]\tau}e^{-\frac{\Delta^{2}}{\omega^{2}}\left[\left(\coth(\beta\omega/2)\right)\left(1-\cos\omega\tau\right)+i\frac{\sin\omega\tau}{\omega^{2}}\right]+i\Delta\frac{2c_{1}\left(\sin\omega t_{1}+\sin\omega(\tau-t_{1})\right)}{\omega^{2}}}
\end{align*}
which has similar form as Eq (B. 18) in \href{http://dx.doi.org/10.1021/acs.jctc.6b00236}{dx.doi.org/10.1021/acs.jctc.6b00236}.

At this point, we conclude by writing the expression of $e^{ih_{1}t_{1}}e^{-ih_{2}(t_{1}-t_{2})}e^{-ih_{1}t_{2}}$:
\begin{equation}
=e^{i\left[\left(E_{1}-\frac{c_{1}^{2}}{\omega}\right)-\left(E_{2}-\frac{c_{2}^{2}}{\omega}\right)\right](t_{1}-t_{2})}e^{-i\Delta\left[a^{\dagger}\frac{e^{i\omega t_{1}}-e^{i\omega t_{2}}}{\omega}-a\frac{e^{-i\omega t_{1}}-e^{-i\omega t_{2}}}{\omega}\right]}e^{i\Delta^{2}\frac{-\sin\left(\omega\left(t_{1}-t_{2}\right)\right)}{\omega^{2}}+i\Delta\frac{\left(\sin\omega t_{1}-\sin\omega t_{2}\right)\left(2c_{1}\right)}{\omega^{2}}}.\label{eq:exp-prod1}
\end{equation}
and the expression of $e^{ih_{1}t}e^{-ih_{2}t}:$
\begin{equation}
=e^{i(E_{1}-E_{2})t}e^{i\int_{0}^{t}(c_{1}-c_{2})(a^{\dagger}e^{i\omega\tau}+ae^{-i\omega\tau})d\tau}e^{i\left(\frac{(c_{1}^{2}-c_{2}^{2})\sin(\omega t)-\text{(}c_{1}^{2}-c_{2}^{2})\omega t}{\omega^{2}}\right)}\label{eq:exp-prod2}
\end{equation}
Another useful identity is for $e^{i\int_{0}^{t_{1}}(c_{1}-c_{2})(a^{\dagger}e^{i\omega\tau}+ae^{-i\omega\tau})d\tau}e^{i\int_{0}^{t_{2}}(\tilde{c}_{1}-\tilde{c}_{2})(a^{\dagger}e^{i\omega\tau}+ae^{-i\omega\tau})d\tau}$
\begin{equation}
=e^{i\left[\int_{0}^{t_{1}}(c_{1}-c_{2})(a^{\dagger}e^{i\omega\tau}+ae^{-i\omega\tau})d\tau+\int_{0}^{t_{2}}(\tilde{c}_{1}-\tilde{c}_{2})(a^{\dagger}e^{i\omega\tau}+ae^{-i\omega\tau})d\tau\right]}e^{i(c_{1}-c_{2})(\tilde{c}_{1}-\tilde{c}_{2})\frac{\sin\omega(t_{1}-t_{2})-(\sin\omega t_{1}-\sin\omega t_{2})}{\omega^{2}}.}\label{eq:exp-prod3}
\end{equation}

\section{Polaron-transformation derivation of Fermi's golden rule}

Another way to derive Fermi's golden rule rate is to use polaron transformation
to remove the diagonal linear vibronic couplings (LVC). We start from
the usual LVC Hamiltonian
\begin{align*}
H_{0} & =\left(E_{1}+\omega a^{\dagger}a+c_{1}(a^{\dagger}+a)\right)\left|1\right\rangle \left\langle 1\right|\\
 & +\left(E_{2}+\omega a^{\dagger}a+c_{2}(a^{\dagger}+a)\right)\left|2\right\rangle \left\langle 2\right|\\
 & +\left(E_{3}+\omega a^{\dagger}a+c_{3}(a^{\dagger}+a)\right)\left|3\right\rangle \left\langle 3\right|\\
V & =g_{12}\left|1\right\rangle \left\langle 2\right|+g_{13}\left|1\right\rangle \left\langle 3\right|+g_{23}\left|2\right\rangle \left\langle 3\right|+h.c.
\end{align*}
The polaron transformation uses the unitary operator 
\[
U_{p}=e^{\sum_{n}(a^{\dagger}-a)\frac{c_{n}}{\omega}\left|n\right\rangle \left\langle n\right|}.
\]
Noticing $e^{\lambda(a^{\dagger}-a)}ae^{-\lambda(a^{\dagger}-a)}=a-\lambda$,
and applying $U_{p}$ to $H_{0}$:
\begin{align*}
\tilde{H}_{0}= & U_{p}H_{0}U_{p}^{\dagger}\\
= & \left(E_{1}+\omega(a^{\dagger}-\frac{c_{1}}{\omega})(a-\frac{c_{1}}{\omega})+c_{1}(a^{\dagger}+a-2\frac{c_{1}}{\omega})\right)\left|1\right\rangle \left\langle 1\right|\\
+ & \left(E_{2}+\omega(a^{\dagger}a-\frac{c_{2}}{\omega}(a^{\dagger}+a)+\frac{c_{2}^{2}}{\omega^{2}})+c_{2}(a^{\dagger}+a-2\frac{c_{2}}{\omega})\right)\left|2\right\rangle \left\langle 2\right|\\
+ & \left(E_{3}+\omega a^{\dagger}a-\frac{c_{3}^{2}}{\omega}\right)\left|2\right\rangle \left\langle 2\right|\\
= & \sum_{n}\left(E_{n}+\omega a^{\dagger}a-c_{n}^{2}/\omega\right)\left|n\right\rangle \left\langle n\right|.
\end{align*}
Applying $U_{p}$ to $V$:
\begin{align*}
\tilde{V}= & U_{p}VU_{p}^{\dagger}\\
= & g_{12}\left|1\right\rangle \left\langle 2\right|e^{(a^{\dagger}-a)(c_{1}-c_{2})/\omega}\\
+ & g_{13}\left|1\right\rangle \left\langle 3\right|e^{(a^{\dagger}-a)(c_{1}-c_{3})/\omega}\\
+ & g_{23}\left|2\right\rangle \left\langle 3\right|e^{(a^{\dagger}-a)(c_{2}-c_{3})/\omega}+h.c.
\end{align*}
To use the perturbation theory, first note
\begin{align*}
\tilde{V}_{I}(t) & =e^{i\tilde{H}_{0}t}\tilde{V}e^{-i\tilde{H}_{0}t}\\
 & =g_{12}\left|1\right\rangle \left\langle 2\right|e^{i\left((E_{1}-\frac{c_{1}^{2}}{\omega})-(E_{2}-\frac{c_{2}^{2}}{\omega})\right)t}e^{(a^{\dagger}e^{i\omega t}-ae^{-i\omega t})(c_{1}-c_{2})/\omega}\\
 & +g_{12}^{*}\left|2\right\rangle \left\langle 1\right|e^{-i\left((E_{1}-\frac{c_{1}^{2}}{\omega})-(E_{2}-\frac{c_{2}^{2}}{\omega})\right)t}e^{-(a^{\dagger}e^{i\omega t}-ae^{-i\omega t})(c_{1}-c_{2})/\omega}\\
 & +g_{13}\left|1\right\rangle \left\langle 3\right|e^{i\left((E_{1}-\frac{c_{1}^{2}}{\omega})-(E_{3}-\frac{c_{3}^{2}}{\omega})\right)t}e^{(a^{\dagger}e^{i\omega t}-ae^{-i\omega t})(c_{1}-c_{3})/\omega}\\
 & +g_{13}^{*}\left|3\right\rangle \left\langle 1\right|e^{-i\left((E_{1}-\frac{c_{1}^{2}}{\omega})-(E_{3}-\frac{c_{3}^{2}}{\omega})\right)t}e^{-(a^{\dagger}e^{i\omega t}-ae^{-i\omega t})(c_{1}-c_{3})/\omega}\\
 & +g_{23}\left|2\right\rangle \left\langle 3\right|e^{i\left((E_{2}-\frac{c_{2}^{2}}{\omega})-(E_{3}-\frac{c_{3}^{2}}{\omega})\right)t}e^{(a^{\dagger}e^{i\omega t}-ae^{-i\omega t})(c_{2}-c_{3})/\omega}\\
 & +g_{13}^{*}\left|3\right\rangle \left\langle 2\right|e^{-i\left((E_{2}-\frac{c_{2}^{2}}{\omega})-(E_{3}-\frac{c_{3}^{2}}{\omega})\right)t}e^{-(a^{\dagger}e^{i\omega t}-ae^{-i\omega t})(c_{2}-c_{3})/\omega}.
\end{align*}
Then we consider, as we did in Section 1, the population on $\left|1\right\rangle $
and let the initial nuclear density matrix be $\rho_{n}=e^{\beta a^{\dagger}a}/\mathcal{Z}$.
\begin{align*}
P_{a}(t) & =\mathrm{Tr}\left[\rho_{n}\left|1\right\rangle \left\langle 1\right|\underbrace{U_{p}^{\dagger}\hat{U}_{I}^{\dagger}(t)e^{iH_{0}t}U_{p}}_{U_{S}^{\dagger}(t)}\left|1\right\rangle \left\langle 1\right|\underbrace{U_{p}^{\dagger}e^{-iH_{0}t}\hat{U}_{I}(t)U_{p}}_{U_{S}(t)}\right]\\
 & =\mathrm{Tr}_{n}\left[\rho_{n}\left(e^{-(a^{\dagger}-a)\frac{c_{1}}{\omega}}\left\langle 1\right|\hat{U}_{I}^{\dagger}(t)\left|1\right\rangle e^{(a^{\dagger}-a)\frac{c_{1}}{\omega}}\right)\left(e^{-(a^{\dagger}-a)\frac{c_{1}}{\omega}}\left\langle 1\right|\hat{U}_{I}^{\dagger}(t)\left|1\right\rangle e^{(a^{\dagger}-a)\frac{c_{1}}{\omega}}\right)\right]\\
\\
 & \approx1-2\mathrm{Re}\int_{0}^{t}dt_{1}\int_{0}^{t_{1}}dt_{2}\mathrm{Tr}_{n}\left[\rho_{n}e^{-(a^{\dagger}-a)\frac{c_{1}}{\omega}}\left\langle 1\right|\tilde{V}_{I}(t_{1})\tilde{V}_{I}(t_{2})\left|1\right\rangle e^{(a^{\dagger}-a)\frac{c_{1}}{\omega}}\right]
\end{align*}
We consider the quantity
\begin{align*}
 & \left\langle 1\right|\tilde{V}(t_{1})\tilde{V}(t_{2})\left|1\right\rangle \\
= & g_{12}^{2}\exp\left[(a^{\dagger}e^{i\omega t_{1}}-ae^{-i\omega t_{1}})(c_{1}-c_{2})/\omega\right]\exp\left[-(a^{\dagger}e^{i\omega t_{2}}-ae^{-i\omega t_{2}})(c_{1}-c_{2})/\omega\right]\\
+ & g_{13}^{2}\exp\left[(a^{\dagger}e^{i\omega t_{1}}-ae^{-i\omega t_{1}})(c_{1}-c_{3})/\omega\right]\exp\left[-(a^{\dagger}e^{i\omega t_{2}}-ae^{-i\omega t_{2}})(c_{1}-c_{3})/\omega\right]
\end{align*}
Before averaging over the initial nuclear condition $\rho=e^{-\beta a^{\dagger}a}/\mathcal{Z}$
we need to calculate
\begin{align*}
 & \exp\left[(a^{\dagger}e^{i\omega t_{1}}-ae^{-i\omega t_{1}})(c_{1}-c_{2})/\omega\right]\exp\left[-(a^{\dagger}e^{i\omega t_{2}}-ae^{-i\omega t_{2}})(c_{1}-c_{2})/\omega\right]\\
= & \exp\left[\frac{c_{1}-c_{2}}{\omega}\left(a^{\dagger}(e^{i\omega t_{1}}-e^{i\omega t_{2}})-a(e^{-i\omega t_{1}}-e^{-i\omega t_{2}})\right)\right]\\
\times & \exp\left(\frac{(c_{1}-c_{2})^{2}}{\omega^{2}}\left[a^{\dagger}e^{i\omega t_{1}}-ae^{-i\omega t_{1}},-(a^{\dagger}e^{i\omega t_{2}}-ae^{-i\omega t_{2}})\right]/2\right)\\
= & \exp\left[\frac{c_{1}-c_{2}}{\omega}\left(a^{\dagger}(e^{i\omega t_{1}}-e^{i\omega t_{2}})-a(e^{-i\omega t_{1}}-e^{-i\omega t_{2}})\right)\right]e^{-i\frac{(c_{1}-c_{2})^{2}}{\omega^{2}}\sin\omega(t_{1}-t_{2})}
\end{align*}
and rotate it by sandwiching it with $e^{-(a^{\dagger}-a)\frac{c_{1}}{\omega}}(\cdot)e^{(a^{\dagger}-a)\frac{c_{1}}{\omega}}$.
After rotation, the operator expression above is
\begin{align*}
 & e^{-(a^{\dagger}-a)\frac{c_{1}}{\omega}}\exp\left[\frac{c_{1}-c_{2}}{\omega}\left(a^{\dagger}(e^{i\omega t_{1}}-e^{i\omega t_{2}})-a(e^{-i\omega t_{1}}-e^{-i\omega t_{2}})\right)\right]e^{(a^{\dagger}-a)\frac{c_{1}}{\omega}}\\
= & \exp\left[\frac{c_{1}-c_{2}}{\omega}\left((a^{\dagger}+\frac{c_{1}}{\omega})(e^{i\omega t_{1}}-e^{i\omega t_{2}})-(a+\frac{c_{1}}{\omega})(e^{-i\omega t_{1}}-e^{-i\omega t_{2}})\right)\right].
\end{align*}
 The average value is
\begin{align*}
 & \mathrm{Tr}_{n}\left[\rho_{n}e^{-(a^{\dagger}-a)\frac{c_{1}}{\omega}}\left\langle 1\right|\tilde{V}_{I}(t_{1})\tilde{V}_{I}(t_{2})\left|1\right\rangle e^{(a^{\dagger}-a)\frac{c_{1}}{\omega}}\right]/e^{i\frac{(c_{1}-c_{2})^{2}}{\omega^{2}}\sin\omega(t_{1}-t_{2})}\\
= & \mathrm{Tr}_{n}\left\{ \rho_{n}\exp\left[\frac{c_{1}-c_{2}}{\omega}\left((a^{\dagger}+\frac{c_{1}}{\omega})(e^{i\omega t_{1}}-e^{i\omega t_{2}})-(a+\frac{c_{1}}{\omega})(e^{-i\omega t_{1}}-e^{-i\omega t_{2}})\right)\right]\right\} \\
= & \mathrm{Tr}_{n}\left\{ \rho_{n}\exp\left[\frac{c_{1}-c_{2}}{\omega}\left(a^{\dagger}(e^{i\omega t_{1}}-e^{i\omega t_{2}})-a(e^{-i\omega t_{1}}-e^{-i\omega t_{2}})\right)+\frac{c_{1}-c_{2}}{\omega}\frac{c_{1}}{\omega}\left(2i\sin\omega t_{1}-2i\sin\omega t_{2}\right)\right]\right\} \\
= & e^{\frac{-(c_{1}-c_{2})^{2}}{\omega^{2}}\left[\left(\coth(\beta\omega/2)\right)\left(1-\cos\omega(t_{1}-t_{2})\right)\right]}e^{2i\frac{(c_{1}-c_{2})c_{1}}{\omega^{2}}(\sin\omega t_{1}-\sin\omega t_{2})}
\end{align*}

This derivation is a bit simpler than that in Appendix A.

\section{Vibronic interactions in the rotated electronic basis}

Start from the original expression of the interaction term,
\begin{align*}
\tilde{V}_{I}(t) & =e^{i\tilde{H}_{0}t}\tilde{V}e^{-i\tilde{H}_{0}t}\\
 & =g_{ab}\left|a\right\rangle \left\langle b\right|e^{i\left((E_{a}-\frac{s_{a}^{2}}{\omega})-(E_{b}-\frac{s_{b}^{2}}{\omega})\right)t}e^{(a^{\dagger}e^{i\omega t}-ae^{-i\omega t})(s_{a}-s_{b})/\omega}\\
 & +g_{ba}\left|b\right\rangle \left\langle a\right|e^{-i\left((E_{a}-\frac{s_{a}^{2}}{\omega})-(E_{b}-\frac{s_{b}^{2}}{\omega})\right)t}e^{-(a^{\dagger}e^{i\omega t}-ae^{-i\omega t})(s_{a}-s_{b})/\omega}\\
 & +g_{ac}\left|a\right\rangle \left\langle c\right|e^{i\left((E_{a}-\frac{s_{a}^{2}}{\omega})-(E_{c}-\frac{s_{c}^{2}}{\omega})\right)t}e^{(a^{\dagger}e^{i\omega t}-ae^{-i\omega t})(s_{a}-s_{c})/\omega}\\
 & +g_{ca}\left|c\right\rangle \left\langle a\right|e^{-i\left((E_{a}-\frac{s_{a}^{2}}{\omega})-(E_{c}-\frac{s_{c}^{2}}{\omega})\right)t}e^{-(a^{\dagger}e^{i\omega t}-ae^{-i\omega t})(s_{a}-s_{c})/\omega}\\
 & +g_{bc}\left|b\right\rangle \left\langle c\right|e^{i\left((E_{b}-\frac{s_{b}^{2}}{\omega})-(E_{c}-\frac{s_{c}^{2}}{\omega})\right)t}e^{(a^{\dagger}e^{i\omega t}-ae^{-i\omega t})(s_{b}-s_{c})/\omega}\\
 & +g_{cb}\left|c\right\rangle \left\langle b\right|e^{-i\left((E_{b}-\frac{s_{b}^{2}}{\omega})-(E_{c}-\frac{s_{c}^{2}}{\omega})\right)t}e^{-(a^{\dagger}e^{i\omega t}-ae^{-i\omega t})(s_{b}-s_{c})/\omega}.
\end{align*}
we rotate the basis $\left|b\right\rangle =\frac{\left|\uparrow\right\rangle +\left|\downarrow\right\rangle }{\sqrt{2}}$
and $\left|c\right\rangle =\frac{\left|\uparrow\right\rangle -\left|\downarrow\right\rangle }{\sqrt{2}}$:

\begin{align*}
\tilde{V}_{I}(t) & =e^{i\tilde{H}_{0}t}\tilde{V}e^{-i\tilde{H}_{0}t}\\
 & =g_{ab}\left|a\right\rangle \frac{\left\langle \uparrow\right|+\left\langle \downarrow\right|}{\sqrt{2}}e^{i\left((E_{a}-\frac{s_{a}^{2}}{\omega})-(E_{b}-\frac{s_{b}^{2}}{\omega})\right)t}e^{(a^{\dagger}e^{i\omega t}-ae^{-i\omega t})(s_{a}-s_{b})/\omega}\\
 & +g_{ba}\frac{\left|\uparrow\right\rangle +\left|\downarrow\right\rangle }{\sqrt{2}}\left\langle a\right|e^{-i\left((E_{a}-\frac{s_{a}^{2}}{\omega})-(E_{b}-\frac{s_{b}^{2}}{\omega})\right)t}e^{-(a^{\dagger}e^{i\omega t}-ae^{-i\omega t})(s_{a}-s_{b})/\omega}\\
 & +g_{ac}\left|a\right\rangle \frac{\left\langle \uparrow\right|-\left\langle \downarrow\right|}{\sqrt{2}}e^{i\left((E_{a}-\frac{s_{a}^{2}}{\omega})-(E_{c}-\frac{s_{c}^{2}}{\omega})\right)t}e^{(a^{\dagger}e^{i\omega t}-ae^{-i\omega t})(s_{a}-s_{c})/\omega}\\
 & +g_{ca}\frac{\left|\uparrow\right\rangle -\left|\downarrow\right\rangle }{\sqrt{2}}\left\langle a\right|e^{-i\left((E_{a}-\frac{s_{a}^{2}}{\omega})-(E_{c}-\frac{s_{c}^{2}}{\omega})\right)t}e^{-(a^{\dagger}e^{i\omega t}-ae^{-i\omega t})(s_{a}-s_{c})/\omega}\\
 & +g_{bc}\frac{\left|\uparrow\right\rangle +\left|\downarrow\right\rangle }{\sqrt{2}}\frac{\left\langle \uparrow\right|-\left\langle \downarrow\right|}{\sqrt{2}}e^{i\left((E_{b}-\frac{s_{b}^{2}}{\omega})-(E_{c}-\frac{s_{c}^{2}}{\omega})\right)t}e^{(a^{\dagger}e^{i\omega t}-ae^{-i\omega t})(s_{b}-s_{c})/\omega}\\
 & +g_{cb}\frac{\left|\uparrow\right\rangle -\left|\downarrow\right\rangle }{\sqrt{2}}\frac{\left\langle \uparrow\right|+\left\langle \downarrow\right|}{\sqrt{2}}e^{-i\left((E_{b}-\frac{s_{b}^{2}}{\omega})-(E_{c}-\frac{s_{c}^{2}}{\omega})\right)t}e^{-(a^{\dagger}e^{i\omega t}-ae^{-i\omega t})(s_{b}-s_{c})/\omega}.
\end{align*}
Term for $\left|a\right\rangle \left\langle \uparrow\right|$:
\begin{align*}
 & g_{ab}\frac{\left|a\right\rangle \left\langle \uparrow\right|}{\sqrt{2}}e^{i\left((E_{a}-\frac{s_{a}^{2}}{\omega})-(E_{b}-\frac{s_{b}^{2}}{\omega})\right)t}e^{(a^{\dagger}e^{i\omega t}-ae^{-i\omega t})(s_{a}-s_{b})/\omega}\\
+ & g_{ac}\frac{\left|a\right\rangle \left\langle \uparrow\right|}{\sqrt{2}}e^{i\left((E_{a}-\frac{s_{a}^{2}}{\omega})-(E_{c}-\frac{s_{c}^{2}}{\omega})\right)t}e^{(a^{\dagger}e^{i\omega t}-ae^{-i\omega t})(s_{a}-s_{c})/\omega}
\end{align*}
Term for $\left|\uparrow\right\rangle \left\langle a\right|$
\begin{align*}
 & g_{ba}\frac{\left|\uparrow\right\rangle \left\langle a\right|}{\sqrt{2}}e^{-i\left((E_{a}-\frac{s_{a}^{2}}{\omega})-(E_{b}-\frac{s_{b}^{2}}{\omega})\right)t}e^{-(a^{\dagger}e^{i\omega t}-ae^{-i\omega t})(s_{a}-s_{b})/\omega}\\
+ & g_{ca}\frac{\left|\uparrow\right\rangle \left\langle a\right|}{\sqrt{2}}e^{-i\left((E_{a}-\frac{s_{a}^{2}}{\omega})-(E_{c}-\frac{s_{c}^{2}}{\omega})\right)t}e^{-(a^{\dagger}e^{i\omega t}-ae^{-i\omega t})(s_{a}-s_{c})/\omega}
\end{align*}
Term for $\left|a\right\rangle \left\langle \downarrow\right|$:
\begin{align*}
 & g_{ab}\frac{\left|a\right\rangle \left\langle \downarrow\right|}{\sqrt{2}}e^{i\left((E_{a}-\frac{s_{a}^{2}}{\omega})-(E_{b}-\frac{s_{b}^{2}}{\omega})\right)t}e^{(a^{\dagger}e^{i\omega t}-ae^{-i\omega t})(s_{a}-s_{b})/\omega}\\
- & g_{ac}\frac{\left|a\right\rangle \left\langle \downarrow\right|}{\sqrt{2}}e^{i\left((E_{a}-\frac{s_{a}^{2}}{\omega})-(E_{c}-\frac{s_{c}^{2}}{\omega})\right)t}e^{(a^{\dagger}e^{i\omega t}-ae^{-i\omega t})(s_{a}-s_{c})/\omega}
\end{align*}
Term for $\left|\downarrow\right\rangle \left\langle a\right|$:
\begin{align*}
 & g_{ba}\frac{\left|\downarrow\right\rangle \left\langle a\right|}{\sqrt{2}}e^{-i\left((E_{a}-\frac{s_{a}^{2}}{\omega})-(E_{b}-\frac{s_{b}^{2}}{\omega})\right)t}e^{-(a^{\dagger}e^{i\omega t}-ae^{-i\omega t})(s_{a}-s_{b})/\omega}\\
- & g_{ca}\frac{\left|\downarrow\right\rangle \left\langle a\right|}{\sqrt{2}}e^{-i\left((E_{a}-\frac{s_{a}^{2}}{\omega})-(E_{c}-\frac{s_{c}^{2}}{\omega})\right)t}e^{-(a^{\dagger}e^{i\omega t}-ae^{-i\omega t})(s_{a}-s_{c})/\omega}
\end{align*}
Term for $\left|\uparrow\right\rangle \left\langle \uparrow\right|$:
\begin{align*}
 & g_{bc}\frac{\left|\uparrow\right\rangle \left\langle \uparrow\right|}{2}e^{i\left((E_{b}-\frac{s_{b}^{2}}{\omega})-(E_{c}-\frac{s_{c}^{2}}{\omega})\right)t}e^{(a^{\dagger}e^{i\omega t}-ae^{-i\omega t})(s_{b}-s_{c})/\omega}\\
+ & g_{cb}\left|\uparrow\right\rangle \left\langle \uparrow\right|e^{-i\left((E_{b}-\frac{s_{b}^{2}}{\omega})-(E_{c}-\frac{s_{c}^{2}}{\omega})\right)t}e^{-(a^{\dagger}e^{i\omega t}-ae^{-i\omega t})(s_{b}-s_{c})/\omega}
\end{align*}
Term for $\left|\downarrow\right\rangle \left\langle \downarrow\right|$:
\begin{align*}
 & -g_{bc}\frac{\left|\downarrow\right\rangle \left\langle \downarrow\right|}{2}e^{i\left((E_{b}-\frac{s_{b}^{2}}{\omega})-(E_{c}-\frac{s_{c}^{2}}{\omega})\right)t}e^{(a^{\dagger}e^{i\omega t}-ae^{-i\omega t})(s_{b}-s_{c})/\omega}\\
 & -g_{cb}\frac{\left|\downarrow\right\rangle \left\langle \downarrow\right|}{2}e^{-i\left((E_{b}-\frac{s_{b}^{2}}{\omega})-(E_{c}-\frac{s_{c}^{2}}{\omega})\right)t}e^{-(a^{\dagger}e^{i\omega t}-ae^{-i\omega t})(s_{b}-s_{c})/\omega}
\end{align*}
Term for $\left|\uparrow\right\rangle \left\langle \downarrow\right|$:
\begin{align*}
 & -g_{bc}\frac{\left|\uparrow\right\rangle \left\langle \downarrow\right|}{2}e^{i\left((E_{b}-\frac{s_{b}^{2}}{\omega})-(E_{c}-\frac{s_{c}^{2}}{\omega})\right)t}e^{(a^{\dagger}e^{i\omega t}-ae^{-i\omega t})(s_{b}-s_{c})/\omega}\\
+ & g_{cb}\frac{\left|\uparrow\right\rangle \left\langle \downarrow\right|}{2}e^{-i\left((E_{b}-\frac{s_{b}^{2}}{\omega})-(E_{c}-\frac{s_{c}^{2}}{\omega})\right)t}e^{-(a^{\dagger}e^{i\omega t}-ae^{-i\omega t})(s_{b}-s_{c})/\omega}
\end{align*}
Term for $\left|\downarrow\right\rangle \left\langle \uparrow\right|$:
\begin{align*}
 & g_{bc}\frac{\left|\downarrow\right\rangle \left\langle \uparrow\right|}{\sqrt{2}}e^{i\left((E_{b}-\frac{s_{b}^{2}}{\omega})-(E_{c}-\frac{s_{c}^{2}}{\omega})\right)t}e^{(a^{\dagger}e^{i\omega t}-ae^{-i\omega t})(s_{b}-s_{c})/\omega}\\
- & g_{cb}\frac{\left|\downarrow\right\rangle \left\langle \uparrow\right|}{2}e^{-i\left((E_{b}-\frac{s_{b}^{2}}{\omega})-(E_{c}-\frac{s_{c}^{2}}{\omega})\right)t}e^{-(a^{\dagger}e^{i\omega t}-ae^{-i\omega t})(s_{b}-s_{c})/\omega}
\end{align*}

\section{Vibrnoic Hamiltonian in the rotated electronic basis:}

Start from the original expression of the Hamiltonian $H=H_{0}+V$
\[
\hat{H}_{0}=\hat{h}_{a}\left|a\right\rangle \left\langle a\right|+\hat{h}_{b}\left|b\right\rangle \left\langle b\right|+\hat{h}_{c}\left|c\right\rangle \left\langle c\right|
\]
\[
V=g_{ab}\left|a\right\rangle \left\langle b\right|+g_{ac}\left|a\right\rangle \left\langle c\right|+g_{bc}\left|b\right\rangle \left\langle c\right|+h.c.
\]
we rotate to the basis $\left|b\right\rangle =\frac{\left|\uparrow\right\rangle +\left|\downarrow\right\rangle }{\sqrt{2}}$
and $\left|c\right\rangle =\frac{\left|\uparrow\right\rangle -\left|\downarrow\right\rangle }{\sqrt{2}}$:

\begin{align*}
H_{0} & =\hat{h}_{a}\left|a\right\rangle \left\langle a\right|+\hat{h}_{b}\frac{(\left|\uparrow\right\rangle +\left|\downarrow\right\rangle )(\left\langle \uparrow\right|+\left\langle \downarrow\right|)}{2}+\hat{h}_{c}\frac{(\left|\uparrow\right\rangle -\left|\downarrow\right\rangle )(\left\langle \uparrow\right|-\left\langle \downarrow\right|)}{2}\\
 & =\hat{h}_{a}\left|a\right\rangle \left\langle a\right|+\hat{h}_{b}\frac{1}{2}(\left|\uparrow\right\rangle \left\langle \uparrow\right|+\left|\downarrow\right\rangle \left\langle \downarrow\right|+\left|\uparrow\right\rangle \left\langle \downarrow\right|+\left|\downarrow\right\rangle \left\langle \uparrow\right|)+\hat{h}_{c}\frac{1}{2}(\left|\uparrow\right\rangle \left\langle \uparrow\right|+\left|\downarrow\right\rangle \left\langle \downarrow\right|-\left|\uparrow\right\rangle \left\langle \downarrow\right|-\left|\downarrow\right\rangle \left\langle \uparrow\right|)\\
 & ={\color{red}\hat{h}_{a}\left|a\right\rangle \left\langle a\right|+\frac{\hat{h}_{b}+\hat{h}_{c}}{2}\left|\uparrow\right\rangle \left\langle \uparrow\right|+\frac{\hat{h}_{b}+\hat{h}_{c}}{2}\left|\downarrow\right\rangle \left\langle \downarrow\right|+\frac{\hat{h}_{b}-\hat{h}_{c}}{2}\left|\uparrow\right\rangle \left\langle \downarrow\right|+\frac{\hat{h}_{b}-\hat{h}_{c}}{2}\left|\downarrow\right\rangle \left\langle \uparrow\right|}
\end{align*}
and consider $g_{ac}=g_{bc}$,
\begin{align*}
V & =g_{ab}\left|a\right\rangle \left(\frac{\left\langle \uparrow\right|+\left\langle \downarrow\right|}{\sqrt{2}}\right)+g_{ac}\left|a\right\rangle \left(\frac{\left\langle \uparrow\right|-\left\langle \downarrow\right|}{\sqrt{2}}\right)+g_{bc}\left(\frac{\left|\uparrow\right\rangle +\left|\downarrow\right\rangle }{\sqrt{2}}\right)\left(\frac{\left\langle \uparrow\right|-\left\langle \downarrow\right|}{\sqrt{2}}\right)+h.c.\\
 & =\frac{g_{ab}}{\sqrt{2}}\left|a\right\rangle \left\langle \uparrow\right|+\frac{g_{ab}}{\sqrt{2}}\left|a\right\rangle \left\langle \downarrow\right|+\frac{g_{ac}}{\sqrt{2}}\left|a\right\rangle \left\langle \uparrow\right|-\frac{g_{ac}}{\sqrt{2}}\left|a\right\rangle \left\langle \downarrow\right|+\frac{g_{bc}}{2}\left(\left|\uparrow\right\rangle \left\langle \uparrow\right|-\left|\downarrow\right\rangle \left\langle \downarrow\right|-\left|\uparrow\right\rangle \left\langle \downarrow\right|+\left|\downarrow\right\rangle \left\langle \uparrow\right|\right)+h.c.\\
 & =\frac{g}{\sqrt{2}}\left|a\right\rangle \left\langle \uparrow\right|+{\color{brown}\frac{g}{\sqrt{2}}\left|a\right\rangle \left\langle \downarrow\right|}+\frac{g}{\sqrt{2}}\left|a\right\rangle \left\langle \uparrow\right|{\color{brown}-\frac{g}{\sqrt{2}}\left|a\right\rangle \left\langle \downarrow\right|}+\frac{g_{bc}}{2}\left(\left|\uparrow\right\rangle \left\langle \uparrow\right|-\left|\downarrow\right\rangle \left\langle \downarrow\right|-\left|\uparrow\right\rangle \left\langle \downarrow\right|+\left|\downarrow\right\rangle \left\langle \uparrow\right|\right)+h.c.\\
 & =\sqrt{2}g\left|a\right\rangle \left\langle \uparrow\right|+\frac{g_{bc}}{2}\left(\left|\uparrow\right\rangle \left\langle \uparrow\right|-\left|\downarrow\right\rangle \left\langle \downarrow\right|-\left|\uparrow\right\rangle \left\langle \downarrow\right|+\left|\downarrow\right\rangle \left\langle \uparrow\right|\right)+h.c.\\
 & ={\color{brown}{\color{red}\sqrt{2}g\left(\left|a\right\rangle \left\langle \uparrow\right|+\left|\uparrow\right\rangle \left\langle a\right|\right)+g_{bc}\left(\left|\uparrow\right\rangle \left\langle \uparrow\right|-\left|\downarrow\right\rangle \left\langle \downarrow\right|\right)}}
\end{align*}